\documentclass[11pt,a4paper]{article}
\usepackage[a4paper, left=1in, right=1in, top=1in, bottom=1in, includehead, includefoot, foot = 0pt, head=0pt]{geometry} 
\parskip=\medskipamount 

\usepackage{booktabs} 
\usepackage{enumitem} 

\usepackage{natbib}

\bibpunct{(}{)}{;}{a}{,}{,}
\newcommand{\biblist}{
\bibliographystyle{apalike}
\bibliography{References}  
}
\usepackage[hyphens,spaces,obeyspaces]{url}
\usepackage[pdftex,colorlinks,citecolor=blue,urlcolor=blue,linkcolor=red]{hyperref} 
\DeclareUrlCommand\doi{\def\UrlLeft##1\UrlRight{doi:\hspace{0.1cm} \href{http://dx.doi.org/##1}{##1}}\urlstyle{rm}}

\usepackage{graphicx} 
\usepackage[small,bf,hang]{caption}	

\usepackage{tabularx}
\usepackage[export]{adjustbox}

\usepackage{amssymb} 
\usepackage{amsmath} 
\usepackage{accents}
\usepackage{bbm} 
\usepackage{amsbsy} 
\usepackage{multirow} 
\usepackage[dvipsnames]{xcolor} 
\usepackage{mathtools}
\usepackage{eurosym}

\usepackage[ruled, lined, longend]{algorithm2e}
\usepackage{float}
\usepackage{rotating}
\usepackage[caption = false]{subfig}
\usepackage{pdflscape}
\usepackage[defaultcolor=black]{changes}

\usepackage{authblk} 

\usepackage{fancyhdr}
\DeclareMathOperator*{\argmax}{arg\,max}

\bibpunct[, ]{(}{)}{;}{a}{,}{,}

\usepackage{tikz,pgfplots}
\usetikzlibrary{fit, positioning,chains,shapes.arrows,fit}
\usetikzlibrary{shapes,arrows, arrows.meta}
\usetikzlibrary{decorations.pathreplacing}
\usetikzlibrary{calc} 
\usetikzlibrary{matrix} 
\usepackage{pgf}
\usetikzlibrary{decorations.text}
\tikzset{cross/.style={cross out, draw=black, minimum size=2*(#1-\pgflinewidth), inner sep=0pt, outer sep=0pt},
cross/.default={1pt}}
\usetikzlibrary{shapes.misc}
\usepackage{environ}
\makeatletter
\newsavebox{\measure@tikzpicture}
\NewEnviron{scaletikzpicturetowidth}[1]{%
  \def\tikz@width{#1}%
  \begin{lrbox}{\measure@tikzpicture}%
  \BODY
  \end{lrbox}%
  \pgfmathparse{#1/\wd\measure@tikzpicture}%
  \BODY
}
\makeatother
\tikzdeclarecoordinatesystem{page}{
    \parsecomma#1\endparsecomma
    \pgfpointanchor{current page}{north east}
    \pgf@xc=\pgf@x%
    \pgf@yc=\pgf@y%
    \pgfpointanchor{current page}{south west}
    \pgf@xb=\pgf@x%
    \pgf@yb=\pgf@y%
    \pgfmathparse{(\pgf@xc-\pgf@xb)/2.*\page@x+(\pgf@xc+\pgf@xb)/2.}
    \expandafter\pgf@x\expandafter=\pgfmathresult pt
    \pgfmathparse{(\pgf@yc-\pgf@yb)/2.*\page@y+(\pgf@yc+\pgf@yb)/2.}
    \expandafter\pgf@y\expandafter=\pgfmathresult pt
}
\makeatother
\usetikzlibrary{arrows,automata}
\usepackage{ragged2e}
\definecolor{arrowcolor}{RGB}{201,216,232}
\definecolor{circlecolor}{RGB}{79,129,189}
\colorlet{textcolor}{white}
\colorlet{bordercolor}{white}

\pgfdeclarelayer{background}
\pgfsetlayers{background,main}

\definecolor{airforceblue}{rgb}{0.36, 0.54, 0.66}
\definecolor{forestgreen}{rgb}{0.13, 0.55, 0.13}\definecolor{fulvous}{rgb}{0.86, 0.52, 0.0}
\definecolor{gray}{rgb}{0.5, 0.5, 0.5}
\definecolor{bistre}{rgb}{0.24, 0.17, 0.12}\definecolor{bostonuniversityred}{rgb}{0.8, 0.0, 0.0}
\definecolor{purpleheart}{rgb}{0.41, 0.21, 0.61}
\definecolor{lightsalmonpink}{rgb}{1.0, 0.6, 0.6}\definecolor{arrowcolor}{rgb}{0.92, 0.92, 0.92}
\tikzset{
inner/.style={
  on chain,
  circle,
  inner sep=4pt,
  fill=circlecolor,
  line width=1.5pt,
  draw=bordercolor,
  text width=1.2em,
  align=center,
  text height=1.25ex,
  text depth=0ex
},
on grid
}

\newcommand\drawarrow{

\node[on chain] (f) {};
\begin{pgfonlayer}{background}
\node[
  inner sep=10pt,
  single arrow,
  single arrow head extend=0.6cm,
  draw=none,
  fill=arrowcolor,
  fit= (c1) (f)
] (arrow) {};
\fill[white] 
  (arrow.before tail) -- (c1|-arrow.west) -- (arrow.after tail) -- cycle;
\end{pgfonlayer}
}

\numberwithin{equation}{section}

\begin{document}
\title{Reducing the dimensionality and granularity in hierarchical categorical variables}

\author[1,3,*]{Paul Wilsens}
\author[1,2,3,4]{Katrien Antonio}
\author[1,4,5]{Gerda Claeskens}
\affil[1]{Faculty of Economics and Business, KU Leuven, Belgium.}
\affil[2]{Faculty of Economics and Business, University of Amsterdam, The Netherlands.}
\affil[3]{Leuven Research Center on Insurance and Financial Risk Analysis, KU Leuven, Belgium.}
\affil[4]{Leuven Statistics Research Center, KU Leuven, Belgium.}
\affil[5]{Research Centre for Operations Research and Statistics, KU Leuven, Belgium.}
\affil[*]{Corresponding author: \href{mailto:paul.wilsens@kuleuven.be}{paul.wilsens@kuleuven.be}}
\date{\vspace{-1.25cm}} 
\maketitle
\thispagestyle{empty}

\begin{abstract}
\noindent Hierarchical categorical variables often exhibit many levels (high granularity) and many classes within each level (high dimensionality). This may cause overfitting and estimation issues when including such covariates in a predictive model. In current literature, a hierarchical covariate is often incorporated via nested random effects. However, this does not facilitate the assumption of classes having the same effect on the response variable. In this paper, we propose a methodology to obtain a reduced representation of a hierarchical categorical variable. We show how entity embedding can be applied in a hierarchical setting. Subsequently, we propose a top-down clustering algorithm which leverages the information encoded in the embeddings to reduce both the within-level dimensionality as well as the overall granularity of the hierarchical categorical variable. In simulation experiments, we show that our methodology can effectively approximate the true underlying structure of a hierarchical covariate in terms of the effect on a response variable, and find that incorporating the reduced hierarchy improves \added{the balance between model fit and complexity}. We apply our methodology to a real dataset and find that the reduced hierarchy is an improvement over the original hierarchical structure and reduced structures proposed in the literature.

\end{abstract}

\vspace{-0.25cm}
\vspace{-0.25cm}
\section*{Statements and declarations}
\paragraph{Data and code availability statement:} Data and code are available on \url{https://github.com/PaulWilsens/reducing-hierarchical-cat}.
\vspace{-0.25cm}
\paragraph{Funding statement and acknowledgements:} The authors gratefully acknowledge funding from the FWO and Fonds De La Recherche
Scientifique - FNRS (F.R.S.-FNRS) under the Excellence of Science (EOS) program, project
ASTeRISK Research Foundation Flanders [grant number 40007517]. Katrien Antonio gratefully
acknowledges support from the Chaire DIALog sponsored by CNP Assurances and the FWO network W001021N. We thank the editor and referees for their comments which helped improve the paper substantially.
\vspace{-0.25cm}
\paragraph{Conflict of interest disclosure:} The authors declare no conflict of interest.

\pagebreak


\section{Introduction} \label{section:introduction}
\begin{figure}[H]

\tikzset{every picture/.style={line width=0.75pt}} 

\begin{tikzpicture}[x=0.75pt,y=0.75pt,yscale=-0.82,xscale=0.82]

\draw    (67.33,100.69) -- (67.57,190.97) ;
\draw    (57.57,100.67) -- (27.23,191.3) ;
\draw    (76.9,101.01) -- (107.23,191.3) ;
\draw    (196,101.36) -- (196.23,191.64) ;
\draw    (186.23,101.34) -- (155.9,191.97) ;
\draw    (205.57,101.67) -- (235.9,191.97) ;
\draw    (319.67,101.36) -- (319.9,191.64) ;
\draw    (309.9,101.34) -- (279.57,191.97) ;
\draw    (329.23,101.67) -- (359.57,191.97) ;
\draw    (443,101.02) -- (443.23,191.3) ;
\draw    (433.23,101.01) -- (402.9,191.64) ;
\draw    (452.57,101.34) -- (482.9,191.64) ;
\draw    (567.33,101.02) -- (567.57,191.3) ;
\draw    (557.57,101.01) -- (527.23,191.64) ;
\draw    (576.9,101.34) -- (607.23,191.64) ;
\draw    (692,101.02) -- (692.23,191.3) ;
\draw    (682.23,101.01) -- (651.9,191.64) ;
\draw    (701.57,101.34) -- (731.9,191.64) ;
\draw    (379.56,13.02) -- (195.94,81.02) ;
\draw    (379.56,13.02) -- (67.19,81.02) ;
\draw    (379.56,13.02) -- (319.69,81.02) ;
\draw    (379.56,13.02) -- (442.94,81.02) ;
\draw    (379.56,13.02) -- (567.07,81.02) ;
\draw    (379.56,13.02) -- (692.06,81.27) ;

\draw (43,82.55) node [anchor=north west][inner sep=0.75pt]   [align=left] {Africa};
\draw (14,192) node [anchor=north west][inner sep=0.75pt]   [align=left] {DZ};
\draw (93,192) node [anchor=north west][inner sep=0.75pt]   [align=left] {ZW};
\draw (59,198) node [anchor=north west][inner sep=0.75pt]   [align=left] {...};
\draw (176,83.22) node [anchor=north west][inner sep=0.75pt]   [align=left] {Asia};
\draw (142,192) node [anchor=north west][inner sep=0.75pt]   [align=left] {AF};
\draw (220,192) node [anchor=north west][inner sep=0.75pt]   [align=left] {YE};
\draw (187,198) node [anchor=north west][inner sep=0.75pt]   [align=left] {...};
\draw (291,83.22) node [anchor=north west][inner sep=0.75pt]   [align=left] {Europe};
\draw (266,192) node [anchor=north west][inner sep=0.75pt]   [align=left] {AL};
\draw (346,192) node [anchor=north west][inner sep=0.75pt]   [align=left] {VA};
\draw (311,198) node [anchor=north west][inner sep=0.75pt]   [align=left] {...};
\draw (380,82.89) node [anchor=north west][inner sep=0.75pt]   [align=left] {North-America};
\draw (389,192) node [anchor=north west][inner sep=0.75pt]   [align=left] {AG};
\draw (469,192) node [anchor=north west][inner sep=0.75pt]   [align=left] {US};
\draw (434,198) node [anchor=north west][inner sep=0.75pt]   [align=left] {...};
\draw (510,82.89) node [anchor=north west][inner sep=0.75pt]   [align=left] {South-America};
\draw (513,192) node [anchor=north west][inner sep=0.75pt]   [align=left] {AR};
\draw (593,192) node [anchor=north west][inner sep=0.75pt]   [align=left] {VE};
\draw (558.5,198) node [anchor=north west][inner sep=0.75pt]   [align=left] {...};
\draw (656,82.89) node [anchor=north west][inner sep=0.75pt]   [align=left] {Oceania};
\draw (638,192) node [anchor=north west][inner sep=0.75pt]   [align=left] {AU};
\draw (717,192) node [anchor=north west][inner sep=0.75pt]   [align=left] {VU};
\draw (683,198) node [anchor=north west][inner sep=0.75pt]   [align=left] {...};

\end{tikzpicture}
 
	\centering
        \caption{Example of a categorical hierarchical variable with two levels representing information about the members of the United Nations. The first level represents the continental level and consists of six classes: Africa, Asia, Europe, North-America, South-America and Oceania. The second level consists of 193 classes, with 53 member states located in Africa, 47 in Asia, 44 in Europe, 23 in North-America, 12 in South-America and 14 in Oceania.}
        \label{fig:hierarchyexamplewords}
\end{figure}
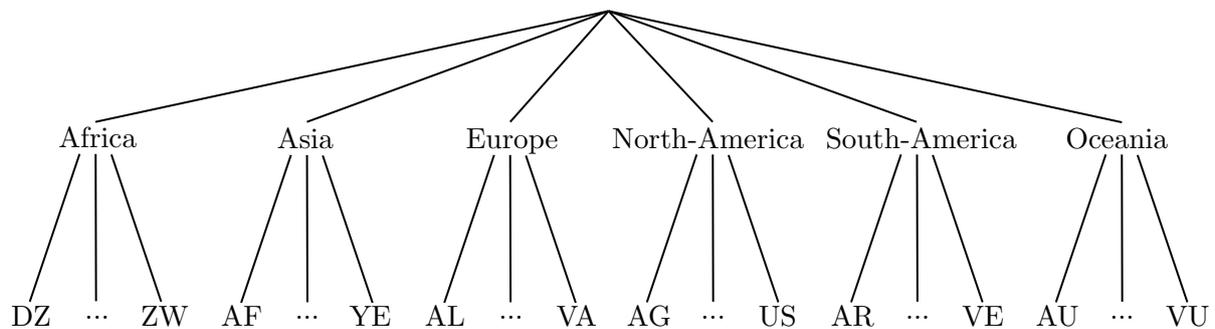

Handling categorical variables in predictive modelling is a challenging task. Nominal class labels cannot immediately be used in a predictive model, instead the variable needs to be numerically encoded, e.g.~via dummy variables in statistics or one-hot encoding in machine learning \citep{suits1957use,goodfellow2016deep}. Additionally, a categorical variable can exhibit an inherent hierarchical structure. A hierarchical structure allows a categorical variable to be represented at different levels of granularity, where each level of the hierarchy consists of classes and the classes in the higher (less granular) levels are a consolidated representation of the information stored in the lower (more granular) levels in the hierarchy. A hierarchical categorical variable with many levels is called highly granular. The more granular levels often consist of many classes, i.e. they exhibit a high dimensionality. Figure~\ref{fig:hierarchyexamplewords} shows an example of a hierarchical categorical variable that represents geographical information at two levels of granularity. The variable can be expressed at low granularity, i.e.~at the continental level, or at a more granular level of detail where we know the specific country within that continent. In this paper, we aim to contribute to the literature on handling categorical variables of this type in predictive models.

Several types of models can deal with hierarchical data. Analysis of variance (ANOVA) models with a nested factor design use a fixed effect for each class at every level in a hierarchy \citep{neter1996applied}. Another common approach to handle hierarchical categorical variables are multilevel models, e.g.~the linear mixed model (LMM), which incorporates information on the hierarchy via nested random effects \citep{gelman2006data}. Multilevel models are widely used in different domains, e.g. in educational sciences to distinguish among schools and classrooms within those schools or in ecology to model the variation across genotypes or species \citep{peugh2010practical,bolker2009generalized}. \added{In actuarial science, hierarchically structured random effects were first studied by \cite{jewell1975use} in the context of rating hierarchically structured risks in a workers' compensation insurance product. \cite{frees1999longitudinal,frees2001case} demonstrate the connection between the so-called actuarial credibility models and linear mixed models and \cite{antonio2007actuarial} extend their study to the use of generalised linear mixed models (GLMMs) for insurance pricing with panel and hierarchically structured data on claim frequencies and severities.}

A hierarchical categorical variable often has many classes, especially at the most granular level. This high dimensionality can lead to overfitting or estimation problems, see \cite{avanzi2024machine}. A solution is to reduce the dimension of the numerical encoding that is used to represent the high-dimensional categorical variable. Several approaches have been proposed in machine learning literature for non-hierarchical categorical variables. For example, \cite{micci2001preprocessing} proposed a preprocessing step that uses summary statistics of the response variable to obtain a numerical encoding. A popular approach to obtain a low-dimensional encoding for a categorical variable is the technique of entity embedding, proposed by \cite{guo2016entity}. This is inspired by word embedding which is an established method in natural language processing (NLP), e.g. see \cite{pennington2014glove}. Entity embedding maps the classes of a categorical variable to a (multi-dimensional) Euclidean space. Classes that are similar to each other in terms of the response variable are placed close to each other in the resulting space. Specifically, the classes' mapping is obtained by learning a feedforward neural network structure with an additional layer, i.e.~the embedding layer, on top of the one-hot encoding of the categorical variable. This embedding layer is concatenated with the other covariates to form the input layer for the network. By predefining an embedding dimension that is smaller than the original dimension of the categorical variable, the information is compressed, forcing the network to learn the essence of the relationship between the categorical variable and the response variable. The resulting entity embedding is an encoding that can subsequently be used in a predictive model of choice \citep{blier2021rethinking}. \added{Recently, \cite{simchoni2023integrating} proposed to extend entity embedding by including random effects in the embedding layer, the so-called random effects entity embedding. The key difference with entity embedding is that the weights in the random effects embedding vector are assumed to come from a distribution instead of being fixed values. Their approach is limited to a normally distributed response variable. \cite{avanzi2024machine} expanded upon this approach by proposing a hybrid of a generalised linear mixed model and a neural network, called GLMMnet, which allows the response variable to follow any distribution from the exponential family. GLMMnet is limited to handling non-hierarchical categorical variables. However, \cite{richman2024high} put forward an extension that can handle hierarchical categorical variables using random effects entity embedding. They interpret the hierarchical categorical variable as a time series, and use a recurrent neural network layer or an attention layer to handle the time series, as laid out by \cite{kuo2021embeddings}.}

As motivated in \cite{campo2024clustering}, some classes of a hierarchical categorical variable might have the same effect on the response variable, allowing those classes to be merged within the hierarchy. By construction, an approach with nested random effects does not directly facilitate this. \cite{carrizosa2022tree} point out that incorporating a reduced representation of a hierarchical categorical variable in a predictive model can not only improve interpretability but also decreases the number of parameters, compared to including the original variable. Some contributions in the literature aim to reduce the dimensionality and/or granularity of a hierarchy, by merging classes within a given level or collapsing classes with their parent class. The former step results in fewer classes while keeping the levels of the original hierarchy, which reduces the dimensionality. The latter reduces the overall granularity of the hierarchical categorical variable. \added{A drawback of this approach is that the resulting clusters might be less intuitive to explain, depending on the semantic meaning of the hierarchy.} \cite{carrizosa2022tree} put forward the tree based linear regression model (TLR) for normally distributed response variables, which balances the predictive accuracy and complexity of the model. The latter is measured by a cost function of the granularity of a hierarchical categorical variable. They only include the most granular level of a hierarchy and aim to increase the predictive accuracy, measured by the mean squared error, when a set of classes with the same parent class is merged with that parent, i.e.~collapsed, resulting in a less granular representation. However, they do not allow classes within the same level of a hierarchy to be merged, nor do they permit the collapse of only a subset of the classes with the same parent class. A second approach is presented by \cite{campo2024clustering}, who apply feature engineering to obtain a profile for all classes within a level. Next, they apply clustering techniques to these engineered features to reduce the number of classes at each level, in a top-down approach. They adhere to the hierarchy by only merging classes with the same (consolidated) parent class. Subsequently, they apply the random effects approach of \cite{campoGLMM} to include the reduced hierarchical variable in a predictive model. Their approach allows for a more flexible distributional assumption compared to \cite{carrizosa2022tree} and allows to merge classes within the same level of a hierarchy. However, while it allows to collapse an entire level, it does not allow for specific levels to be merged with their parent class.

We contribute to the existing literature in multiple ways. \cite{guo2016entity} showed that the Euclidean space resulting from entity embedding allows for the clustering of classes in the case of a non-hierarchical categorical variable. In this paper, we extend their strategy to a hierarchical categorical variable. We construct meaningful embeddings for each class in a hierarchy by first applying entity embedding at the most granular level, and subsequently using a summary measure to obtain embeddings for classes at higher levels. We then propose a top-down algorithm that applies clustering techniques to the hierarchical categorical variable's embedding space. Compared to the literature, our algorithm is the first to both merge classes within the same level of a hierarchy and to allow individual classes or clusters of classes to be merged with their respective parent class. We obtain a new hierarchical categorical variable that is a reduced representation of the original one. The reduced hierarchical variable can subsequently be used in a predictive model of choice. Our methodology is not limited to linear models and allows flexible assumptions on the distribution of the response variable. We cluster directly on the embeddings and, therefore, do not require additional feature engineering. Using simulated data, we demonstrate the added value by showing that our methodology accurately retrieves the essential structure in terms of the effect on the response variable. Lastly, we illustrate the application of our methodology to a real dataset.

In the next section, the setup and notation are introduced. Section~\ref{sec:embedding} explains how we obtain entity embeddings for every class at each level in a hierarchy. In Section~\ref{sec:collapsing}, the algorithm to collapse the hierarchical structure is introduced. Next, the performance of the methodology is assessed in Section~\ref{sec:simulating} using simulation experiments. In Section~\ref{reallife}, we apply our method to a real dataset. Finally, Section~\ref{sec:conclus} concludes our paper.


\section{Notation and setup}
Suppose we have a hierarchical categorical variable $\boldsymbol{h}=(h_{1},\ldotp \ldotp,h_{R})$ with $R$ levels that denotes the class label at each level of the hierarchy. For $r=1,\ldotp\ldotp,R$, we have a (non-hierarchical) categorical variable $h_{r} \in H_{r}=\{h_{r,1},\ldotp \ldotp,h_{r,n_{r}} \}$ with $n_{r}$ distinct classes that contains the information at the granularity of the $r$th level. We define the set of direct descendants of $h_{r,s}$, i.e.~the classes at level $r+1$ of which $h_{r,s}$ is the parent class, as $\mathcal{H}_{r,s}$. We also make the assumption that each class can only have a single parent class. Consequently, we have that
\begin{align*}
    \bigcup_{s=1}^{n_{r}} \mathcal{H}_{r,s} = H_{r+1} \,\, \text{and} \,\, \bigcap_{s=1}^{n_{r}} \mathcal{H}_{r,s} = \varnothing \,\,\,\, \text{for} \,\, r=1,\ldotp \ldotp,R.
\end{align*}

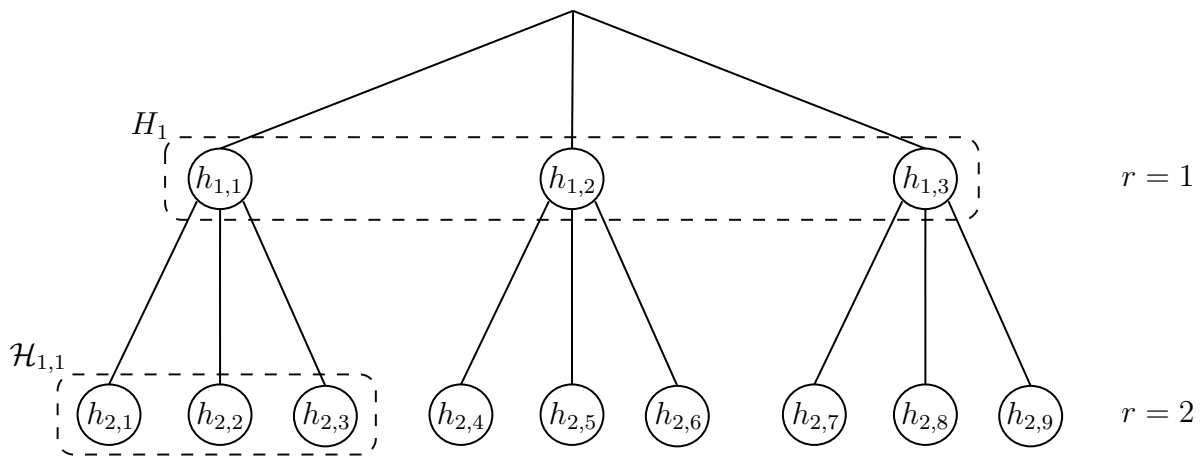
\begin{figure}[H]
	\tikzset{every picture/.style={line width=0.75pt}} 

\begin{tikzpicture}[x=0.54pt,y=0.54pt,yscale=-1,xscale=1]

\draw    (407.79,18) -- (163.09,114.26) ;
\draw    (407.79,18) -- (407.09,114.26) ;
\draw    (407.79,18) -- (652.09,114.26) ;
\draw    (147.07,151.02) -- (86.07,279.02) ;
\draw    (163.09,156.51) -- (163.04,191.02) -- (163.07,279.02) ;
\draw    (179.07,151.02) -- (236.07,280.02) ;
\draw  [fill={rgb, 255:red, 255; green, 255; blue, 255 }  ,fill opacity=1 ] (141.29,135.38) .. controls (141.29,123.72) and (151.05,114.26) .. (163.09,114.26) .. controls (175.13,114.26) and (184.9,123.72) .. (184.9,135.38) .. controls (184.9,147.05) and (175.13,156.51) .. (163.09,156.51) .. controls (151.05,156.51) and (141.29,147.05) .. (141.29,135.38) -- cycle ;
\draw    (391.07,151.02) -- (330.07,279.02) ;
\draw    (407.09,156.51) -- (407.04,191.02) -- (407.07,279.02) ;
\draw    (423.07,151.02) -- (480.07,280.02) ;
\draw  [fill={rgb, 255:red, 255; green, 255; blue, 255 }  ,fill opacity=1 ] (385.29,135.38) .. controls (385.29,123.72) and (395.05,114.26) .. (407.09,114.26) .. controls (419.13,114.26) and (428.9,123.72) .. (428.9,135.38) .. controls (428.9,147.05) and (419.13,156.51) .. (407.09,156.51) .. controls (395.05,156.51) and (385.29,147.05) .. (385.29,135.38) -- cycle ;
\draw  [fill={rgb, 255:red, 255; green, 255; blue, 255 }  ,fill opacity=1 ] (64.27,300.14) .. controls (64.27,288.48) and (74.03,279.02) .. (86.07,279.02) .. controls (98.11,279.02) and (107.87,288.48) .. (107.87,300.14) .. controls (107.87,311.81) and (98.11,321.27) .. (86.07,321.27) .. controls (74.03,321.27) and (64.27,311.81) .. (64.27,300.14) -- cycle ;
\draw  [fill={rgb, 255:red, 255; green, 255; blue, 255 }  ,fill opacity=1 ] (214.27,301.14) .. controls (214.27,289.48) and (224.03,280.02) .. (236.07,280.02) .. controls (248.11,280.02) and (257.87,289.48) .. (257.87,301.14) .. controls (257.87,312.81) and (248.11,322.27) .. (236.07,322.27) .. controls (224.03,322.27) and (214.27,312.81) .. (214.27,301.14) -- cycle ;
\draw  [fill={rgb, 255:red, 255; green, 255; blue, 255 }  ,fill opacity=1 ] (141.27,300.14) .. controls (141.27,288.48) and (151.03,279.02) .. (163.07,279.02) .. controls (175.11,279.02) and (184.87,288.48) .. (184.87,300.14) .. controls (184.87,311.81) and (175.11,321.27) .. (163.07,321.27) .. controls (151.03,321.27) and (141.27,311.81) .. (141.27,300.14) -- cycle ;
\draw  [fill={rgb, 255:red, 255; green, 255; blue, 255 }  ,fill opacity=1 ] (308.27,300.14) .. controls (308.27,288.48) and (318.03,279.02) .. (330.07,279.02) .. controls (342.11,279.02) and (351.87,288.48) .. (351.87,300.14) .. controls (351.87,311.81) and (342.11,321.27) .. (330.07,321.27) .. controls (318.03,321.27) and (308.27,311.81) .. (308.27,300.14) -- cycle ;
\draw  [fill={rgb, 255:red, 255; green, 255; blue, 255 }  ,fill opacity=1 ] (385.27,300.14) .. controls (385.27,288.48) and (395.03,279.02) .. (407.07,279.02) .. controls (419.11,279.02) and (428.87,288.48) .. (428.87,300.14) .. controls (428.87,311.81) and (419.11,321.27) .. (407.07,321.27) .. controls (395.03,321.27) and (385.27,311.81) .. (385.27,300.14) -- cycle ;
\draw  [fill={rgb, 255:red, 255; green, 255; blue, 255 }  ,fill opacity=1 ] (458.27,301.14) .. controls (458.27,289.48) and (468.03,280.02) .. (480.07,280.02) .. controls (492.11,280.02) and (501.87,289.48) .. (501.87,301.14) .. controls (501.87,312.81) and (492.11,322.27) .. (480.07,322.27) .. controls (468.03,322.27) and (458.27,312.81) .. (458.27,301.14) -- cycle ;
\draw    (636.07,151.02) -- (575.07,279.02) ;
\draw    (652.09,156.51) -- (652.04,191.02) -- (652.07,279.02) ;
\draw    (668.07,151.02) -- (725.07,280.02) ;
\draw  [fill={rgb, 255:red, 255; green, 255; blue, 255 }  ,fill opacity=1 ] (630.29,135.38) .. controls (630.29,123.72) and (640.05,114.26) .. (652.09,114.26) .. controls (664.13,114.26) and (673.9,123.72) .. (673.9,135.38) .. controls (673.9,147.05) and (664.13,156.51) .. (652.09,156.51) .. controls (640.05,156.51) and (630.29,147.05) .. (630.29,135.38) -- cycle ;
\draw  [fill={rgb, 255:red, 255; green, 255; blue, 255 }  ,fill opacity=1 ] (553.27,300.14) .. controls (553.27,288.48) and (563.03,279.02) .. (575.07,279.02) .. controls (587.11,279.02) and (596.87,288.48) .. (596.87,300.14) .. controls (596.87,311.81) and (587.11,321.27) .. (575.07,321.27) .. controls (563.03,321.27) and (553.27,311.81) .. (553.27,300.14) -- cycle ;
\draw  [fill={rgb, 255:red, 255; green, 255; blue, 255 }  ,fill opacity=1 ] (630.27,300.14) .. controls (630.27,288.48) and (640.03,279.02) .. (652.07,279.02) .. controls (664.11,279.02) and (673.87,288.48) .. (673.87,300.14) .. controls (673.87,311.81) and (664.11,321.27) .. (652.07,321.27) .. controls (640.03,321.27) and (630.27,311.81) .. (630.27,300.14) -- cycle ;
\draw  [fill={rgb, 255:red, 255; green, 255; blue, 255 }  ,fill opacity=1 ] (703.27,301.14) .. controls (703.27,289.48) and (713.03,280.02) .. (725.07,280.02) .. controls (737.11,280.02) and (746.87,289.48) .. (746.87,301.14) .. controls (746.87,312.81) and (737.11,322.27) .. (725.07,322.27) .. controls (713.03,322.27) and (703.27,312.81) .. (703.27,301.14) -- cycle ;
\draw  [color={rgb, 255:red, 0; green, 0; blue, 0 }  ,draw opacity=0.96 ][dash pattern={on 4.5pt off 4.5pt}] (124.96,117.83) .. controls (124.96,111.45) and (130.13,106.29) .. (136.51,106.29) -- (677.53,106.29) .. controls (683.9,106.29) and (689.07,111.45) .. (689.07,117.83) -- (689.07,152.46) .. controls (689.07,158.83) and (683.9,164) .. (677.53,164) -- (136.51,164) .. controls (130.13,164) and (124.96,158.83) .. (124.96,152.46) -- cycle ;
\draw  [color={rgb, 255:red, 0; green, 0; blue, 0 }  ,draw opacity=0.96 ][dash pattern={on 4.5pt off 4.5pt}] (50.45,283.2) .. controls (50.45,277.01) and (55.46,272) .. (61.65,272) -- (259.87,272) .. controls (266.06,272) and (271.07,277.01) .. (271.07,283.2) -- (271.07,316.8) .. controls (271.07,322.99) and (266.06,328) .. (259.87,328) -- (61.65,328) .. controls (55.46,328) and (50.45,322.99) .. (50.45,316.8) -- cycle ;

\draw (786,124.08) node [anchor=north west][inner sep=0.75pt]  [font=\large] [align=left] {$\displaystyle r=1$};
\draw (786,289.13) node [anchor=north west][inner sep=0.75pt]  [font=\large] [align=left] {$\displaystyle r=2$};
\draw (143.64,124.08) node [anchor=north west][inner sep=0.75pt]  [font=\large] [align=left] {$ $$\displaystyle h_{1,1}$};
\draw (388.53,124.08) node [anchor=north west][inner sep=0.75pt]  [font=\large] [align=left] {$\displaystyle h_{1,2}$};
\draw (633.86,124.08) node [anchor=north west][inner sep=0.75pt]  [font=\large] [align=left] {$\displaystyle h_{1,3}$};
\draw (68.29,289.13) node [anchor=north west][inner sep=0.75pt]  [font=\large] [align=left] {$\displaystyle h_{2,1}$};
\draw (144.45,289.13) node [anchor=north west][inner sep=0.75pt]  [font=\large] [align=left] {$\displaystyle h_{2,2}$};
\draw (217.98,289.13) node [anchor=north west][inner sep=0.75pt]  [font=\large] [align=left] {$\displaystyle h_{2,3}$};
\draw (311.25,289.13) node [anchor=north west][inner sep=0.75pt]  [font=\large] [align=left] {$\displaystyle h_{2,4}$};
\draw (389.03,289.13) node [anchor=north west][inner sep=0.75pt]  [font=\large] [align=left] {$\displaystyle h_{2,5}$};
\draw (462.18,289.13) node [anchor=north west][inner sep=0.75pt]  [font=\large] [align=left] {$\displaystyle h_{2,6}$};
\draw (557.33,289.13) node [anchor=north west][inner sep=0.75pt]  [font=\large] [align=left] {$\displaystyle h_{2,7}$};
\draw (633.11,289.13) node [anchor=north west][inner sep=0.75pt]  [font=\large] [align=left] {$\displaystyle h_{2,8}$};
\draw (705.89,289.13) node [anchor=north west][inner sep=0.75pt]  [font=\large] [align=left] {$\displaystyle h_{2,9}$};
\draw (99,86.86) node [anchor=north west][inner sep=0.75pt]  [font=\large] [align=left] {$\displaystyle H_{1}$};
\draw (15,247.86) node [anchor=north west][inner sep=0.75pt]  [font=\large] [align=left] {$\displaystyle \mathcal{H}_{1,1}$};

\end{tikzpicture}
	\centering
        \caption{Example of the structure of a hierarchical categorical variable with two levels, thus $R=2$. $H_{1}=\{h_{1,1},h_{1,2},h_{1,3} \}$ denotes the first level's set of classes. $\mathcal{H}_{1,1}= \{ h_{2,1},h_{2,2},h_{2,3} \}$ denotes the set of direct descendants of $h_{1,1}$.}
        \label{fig:hierarchyexample}
\end{figure}

Figure~\ref{fig:hierarchyexample} visualises the hierarchy for an example with $R=2$ levels. Since the hierarchy is considered known and because of the assumption stated above, having information about the class label at the lowest level $h_{R}$ immediately implies $\boldsymbol{h}$ since there exists a unique path from each class at the lowest level to a class at the upper level of the hierarchy. For class $h_{2,7}$, for example, we have $\boldsymbol{h}=(h_{1,3},h_{2,7})$.
\vspace{-0.8cm}
\paragraph{Predictive modelling with a hierarchical covariate}

Assume the availability of a dataset $\mathcal{D}= (y_{i},\boldsymbol{h}_{i},\boldsymbol{x}_{i})_{i=1}^{n}$ of $n$ observations where $i$ refers to the observation, $y_{i}$ is the response variable, $\boldsymbol{h}_{i}$ is the hierarchical categorical variable and $\boldsymbol{x}_{i}$  denotes the vector containing all other types of covariates, e.g.~continuous, spatial and non-hierarchical categorical variables. Our aim is to learn a reduced representation of $\boldsymbol{h}$, which we denote by $\boldsymbol{\widetilde{h}}=(\widetilde{h}_{1},\ldotp \ldotp,\widetilde{h}_{\widetilde{R}})$ with $\widetilde{R}$ levels, where the overall granularity as well as the within-level dimensionality are reduced, i.e.~$\widetilde{R} \le R$ and $\widetilde{n}_{r} \le n_{r}$ for $r=1,\ldotp\ldotp,\widetilde{R}$. We then learn a predictive model $f_{\boldsymbol{\Theta}}(\boldsymbol{\widetilde{h}}_{i},\boldsymbol{x}_{i})$ for the data sample with parameter vector $\boldsymbol{\Theta}$ by minimising
\begin{equation} \label{eq:modelling}
    \sum_{i=1}^{n} \mathcal{L}(y_{i},f_{\boldsymbol{\Theta}}(\boldsymbol{\widetilde{h}}_{i},\boldsymbol{x}_{i}))
\end{equation}
over the parameter vector $\boldsymbol{\Theta}$, where $\mathcal{L}(y_{i},f_{\boldsymbol{\Theta}}(\boldsymbol{\widetilde{h}}_{i},\boldsymbol{x}_{i}))$ is a suitable loss function. The model $f_{\boldsymbol{\Theta}}(\boldsymbol{\widetilde{h}}_{i},\boldsymbol{x}_{i})$ can refer to a model of any complexity, for example a generalised linear model or an advanced machine learning technique. Our objective is to construct a reduced representation $\boldsymbol{\widetilde{h}}$ that results \added{in an improved balance between model fit and complexity.} In the remainder, the index $i$ is omitted for brevity.


\section{Embedding a hierarchy} \label{sec:embedding}
\subsection{Entity embeddings in feedforward neural networks } \label{sec:network}
As a first step, we learn an entity embedding mapping $\boldsymbol{\theta}^e: \mathbb{R}^{n_{R}} \mapsto \mathbb{R}^{q_{e}}$ for $h_{R}$, which provides us with a unique encoding in $\mathbb{R}^{q_{e}}$ for each of the classes $h_{R,1},\ldotp\ldotp,h_{R,n_{R}}$ at the lowest level (i.e. level $R$) in the hierarchy. Following \cite{guo2016entity}, we fit a feedforward neural network (FFNN) with an embedding layer to learn this representation. A feedforward network consists of interconnected layers, with information flowing through the network without any feedback loops.

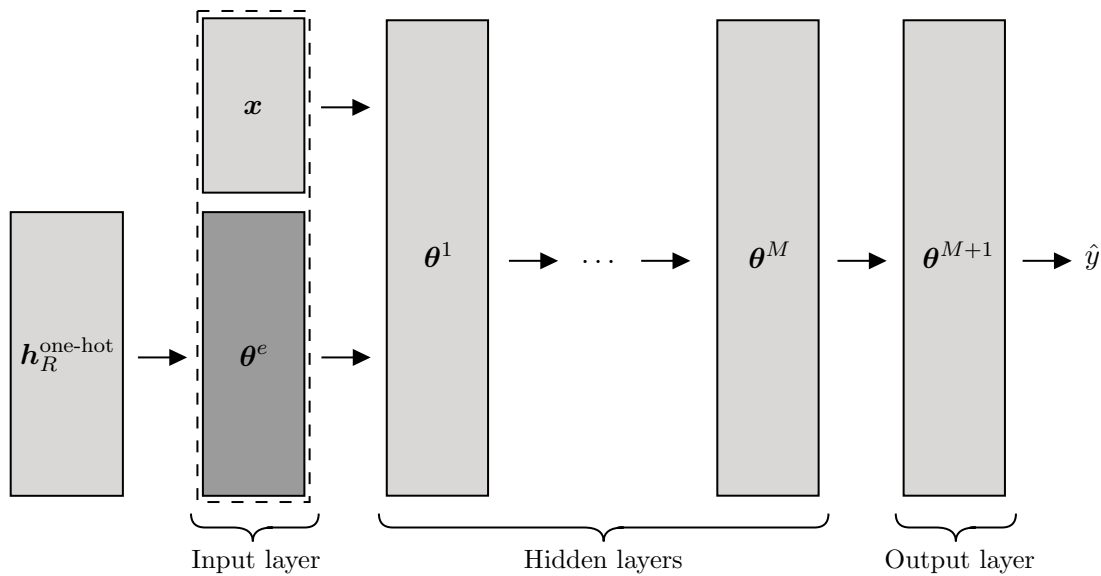
\begin{figure}[H]
	\tikzset{every picture/.style={line width=0.75pt}} 

\begin{tikzpicture}[x=0.75pt,y=0.75pt,yscale=-1,xscale=1]

\draw  [fill={rgb, 255:red, 218; green, 215; blue, 215 }  ,fill opacity=1 ][line width=0.75]  (198,23) -- (248.57,23) -- (248.57,262.14) -- (198,262.14) -- cycle ;
\draw  [fill={rgb, 255:red, 218; green, 215; blue, 215 }  ,fill opacity=1 ][line width=0.75]  (363,23) -- (413.57,23) -- (413.57,262.14) -- (363,262.14) -- cycle ;
\draw    (165.4,192.6) -- (186.97,192.6) ;
\draw [shift={(189.97,192.6)}, rotate = 180] [fill={rgb, 255:red, 0; green, 0; blue, 0 }  ][line width=0.08]  [draw opacity=0] (8.93,-4.29) -- (0,0) -- (8.93,4.29) -- cycle    ;
\draw    (259,144) -- (280.57,144) ;
\draw [shift={(283.57,144)}, rotate = 180] [fill={rgb, 255:red, 0; green, 0; blue, 0 }  ][line width=0.08]  [draw opacity=0] (8.93,-4.29) -- (0,0) -- (8.93,4.29) -- cycle    ;
\draw    (325,144) -- (346.57,144) ;
\draw [shift={(349.57,144)}, rotate = 180] [fill={rgb, 255:red, 0; green, 0; blue, 0 }  ][line width=0.08]  [draw opacity=0] (8.93,-4.29) -- (0,0) -- (8.93,4.29) -- cycle    ;
\draw  [fill={rgb, 255:red, 218; green, 215; blue, 215 }  ,fill opacity=1 ][line width=0.75]  (456,23) -- (506.57,23) -- (506.57,262.14) -- (456,262.14) -- cycle ;
\draw    (423,144) -- (444.57,144) ;
\draw [shift={(447.57,144)}, rotate = 180] [fill={rgb, 255:red, 0; green, 0; blue, 0 }  ][line width=0.08]  [draw opacity=0] (8.93,-4.29) -- (0,0) -- (8.93,4.29) -- cycle    ;
\draw   (194.08,271.14) .. controls (194.08,275.81) and (196.41,278.14) .. (201.08,278.14) -- (296.28,278.14) .. controls (302.95,278.14) and (306.28,280.47) .. (306.27,285.14) .. controls (306.28,280.47) and (309.61,278.14) .. (316.28,278.14)(313.28,278.14) -- (411.48,278.14) .. controls (416.15,278.14) and (418.48,275.81) .. (418.48,271.14) ;
\draw   (452.07,271.14) .. controls (452.07,275.81) and (454.4,278.14) .. (459.07,278.14) -- (472.07,278.14) .. controls (478.74,278.14) and (482.07,280.47) .. (482.07,285.14) .. controls (482.07,280.47) and (485.4,278.14) .. (492.07,278.14)(489.07,278.14) -- (505.07,278.14) .. controls (509.74,278.14) and (512.07,275.81) .. (512.07,271.14) ;
\draw   (100.28,271.14) .. controls (100.3,275.81) and (102.64,278.14) .. (107.31,278.14) -- (121.11,278.14) .. controls (127.78,278.14) and (131.12,280.47) .. (131.14,285.14) .. controls (131.12,280.47) and (134.44,278.14) .. (141.11,278.14)(138.11,278.14) -- (157.04,278.14) .. controls (161.71,278.14) and (164.03,275.81) .. (164.02,271.14) ;
\draw  [fill={rgb, 255:red, 218; green, 215; blue, 215 }  ,fill opacity=1 ][line width=0.75]  (106.22,22) -- (156.88,22) -- (156.88,109.71) -- (106.22,109.71) -- cycle ;
\draw  [color={rgb, 255:red, 0; green, 0; blue, 0 }  ,draw opacity=1 ][fill={rgb, 255:red, 155; green, 155; blue, 155 }  ,fill opacity=1 ] (106.22,119.51) -- (156.88,119.51) -- (156.88,262.2) -- (106.22,262.2) -- cycle ;
\draw  [color={rgb, 255:red, 0; green, 0; blue, 0 }  ,draw opacity=1 ][fill={rgb, 255:red, 218; green, 215; blue, 215 }  ,fill opacity=1 ] (10.28,119.51) -- (66.28,119.51) -- (66.28,262.2) -- (10.28,262.2) -- cycle ;
\draw    (165.2,66.8) -- (186.77,66.8) ;
\draw [shift={(189.77,66.8)}, rotate = 180] [fill={rgb, 255:red, 0; green, 0; blue, 0 }  ][line width=0.08]  [draw opacity=0] (8.93,-4.29) -- (0,0) -- (8.93,4.29) -- cycle    ;
\draw    (73.8,192.6) -- (95.37,192.6) ;
\draw [shift={(98.37,192.6)}, rotate = 180] [fill={rgb, 255:red, 0; green, 0; blue, 0 }  ][line width=0.08]  [draw opacity=0] (8.93,-4.29) -- (0,0) -- (8.93,4.29) -- cycle    ;
\draw  [dash pattern={on 4.5pt off 4.5pt}] (103.49,18.33) -- (159.49,18.33) -- (159.49,265.33) -- (103.49,265.33) -- cycle ;
\draw    (515.4,144) -- (536.97,144) ;
\draw [shift={(539.97,144)}, rotate = 180] [fill={rgb, 255:red, 0; green, 0; blue, 0 }  ][line width=0.08]  [draw opacity=0] (8.93,-4.29) -- (0,0) -- (8.93,4.29) -- cycle    ;

\draw (125,62) node [anchor=north west][inner sep=0.75pt]   [align=left] {$\boldsymbol{x}$};
\draw (293.5,142) node [anchor=north west][inner sep=0.75pt]   [align=left] {$\ldots$};
\draw (545,134) node [anchor=north west][inner sep=0.75pt]   [align=left] {$\hat{y}$};
\draw (123,185) node [anchor=north west][inner sep=0.75pt]   [align=left] {$\boldsymbol{\theta }^{e}$};
\draw (215,133) node [anchor=north west][inner sep=0.75pt]   [align=left] {$\boldsymbol{\theta }^{1}$};
\draw (377,133) node [anchor=north west][inner sep=0.75pt]   [align=left] {$\boldsymbol{\theta }^{M}$};
\draw (464,133) node [anchor=north west][inner sep=0.75pt]   [align=left] {$\boldsymbol{\theta }^{M+1}$};
\draw (265,288) node [anchor=north west][inner sep=0.75pt]  [font=\small] [align=left] {Hidden layers};
\draw (445,288) node [anchor=north west][inner sep=0.75pt]  [font=\small] [align=left] {Output layer};
\draw (13.5,181) node [anchor=north west][inner sep=0.75pt]   [align=left] {$\boldsymbol{h}_{R}^{\text{one-hot}}$};
\draw (99,288) node [anchor=north west][inner sep=0.75pt]  [font=\small] [align=left] {Input layer};

\end{tikzpicture}
\centering
        \caption{Feedforward neural network structure to learn an embedding mapping for $h_{R}$ starting from the one-hot encoding $\boldsymbol{h}_{R}^{\text{one-hot}}$. The embedding layer $\boldsymbol{\theta}^e$ consists of $q_{e}$ neurons and forms the input layer together with the covariate vector $\boldsymbol{x}$. $\boldsymbol{\theta}^1,\ldotp\ldotp,\boldsymbol{\theta}^M$ are the hidden layers composed of $q_1,\ldotp\ldotp,q_M$ neurons, respectively. The output layer $\boldsymbol{\theta}^{M+1}$ consists of $q_{M+1}$ neurons.}
        \label{fig:network}
\end{figure}

Figure~\ref{fig:network} pictures the general structure of our network. The input layer consists of the vector with the non-hierarchical covariates $\boldsymbol{x}$ concatenated with the embedding layer $\boldsymbol{\theta}^e$. The embedding layer is a layer composed of $q_{e}$ neurons, which uses the one-hot encoding of a class at the most granular level of the hierarchical categorical variable, i.e.~$\boldsymbol{h}_{R}^{\text{one-hot}} = (\boldsymbol{1}_{ \{h_{R}=h_{R,1} \} },\ldotp \ldotp,\boldsymbol{1}_{\{h_{R}=h_{R,n_{R}}\}} )^{'}$, as input. We can express the embedding layer as $\boldsymbol{\theta}^{e}(\boldsymbol{h}_{R}^{\text{one-hot}}) = \left(\theta_1^e(\boldsymbol{h}_{R}^{\text{one-hot}}), \ldotp\ldotp, \theta_{q_e}^e(\boldsymbol{h}_{R}^{\text{one-hot}})\right)^{\prime}$. The embedding dimension $q_{e}$ is a hyperparameter that needs to be predefined and has to be smaller than the number of classes $n_{R}$ at the most granular level in the original hierarchical structure, which is the input dimension of the embedding layer. The $u$th neuron of the embedding layer can be defined as
\begin{align*}
\theta_u^e(\boldsymbol{h}_{R}^{\text{one-hot}})=\left\langle\boldsymbol{w}_u^e, \boldsymbol{h}_{R}^{\text{one-hot}}\right\rangle \,\, \text{for} \,\, u=1,\ldotp \ldotp,q_{e},
\end{align*}
where $\boldsymbol{w}_u^e= (w_{u,1}^e,\ldotp\ldotp,w_{u,n_{R}}^e) \in \mathbb{R}^{n_{R}}$ refers to the weight vector which needs to be estimated for that neuron. In Figure~\ref{fig:network}, the layers between the input and output layer $\boldsymbol{\theta}^1,\ldotp\ldotp,\boldsymbol{\theta}^M$ are referred to as the hidden layers. Following the notation of \cite{delong2023use} and \cite{schelldorfer2019nesting}, the $m$th hidden layer $\boldsymbol{\theta}^m(\boldsymbol{z})=\left(\theta_1^m(\boldsymbol{z}), \ldots, \theta_{q_m}^m(\boldsymbol{z})\right)^{\prime} \in \mathbb{R}^{q_m}$ consists of $q_{m}$ neurons, which can be interpreted as regression models where the input is the output of all neurons of the previous layer. The $u$th neuron is in itself a mapping and is defined as
\begin{align*}
    \theta_u^m(\boldsymbol{z})=\varphi^{m}\left(b_u^m+\left\langle\boldsymbol{w}_u^m, \boldsymbol{z}\right\rangle\right) \,\, \text{for} \,\, u=1,\ldotp \ldotp,q_{m},
\end{align*}
where $\boldsymbol{z}$ is $q_{m-1}$-dimensional input and $b_u^m \in \mathbb{R}$ and $\boldsymbol{w}_u^m = (w_{u,1}^m,\ldotp\ldotp,w_{u,q_{m-1}}^m) \in \mathbb{R}^{q_{m-1}}$ refer to the bias and weight vector respectively, which are the parameters to be estimated for the $m$th hidden layer. $\varphi^m(.)$ is commonly known as an activation function in layer $m$, and allows the network to handle non-linearity by transforming the output of each neuron in the $m$th hidden layer before it is passed on to the next layer. The number of neurons $q_{m}$ and the activation function $\varphi^{m}(.)$ as well as the number of hidden layers $M$ are hyperparameters that need to be tuned when calibrating the network. \added{The choice of these parameters impacts the model performance as well as the computational cost. For tuning practices, we refer to Section 11 in \cite{goodfellow2016deep}.}

\begin{figure}[H]
	\tikzset{every picture/.style={line width=0.75pt}} 

\begin{tikzpicture}[x=0.75pt,y=0.75pt,yscale=-1,xscale=1]

\draw [line width=1.5]    (257,83) -- (292,83) ;
\draw [shift={(299,83)}, rotate = 180] [fill={rgb, 255:red, 0; green, 0; blue, 0 }  ][line width=0.08]  [draw opacity=0] (11.61,-5.58) -- (0,0) -- (11.61,5.58) -- cycle    ;
\draw [line width=1.5]    (395,83) -- (430,83) ;
\draw [shift={(437,83)}, rotate = 180] [fill={rgb, 255:red, 0; green, 0; blue, 0 }  ][line width=0.08]  [draw opacity=0] (11.61,-5.58) -- (0,0) -- (11.61,5.58) -- cycle    ;
\draw  [color={rgb, 255:red, 75; green, 73; blue, 73 }  ,draw opacity=1 ][fill={rgb, 255:red, 155; green, 155; blue, 155 }  ,fill opacity=1 ] (322,15) -- (365,15) -- (365,158) -- (322,158) -- cycle ;

\draw (30,25) node [anchor=north west][inner sep=0.75pt]   [align=left] {$ \boldsymbol{h}_{2,1}^{\text{one-hot}} =\ ( 1,0,0,0,0,0,0,0,0) '$};
\draw (92,73.04) node [anchor=north west][inner sep=0.75pt]  [font=\Large,rotate=-90.26] [align=left] {...};
\draw (30,114) node [anchor=north west][inner sep=0.75pt]   [align=left] {$ \boldsymbol{h}_{2,9}^{\text{one-hot}} =\ ( 0,0,0,0,0,0,0,0,1) '$};
\draw (336,75) node [anchor=north west][inner sep=0.75pt]   [align=left] {$ \boldsymbol{\theta }^{e}$};
\draw (450,25) node [anchor=north west][inner sep=0.75pt]   [align=left] {$ \boldsymbol{e}_{2,1} = \left( w_{1,1}^{e} ,\dotsc ,w_{q_{e},1}^{e}\right)$};
\draw (487,73.04) node [anchor=north west][inner sep=0.75pt]  [font=\Large,rotate=-90.26] [align=left] {...};
\draw (450,113) node [anchor=north west][inner sep=0.75pt]   [align=left] {$ \boldsymbol{e}_{2,9}= \left( w_{1,9}^{e} ,\dotsc ,w_{q_{e} ,9}^{e}\right)$};

\end{tikzpicture}
    \centering
        \caption{For the classes $h_{2,1},\ldotp\ldotp,h_{2,9}$ at the lowest level in Figure~\ref{fig:hierarchyexample}, different embedding vectors $\boldsymbol{e}_{2,1},\ldotp\ldotp,\boldsymbol{e}_{2,9}$ are learned by training the FFNN using the one-hot encodings as input for the embedding mapping $\boldsymbol{\theta}^{e}$.}
        \label{fig:mapping}
\end{figure}
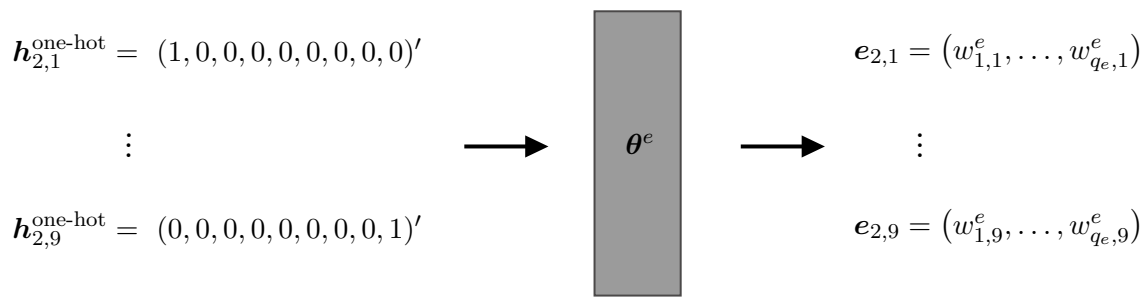

 The dimension $q_{M+1}$ of the output layer depends on the type of response variable. In our examples, we will work with $y \in \mathbb{R}$. Therefore, we specify $q_{M+1}=1$. Ultimately, the network prediction is given by
\begin{align*}
     \hat{y} =\theta^{M+1}(\boldsymbol{x},\boldsymbol{\theta}^{e}(\boldsymbol{h}_{R}^{\text{one-hot}}))= \varphi^{M+1}(b^{M+1}+\left\langle\boldsymbol{w}^{M+1},\left(\theta^M \circ \cdots \circ \theta^1\right)(\boldsymbol{x},\boldsymbol{\theta}^{e}(\boldsymbol{h}_{R}^{\text{one-hot}}))\right\rangle). 
\end{align*}
The activation function $\varphi^{M+1}(.)$ should take into account the type of response variable, for example, if the response variable is strictly positive this can be enforced by using the exponential function as the activation function in the output layer. The network is calibrated by minimising
\begin{align*}
    \sum_{i=1}^{n} \mathcal{L}_{NN}(y_{i},\hat{y}_{i})
\end{align*}
with respect to the weights and biases, where $\mathcal{L}_{NN}(.,.)$ refers to a loss function appropriate for the response variable. The embedding mapping $\boldsymbol{\theta}^{e}(\boldsymbol{h}_{R}^{\text{one-hot}})$ then provides us with a different embedding vector in $\mathbb{R}^{q_{e}}$, for each of $h_{R}$'s classes. For ease of notation, we define
\begin{align*}
    \boldsymbol{e}_{R,s} = \boldsymbol{\theta}^{e}(\boldsymbol{h}_{R,s}^{\text{one-hot}}) = (w_{1,s}^{e},\ldotp\ldotp,w_{q_{e},s}^{e}) \in \mathbb{R}^{q_{e}} \, \, \text{for} \, \, s=1,\ldotp\ldotp,n_{R} 
\end{align*}
as the embedding for class $h_{R,s}$. Figure~\ref{fig:mapping} illustrates this for the hierarchical variable of Figure~\ref{fig:hierarchyexample}. 
%


\subsection{Entity embedding for a hierarchy}
Using the approach from Section~\ref{sec:network}, we obtain an embedding for each of the classes at the most granular level in the hierarchy. \added{Next, we use the information encoded in these embeddings} to obtain representations for all classes at each level in $\boldsymbol{h}$'s hierarchical structure, i.e.~for $h_{r,s}$ with $r=1,\ldotp\ldotp,R , \, \text{and} \, s=1,\ldotp\ldotp,n_{r}$. Moreover, we want these to have dimension $q_{e}$ as well, to enable the comparison of embeddings at different levels in Section~\ref{sec:collapsing}. \cite{mumtaz2022hierarchy} aggregated embedding vectors representing a single class to obtain an embedding vector for the case that an observation can belong to multiple classes at the same time, i.e.~in the context of a so-called multi-valued categorical variable. Here, we apply a similar idea and use a summary measure to the embedding vectors over the set of classes with the same parent class in the hierarchical structure. \added{Each class' embedding at the most granular level differs. To enable between-level comparison of classes, see Section~\ref{sec:collapsing}, we require a summary measure that results in an embedding for a parent class that is different from its descendants. Therefore, and because of its convenience in computation, we opt for the mean over alternatives such as the median which can select the embedding vector of a descendant class as the representation for the parent.} For example, the embedding vector $\boldsymbol{e}_{1,1}$ for class $h_{1,1}$ at the first level of the hierarchy, pictured in Figure~\ref{fig:hierarchyexample}, is constructed as the average of the embedding vectors of the classes in $\mathcal{H}_{1,1}$, i.e. $\boldsymbol{e}_{2,1}$, $\boldsymbol{e}_{2,2}$ and $\boldsymbol{e}_{2,3}$. We construct the embeddings at the $R-1$ levels higher up in the hierarchy as follows:
\begin{align*}
    \boldsymbol{e}_{r,s} =  \frac{1}{\text{dim}(\mathcal{H}_{r,s})}  \sum_{l | h_{r+1,l} \in \mathcal{H}_{r,s}}{}  \boldsymbol{e}_{r+1,l} 
\,\,\, \in \mathbb{R}^{q_{e}} \,\,\, \text{for} \, r=1,\ldotp\ldotp,R-1 , \, \text{and} \, s=1,\ldotp\ldotp,n_{r}.
\end{align*}
Since we already have embedding vectors for each class $h_{R,s}$ at the most granular level $R$ in the hierarchy, we immediately obtain the embedding $e_{R-1,s}$ for each class $h_{R-1,s}$ at level $R-1$. \added{As a result, we now have an embedding space at level $R-1$ where each class is represented using a numerical encoding that takes into account that class' descendants.} We can then use these to compute the embeddings at level $R-2$ and so forth, by moving bottom-up through the hierarchy. Consequently, we have a $q_{e}$-dimensional embedding vector for each class at every level in the hierarchy of $\boldsymbol{h}$\added{, enabling within-level as well as between-level comparison}. In the next section, we apply clustering techniques to the embeddings to learn a reduced representation $\widetilde{\boldsymbol{h}}$ of the original hierarchy with respect to the response variable $y$.


\section{Reducing a hierarchy } \label{sec:collapsing}
We propose a top-down clustering algorithm that aims to reduce the dimensionality and granularity of a hierarchical categorical variable $\boldsymbol{h}$ by merging classes that have a sufficiently similar effect on the response variable. Our algorithm consists of two steps: a horizontal (within-level) clustering step and a vertical (between-level) clustering step. Given a level $r$, we first perform clustering on each of the subsets of classes within level $r$ that have a common parent class, resulting in $\widetilde{n}_{r}$ clusters. Next, for each of the formed clusters, we collapse its descendant classes at level $r+1$ that are sufficiently close within the embedding space. The term `collapsing' refers to merging descendants with their parent class (or cluster). Both steps are repeated until the lowest level in the hierarchy is reached, to which only the first step is applied. The result is a reduced representation of the hierarchy, which we denote by $\boldsymbol{\widetilde{h}}$. The full details of the algorithm are given in Appendix~\ref{append:algo}.

Our algorithm does not directly cluster the classes themselves, but instead uses their embeddings. \cite{guo2016entity} showed that entity embedding puts similar classes of a categorical variable close to each other in an Euclidean space. As a result, any clustering technique can be applied to the embedding vectors. In Figure~\ref{fig:hierarchyexample} for example, if we want to cluster the classes in the set $H_{1}=\{h_{1,1},h_{1,2},h_{1,3} \}$, we apply a clustering technique to the set $\{ \boldsymbol{e}_{r,s} \: | \:  h_{r,s} \in H_{1} \}$. The classes $h_{1,1}$, $h_{1,2}$ and $h_{1,3}$ are subsequently merged according to the clustering of their respective embeddings $\boldsymbol{e}_{1,1}$, $\boldsymbol{e}_{1,2}$ and $\boldsymbol{e}_{1,3}$. When a cluster is formed, we define its embedding as the average of the cluster members' embedding. 

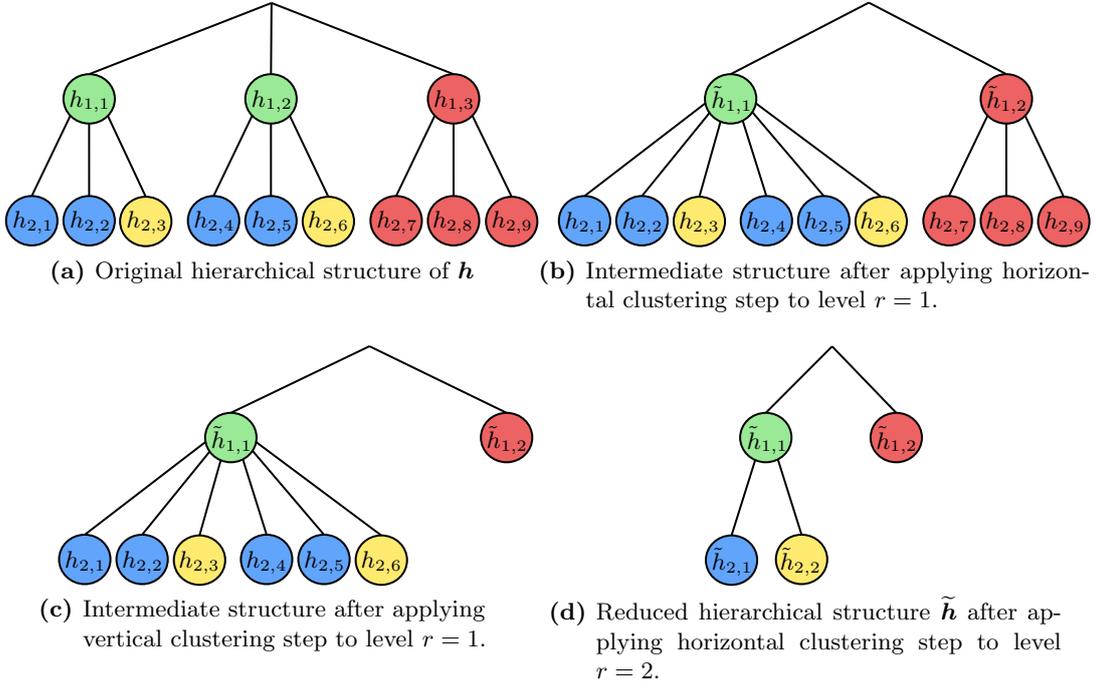
\begin{figure}[H] 
\centering
\subfloat[Original hierarchical structure of $\boldsymbol{h}$]{
\label{fig:collapse:a}
\tikzset{every picture/.style={line width=0.75pt}} 

\begin{tikzpicture}[x=0.45pt,y=0.45pt,yscale=-0.62,xscale=0.62]

\draw    (360.5,6.17) -- (115.8,102.43) ;
\draw    (360.5,6.17) -- (359.8,102.43) ;
\draw    (360.5,6.17) -- (604.8,102.43) ;
\draw    (99.78,139.19) -- (38.78,267.19) ;
\draw    (115.8,144.68) -- (115.75,179.19) -- (115.78,267.19) ;
\draw    (131.78,139.19) -- (191.78,268.19) ;
\draw    (343.78,139.19) -- (282.78,267.19) ;
\draw    (359.8,144.68) -- (359.75,179.19) -- (359.78,267.19) ;
\draw    (375.78,139.19) -- (436.78,268.19) ;
\draw    (588.78,139.19) -- (527.78,267.19) ;
\draw    (604.8,144.68) -- (604.75,179.19) -- (604.78,267.19) ;
\draw    (620.78,139.19) -- (682.78,268.19) ;
\draw  [fill={rgb, 255:red, 97; green, 164; blue, 252 }  ,fill opacity=1 ] (4.3,300.59) .. controls (4.3,282.14) and (19.74,267.19) .. (38.78,267.19) .. controls (57.82,267.19) and (73.26,282.14) .. (73.26,300.59) .. controls (73.26,319.04) and (57.82,334) .. (38.78,334) .. controls (19.74,334) and (4.3,319.04) .. (4.3,300.59) -- cycle ;
\draw  [fill={rgb, 255:red, 97; green, 164; blue, 252 }  ,fill opacity=1 ] (81.3,300.59) .. controls (81.3,282.14) and (96.74,267.19) .. (115.78,267.19) .. controls (134.82,267.19) and (150.26,282.14) .. (150.26,300.59) .. controls (150.26,319.04) and (134.82,334) .. (115.78,334) .. controls (96.74,334) and (81.3,319.04) .. (81.3,300.59) -- cycle ;
\draw  [fill={rgb, 255:red, 97; green, 164; blue, 252 }  ,fill opacity=1 ] (248.3,300.59) .. controls (248.3,282.14) and (263.74,267.19) .. (282.78,267.19) .. controls (301.82,267.19) and (317.26,282.14) .. (317.26,300.59) .. controls (317.26,319.04) and (301.82,334) .. (282.78,334) .. controls (263.74,334) and (248.3,319.04) .. (248.3,300.59) -- cycle ;
\draw  [fill={rgb, 255:red, 97; green, 164; blue, 252 }  ,fill opacity=1 ] (325.3,300.59) .. controls (325.3,282.14) and (340.74,267.19) .. (359.78,267.19) .. controls (378.82,267.19) and (394.26,282.14) .. (394.26,300.59) .. controls (394.26,319.04) and (378.82,334) .. (359.78,334) .. controls (340.74,334) and (325.3,319.04) .. (325.3,300.59) -- cycle ;
\draw  [fill={rgb, 255:red, 254; green, 234; blue, 111 }  ,fill opacity=1 ] (402.3,301.59) .. controls (402.3,283.14) and (417.74,268.19) .. (436.78,268.19) .. controls (455.82,268.19) and (471.26,283.14) .. (471.26,301.59) .. controls (471.26,320.04) and (455.82,335) .. (436.78,335) .. controls (417.74,335) and (402.3,320.04) .. (402.3,301.59) -- cycle ;
\draw  [fill={rgb, 255:red, 254; green, 234; blue, 111 }  ,fill opacity=1 ] (157.3,301.59) .. controls (157.3,283.14) and (172.74,268.19) .. (191.78,268.19) .. controls (210.82,268.19) and (226.26,283.14) .. (226.26,301.59) .. controls (226.26,320.04) and (210.82,335) .. (191.78,335) .. controls (172.74,335) and (157.3,320.04) .. (157.3,301.59) -- cycle ;
\draw  [fill={rgb, 255:red, 237; green, 98; blue, 98 }  ,fill opacity=1 ] (570.3,300.59) .. controls (570.3,282.14) and (585.74,267.19) .. (604.78,267.19) .. controls (623.82,267.19) and (639.26,282.14) .. (639.26,300.59) .. controls (639.26,319.04) and (623.82,334) .. (604.78,334) .. controls (585.74,334) and (570.3,319.04) .. (570.3,300.59) -- cycle ;
\draw  [fill={rgb, 255:red, 237; green, 98; blue, 98 }  ,fill opacity=1 ] (570.33,135.83) .. controls (570.33,117.38) and (585.76,102.43) .. (604.8,102.43) .. controls (623.84,102.43) and (639.28,117.38) .. (639.28,135.83) .. controls (639.28,154.28) and (623.84,169.24) .. (604.8,169.24) .. controls (585.76,169.24) and (570.33,154.28) .. (570.33,135.83) -- cycle ;
\draw  [fill={rgb, 255:red, 237; green, 98; blue, 98 }  ,fill opacity=1 ] (493.3,300.59) .. controls (493.3,282.14) and (508.74,267.19) .. (527.78,267.19) .. controls (546.82,267.19) and (562.26,282.14) .. (562.26,300.59) .. controls (562.26,319.04) and (546.82,334) .. (527.78,334) .. controls (508.74,334) and (493.3,319.04) .. (493.3,300.59) -- cycle ;
\draw  [fill={rgb, 255:red, 237; green, 98; blue, 98 }  ,fill opacity=1 ] (648.3,301.59) .. controls (648.3,283.14) and (663.74,268.19) .. (682.78,268.19) .. controls (701.82,268.19) and (717.26,283.14) .. (717.26,301.59) .. controls (717.26,320.04) and (701.82,335) .. (682.78,335) .. controls (663.74,335) and (648.3,320.04) .. (648.3,301.59) -- cycle ;
\draw  [fill={rgb, 255:red, 153; green, 233; blue, 150 }  ,fill opacity=1 ] (81.33,135.83) .. controls (81.33,117.38) and (96.76,102.43) .. (115.8,102.43) .. controls (134.84,102.43) and (150.28,117.38) .. (150.28,135.83) .. controls (150.28,154.28) and (134.84,169.24) .. (115.8,169.24) .. controls (96.76,169.24) and (81.33,154.28) .. (81.33,135.83) -- cycle ;
\draw  [fill={rgb, 255:red, 153; green, 233; blue, 150 }  ,fill opacity=1 ] (325.33,135.83) .. controls (325.33,117.38) and (340.76,102.43) .. (359.8,102.43) .. controls (378.84,102.43) and (394.28,117.38) .. (394.28,135.83) .. controls (394.28,154.28) and (378.84,169.24) .. (359.8,169.24) .. controls (340.76,169.24) and (325.33,154.28) .. (325.33,135.83) -- cycle ;

\draw (85,120) node [anchor=north west][inner sep=0.75pt]  [font=\footnotesize] [align=left] {$ $$\displaystyle h_{1,1}$};
\draw (329.5,120) node [anchor=north west][inner sep=0.75pt]  [font=\footnotesize] [align=left] {$\displaystyle h_{1,2}$};
\draw (574,120) node [anchor=north west][inner sep=0.75pt]  [font=\footnotesize] [align=left] {$\displaystyle h_{1,3}$};
\draw (9,284) node [anchor=north west][inner sep=0.75pt]  [font=\footnotesize] [align=left] {$\displaystyle h_{2,1}$};
\draw (86,284) node [anchor=north west][inner sep=0.75pt]  [font=\footnotesize] [align=left] {$\displaystyle h_{2,2}$};
\draw (162,284) node [anchor=north west][inner sep=0.75pt]  [font=\footnotesize] [align=left] {$\displaystyle h_{2,3}$};
\draw (252,284) node [anchor=north west][inner sep=0.75pt]  [font=\footnotesize] [align=left] {$\displaystyle h_{2,4}$};
\draw (330,284) node [anchor=north west][inner sep=0.75pt]  [font=\footnotesize] [align=left] {$\displaystyle h_{2,5}$};
\draw (405.5,284) node [anchor=north west][inner sep=0.75pt]  [font=\footnotesize] [align=left] {$\displaystyle h_{2,6}$};
\draw (497.5,284) node [anchor=north west][inner sep=0.75pt]  [font=\footnotesize] [align=left] {$\displaystyle h_{2,7}$};
\draw (572.5,284) node [anchor=north west][inner sep=0.75pt]  [font=\footnotesize] [align=left] {$\displaystyle h_{2,8}$};
\draw (652,284) node [anchor=north west][inner sep=0.75pt]  [font=\footnotesize] [align=left] {$\displaystyle h_{2,9}$};

\end{tikzpicture}} 
\subfloat[Intermediate structure after applying horizontal clustering step to level $r=1$.]{
\label{fig:collapse:b}
\tikzset{every picture/.style={line width=0.75pt}} 

\begin{tikzpicture}[x=0.45pt,y=0.45pt,yscale=-0.62,xscale=0.62]

\draw    (421.09,4.62) -- (235.69,101.35) ;
\draw    (421.09,4.62) -- (605.69,101.35) ;
\draw    (216.67,129.11) -- (39.67,266.11) ;
\draw    (228.67,142.11) -- (193.67,267.11) ;
\draw    (244.67,141.11) -- (283.67,266.11) ;
\draw    (256.67,131.11) -- (437.67,267.11) ;
\draw    (589.67,138.11) -- (528.67,266.11) ;
\draw    (605.69,143.6) -- (605.63,178.11) -- (605.67,266.11) ;
\draw    (621.67,138.11) -- (682.67,267.11) ;
\draw    (220.67,137.11) -- (116.67,266.11) ;
\draw    (251.67,137.11) -- (360.67,266.11) ;
\draw  [fill={rgb, 255:red, 97; green, 164; blue, 252 }  ,fill opacity=1 ] (5.19,299.51) .. controls (5.19,281.06) and (20.62,266.11) .. (39.67,266.11) .. controls (58.71,266.11) and (74.14,281.06) .. (74.14,299.51) .. controls (74.14,317.96) and (58.71,332.92) .. (39.67,332.92) .. controls (20.62,332.92) and (5.19,317.96) .. (5.19,299.51) -- cycle ;
\draw  [fill={rgb, 255:red, 97; green, 164; blue, 252 }  ,fill opacity=1 ] (82.19,299.51) .. controls (82.19,281.06) and (97.62,266.11) .. (116.67,266.11) .. controls (135.71,266.11) and (151.14,281.06) .. (151.14,299.51) .. controls (151.14,317.96) and (135.71,332.92) .. (116.67,332.92) .. controls (97.62,332.92) and (82.19,317.96) .. (82.19,299.51) -- cycle ;
\draw  [fill={rgb, 255:red, 97; green, 164; blue, 252 }  ,fill opacity=1 ] (249.19,299.51) .. controls (249.19,281.06) and (264.62,266.11) .. (283.67,266.11) .. controls (302.71,266.11) and (318.14,281.06) .. (318.14,299.51) .. controls (318.14,317.96) and (302.71,332.92) .. (283.67,332.92) .. controls (264.62,332.92) and (249.19,317.96) .. (249.19,299.51) -- cycle ;
\draw  [fill={rgb, 255:red, 97; green, 164; blue, 252 }  ,fill opacity=1 ] (326.19,299.51) .. controls (326.19,281.06) and (341.62,266.11) .. (360.67,266.11) .. controls (379.71,266.11) and (395.14,281.06) .. (395.14,299.51) .. controls (395.14,317.96) and (379.71,332.92) .. (360.67,332.92) .. controls (341.62,332.92) and (326.19,317.96) .. (326.19,299.51) -- cycle ;
\draw  [fill={rgb, 255:red, 254; green, 234; blue, 111 }  ,fill opacity=1 ] (159.19,300.51) .. controls (159.19,282.06) and (174.62,267.11) .. (193.67,267.11) .. controls (212.71,267.11) and (228.14,282.06) .. (228.14,300.51) .. controls (228.14,318.96) and (212.71,333.92) .. (193.67,333.92) .. controls (174.62,333.92) and (159.19,318.96) .. (159.19,300.51) -- cycle ;
\draw  [fill={rgb, 255:red, 254; green, 234; blue, 111 }  ,fill opacity=1 ] (403.19,300.51) .. controls (403.19,282.06) and (418.62,267.11) .. (437.67,267.11) .. controls (456.71,267.11) and (472.14,282.06) .. (472.14,300.51) .. controls (472.14,318.96) and (456.71,333.92) .. (437.67,333.92) .. controls (418.62,333.92) and (403.19,318.96) .. (403.19,300.51) -- cycle ;
\draw  [fill={rgb, 255:red, 237; green, 98; blue, 98 }  ,fill opacity=1 ] (571.21,134.75) .. controls (571.21,116.3) and (586.65,101.35) .. (605.69,101.35) .. controls (624.73,101.35) and (640.16,116.3) .. (640.16,134.75) .. controls (640.16,153.2) and (624.73,168.16) .. (605.69,168.16) .. controls (586.65,168.16) and (571.21,153.2) .. (571.21,134.75) -- cycle ;
\draw  [fill={rgb, 255:red, 237; green, 98; blue, 98 }  ,fill opacity=1 ] (494.19,299.51) .. controls (494.19,281.06) and (509.62,266.11) .. (528.67,266.11) .. controls (547.71,266.11) and (563.14,281.06) .. (563.14,299.51) .. controls (563.14,317.96) and (547.71,332.92) .. (528.67,332.92) .. controls (509.62,332.92) and (494.19,317.96) .. (494.19,299.51) -- cycle ;
\draw  [fill={rgb, 255:red, 237; green, 98; blue, 98 }  ,fill opacity=1 ] (571.19,299.51) .. controls (571.19,281.06) and (586.62,266.11) .. (605.67,266.11) .. controls (624.71,266.11) and (640.14,281.06) .. (640.14,299.51) .. controls (640.14,317.96) and (624.71,332.92) .. (605.67,332.92) .. controls (586.62,332.92) and (571.19,317.96) .. (571.19,299.51) -- cycle ;
\draw  [fill={rgb, 255:red, 237; green, 98; blue, 98 }  ,fill opacity=1 ] (648.19,300.51) .. controls (648.19,282.06) and (663.62,267.11) .. (682.67,267.11) .. controls (701.71,267.11) and (717.14,282.06) .. (717.14,300.51) .. controls (717.14,318.96) and (701.71,333.92) .. (682.67,333.92) .. controls (663.62,333.92) and (648.19,318.96) .. (648.19,300.51) -- cycle ;
\draw  [fill={rgb, 255:red, 153; green, 233; blue, 150 }  ,fill opacity=1 ] (201.21,134.75) .. controls (201.21,116.3) and (216.65,101.35) .. (235.69,101.35) .. controls (254.73,101.35) and (270.16,116.3) .. (270.16,134.75) .. controls (270.16,153.2) and (254.73,168.16) .. (235.69,168.16) .. controls (216.65,168.16) and (201.21,153.2) .. (201.21,134.75) -- cycle ;

\draw (206,113) node [anchor=north west][inner sep=0.75pt]  [font=\footnotesize] [align=left] {$ $$\displaystyle \tilde{h}_{1,1}$};
\draw (9,284) node [anchor=north west][inner sep=0.75pt]  [font=\footnotesize] [align=left] {$\displaystyle h_{2,1}$};
\draw (86,284) node [anchor=north west][inner sep=0.75pt]  [font=\footnotesize] [align=left] {$\displaystyle h_{2,2}$};
\draw (162,284) node [anchor=north west][inner sep=0.75pt]  [font=\footnotesize] [align=left] {$\displaystyle h_{2,3}$};
\draw (252,284) node [anchor=north west][inner sep=0.75pt]  [font=\footnotesize] [align=left] {$\displaystyle h_{2,4}$};
\draw (330,284) node [anchor=north west][inner sep=0.75pt]  [font=\footnotesize] [align=left] {$\displaystyle h_{2,5}$};
\draw (405.5,284) node [anchor=north west][inner sep=0.75pt]  [font=\footnotesize] [align=left] {$\displaystyle h_{2,6}$};
\draw (497.5,284) node [anchor=north west][inner sep=0.75pt]  [font=\footnotesize] [align=left] {$\displaystyle h_{2,7}$};
\draw (573.5,284) node [anchor=north west][inner sep=0.75pt]  [font=\footnotesize] [align=left] {$\displaystyle h_{2,8}$};
\draw (652,284) node [anchor=north west][inner sep=0.75pt]  [font=\footnotesize] [align=left] {$\displaystyle h_{2,9}$};
\draw (575,113) node [anchor=north west][inner sep=0.75pt]  [font=\footnotesize] [align=left] {$ $$\displaystyle \tilde{h}_{1,2}$};

\end{tikzpicture}}\\
\hspace{0.5em}
\subfloat[Intermediate structure after applying\\ vertical clustering step to level $r=1$.]{
\label{fig:collapse:c}
\tikzset{every picture/.style={line width=0.75pt}} 

\begin{tikzpicture}[x=0.45pt,y=0.45pt,yscale=-0.62,xscale=0.62]

\draw    (216.78,122.33) -- (39.78,259.33) ;
\draw    (228.78,135.33) -- (193.78,260.33) ;
\draw    (244.78,134.33) -- (283.78,259.33) ;
\draw    (256.78,124.33) -- (437.78,260.33) ;
\draw    (220.78,130.33) -- (116.78,259.33) ;
\draw    (251.78,130.33) -- (360.78,259.33) ;
\draw    (421.78,4.49) -- (235.8,94.57) ;
\draw    (421.78,4.49) -- (605.8,94.57) ;
\draw  [fill={rgb, 255:red, 254; green, 234; blue, 111 }  ,fill opacity=1 ] (159.3,293.74) .. controls (159.3,275.29) and (174.74,260.33) .. (193.78,260.33) .. controls (212.82,260.33) and (228.26,275.29) .. (228.26,293.74) .. controls (228.26,312.19) and (212.82,327.14) .. (193.78,327.14) .. controls (174.74,327.14) and (159.3,312.19) .. (159.3,293.74) -- cycle ;
\draw  [fill={rgb, 255:red, 97; green, 164; blue, 252 }  ,fill opacity=1 ] (5.3,292.74) .. controls (5.3,274.29) and (20.74,259.33) .. (39.78,259.33) .. controls (58.82,259.33) and (74.26,274.29) .. (74.26,292.74) .. controls (74.26,311.19) and (58.82,326.14) .. (39.78,326.14) .. controls (20.74,326.14) and (5.3,311.19) .. (5.3,292.74) -- cycle ;
\draw  [fill={rgb, 255:red, 97; green, 164; blue, 252 }  ,fill opacity=1 ] (82.3,292.74) .. controls (82.3,274.29) and (97.74,259.33) .. (116.78,259.33) .. controls (135.82,259.33) and (151.26,274.29) .. (151.26,292.74) .. controls (151.26,311.19) and (135.82,326.14) .. (116.78,326.14) .. controls (97.74,326.14) and (82.3,311.19) .. (82.3,292.74) -- cycle ;
\draw  [fill={rgb, 255:red, 97; green, 164; blue, 252 }  ,fill opacity=1 ] (249.3,292.74) .. controls (249.3,274.29) and (264.74,259.33) .. (283.78,259.33) .. controls (302.82,259.33) and (318.26,274.29) .. (318.26,292.74) .. controls (318.26,311.19) and (302.82,326.14) .. (283.78,326.14) .. controls (264.74,326.14) and (249.3,311.19) .. (249.3,292.74) -- cycle ;
\draw  [fill={rgb, 255:red, 97; green, 164; blue, 252 }  ,fill opacity=1 ] (326.3,292.74) .. controls (326.3,274.29) and (341.74,259.33) .. (360.78,259.33) .. controls (379.82,259.33) and (395.26,274.29) .. (395.26,292.74) .. controls (395.26,311.19) and (379.82,326.14) .. (360.78,326.14) .. controls (341.74,326.14) and (326.3,311.19) .. (326.3,292.74) -- cycle ;
\draw  [fill={rgb, 255:red, 254; green, 234; blue, 111 }  ,fill opacity=1 ] (403.3,293.74) .. controls (403.3,275.29) and (418.74,260.33) .. (437.78,260.33) .. controls (456.82,260.33) and (472.26,275.29) .. (472.26,293.74) .. controls (472.26,312.19) and (456.82,327.14) .. (437.78,327.14) .. controls (418.74,327.14) and (403.3,312.19) .. (403.3,293.74) -- cycle ;
\draw  [fill={rgb, 255:red, 237; green, 98; blue, 98 }  ,fill opacity=1 ] (571.33,127.98) .. controls (571.33,109.53) and (586.76,94.57) .. (605.8,94.57) .. controls (624.84,94.57) and (640.28,109.53) .. (640.28,127.98) .. controls (640.28,146.43) and (624.84,161.38) .. (605.8,161.38) .. controls (586.76,161.38) and (571.33,146.43) .. (571.33,127.98) -- cycle ;
\draw  [fill={rgb, 255:red, 153; green, 233; blue, 150 }  ,fill opacity=1 ] (201.33,127.98) .. controls (201.33,109.53) and (216.76,94.57) .. (235.8,94.57) .. controls (254.84,94.57) and (270.28,109.53) .. (270.28,127.98) .. controls (270.28,146.43) and (254.84,161.38) .. (235.8,161.38) .. controls (216.76,161.38) and (201.33,146.43) .. (201.33,127.98) -- cycle ;

\draw (575,108) node [anchor=north west][inner sep=0.75pt]  [font=\footnotesize] [align=left] {$ $$\displaystyle \tilde{h}_{1,2}$};
\draw (205.5,108) node [anchor=north west][inner sep=0.75pt]  [font=\footnotesize] [align=left] {$ $$\displaystyle \tilde{h}_{1,1}$};
\draw (9,278) node [anchor=north west][inner sep=0.75pt]  [font=\footnotesize] [align=left] {$\displaystyle h_{2,1}$};
\draw (86,278) node [anchor=north west][inner sep=0.75pt]  [font=\footnotesize] [align=left] {$\displaystyle h_{2,2}$};
\draw (162,278) node [anchor=north west][inner sep=0.75pt]  [font=\footnotesize] [align=left] {$\displaystyle h_{2,3}$};
\draw (252,278) node [anchor=north west][inner sep=0.75pt]  [font=\footnotesize] [align=left] {$\displaystyle h_{2,4}$};
\draw (330,278) node [anchor=north west][inner sep=0.75pt]  [font=\footnotesize] [align=left] {$\displaystyle h_{2,5}$};
\draw (406.5,278) node [anchor=north west][inner sep=0.75pt]  [font=\footnotesize] [align=left] {$\displaystyle h_{2,6}$};

\end{tikzpicture}}
\hspace{0.2em}
\subfloat[Reduced hierarchical structure $\boldsymbol{\widetilde{h}}$ after applying horizontal clustering step to level $r=2$.]{
\label{fig:collapse:d}
\tikzset{every picture/.style={line width=0.75pt}} 

\tikzset{every picture/.style={line width=0.75pt}} 

\tikzset{every picture/.style={line width=0.75pt}} 

\begin{tikzpicture}[x=0.45pt,y=0.45pt,yscale=-0.62,xscale=0.62]

\draw    (330.43,3.78) -- (241.46,93.66) ;
\draw    (330.43,3.78) -- (416.46,93.66) ;
\draw    (234.43,134.42) -- (195.43,259.42) ;
\draw    (250.43,133.42) -- (289.43,258.42) ;
\draw  [fill={rgb, 255:red, 97; green, 164; blue, 252 }  ,fill opacity=1 ] (160.96,292.82) .. controls (160.96,274.38) and (176.39,259.42) .. (195.43,259.42) .. controls (214.47,259.42) and (229.91,274.38) .. (229.91,292.82) .. controls (229.91,311.27) and (214.47,326.23) .. (195.43,326.23) .. controls (176.39,326.23) and (160.96,311.27) .. (160.96,292.82) -- cycle ;
\draw  [fill={rgb, 255:red, 254; green, 234; blue, 111 }  ,fill opacity=1 ] (254.96,291.82) .. controls (254.96,273.38) and (270.39,258.42) .. (289.43,258.42) .. controls (308.47,258.42) and (323.91,273.38) .. (323.91,291.82) .. controls (323.91,310.27) and (308.47,325.23) .. (289.43,325.23) .. controls (270.39,325.23) and (254.96,310.27) .. (254.96,291.82) -- cycle ;
\draw  [fill={rgb, 255:red, 237; green, 98; blue, 98 }  ,fill opacity=1 ] (381.98,127.07) .. controls (381.98,108.62) and (397.41,93.66) .. (416.46,93.66) .. controls (435.5,93.66) and (450.93,108.62) .. (450.93,127.07) .. controls (450.93,145.52) and (435.5,160.47) .. (416.46,160.47) .. controls (397.41,160.47) and (381.98,145.52) .. (381.98,127.07) -- cycle ;
\draw  [fill={rgb, 255:red, 153; green, 233; blue, 150 }  ,fill opacity=1 ] (206.98,127.07) .. controls (206.98,108.62) and (222.41,93.66) .. (241.46,93.66) .. controls (260.5,93.66) and (275.93,108.62) .. (275.93,127.07) .. controls (275.93,145.52) and (260.5,160.47) .. (241.46,160.47) .. controls (222.41,160.47) and (206.98,145.52) .. (206.98,127.07) -- cycle ;

\draw (165,272) node [anchor=north west][inner sep=0.75pt]  [font=\footnotesize] [align=left] {$ $$\displaystyle \tilde{h}_{2,1}$};
\draw (256.5,271.5) node [anchor=north west][inner sep=0.75pt]  [font=\footnotesize] [align=left] {$ $$\displaystyle \tilde{h}_{2,2}$};
\draw (212,108) node [anchor=north west][inner sep=0.75pt]  [font=\footnotesize] [align=left] {$ $$\displaystyle \tilde{h}_{1,1}$};
\draw (385.5,108) node [anchor=north west][inner sep=0.75pt]  [font=\footnotesize] [align=left] {$ $$\displaystyle \tilde{h}_{1,2}$};
\draw (4,297.89) node [anchor=north west][inner sep=0.75pt]   [align=left] {\textcolor[rgb]{1,1,1}{.}};
\draw (621,283.89) node [anchor=north west][inner sep=0.75pt]   [align=left] {\textcolor[rgb]{1,1,1}{.}};

\end{tikzpicture}}
\caption{Toy example of applying the top-down clustering algorithm in Appendix~\ref{append:algo} to a hierarchical variable $\boldsymbol{h}$ with $R=2$ levels to obtain a reduced representation $\boldsymbol{\widetilde{h}}$. Classes with the same effect on the response variable are represented using the same colour.} 
\label{fig:collapse}
\end{figure}

\paragraph{Toy example}Figure~\ref{fig:collapse} gives a step-by-step toy example of the algorithm's working principles. We consider a hierarchical variable $\boldsymbol{h}$ with $h_{1} \in H_{1}=\{h_{1,1},h_{1,2},h_{1,3} \}$, $h_{2} \in H_{2}=\{h_{2,1},\ldotp\ldotp,h_{2,9}\}$, $\mathcal{H}_{1,1}=\{h_{2,1},h_{2,2},h_{2,3}\}$, $\mathcal{H}_{1,2}=\{h_{2,4},h_{2,5},h_{2,6}\}$, and $\mathcal{H}_{1,3}=\{h_{2,7},h_{2,8},h_{2,9}\}$, as shown in Figure~\ref{fig:collapse:a}. For illustration purposes, we assume that we know which classes have the same effect on the response variable. Hence, we know in this toy example the ultimately reduced specification of the hierarchical covariate, pictured in Figure~\ref{fig:collapse:d}. The objective of the algorithm is to construct a new hierarchical categorical variable $\boldsymbol{\widetilde{h}}$ that closely approximates the hierarchical structure as illustrated in Figure~\ref{fig:collapse:d}.

First, we perform a horizontal clustering step and aim to cluster classes at the upper level $r=1$ using their embeddings. The classes $h_{1,1}$ and $h_{1,2}$ are merged while $h_{1,3}$ is not, resulting in $C_{1,1}=\{h_{1,1},h_{1,2}\}$ and $C_{1,2}=\{h_{1,3}\}$. Consequently, we have $\mathcal{C}^{(1)}_{1,1}=\{h_{2,1},\ldotp\ldotp,h_{2,6}\}$ which is the set of classes that are descendants of the classes in $C_{1,1}$, $\mathcal{C}^{(1)}_{1,2}=\{h_{2,7},h_{2,8},h_{2,9}\}$ and $\widetilde{H}_{1}=\{\widetilde{h}_{1,1},\widetilde{h}_{1,2} \}$. The latter denotes the set of classes at the first level in the reduced representation $\boldsymbol{\widetilde{h}}$. The intermediate structure resulting from this step is pictured in Figure~\ref{fig:collapse:b}. 

Secondly, we apply a vertical clustering step to level $r=1$. Since we have $\widetilde{n}_1=2$ classes at level $r=1$ resulting from the previous step, we consider two clustering tasks. We apply a clustering technique to the embeddings of the classes in $\widetilde{h}_{1,1} \cup \mathcal{C}_{1,1}^{(1)}$ as well as to those in $\widetilde{h}_{1,2} \cup \mathcal{C}_{1,2}^{(1)}$. As pictured in Figure~\ref{fig:collapse:c}, we collapse $h_{2,7}$, $h_{2,8}$ and $h_{2,9}$ into their parent class $\widetilde{h}_{1,2}$. As a result, we define $\mathcal{C}_{1,1}^{(2)}=\mathcal{C}_{1,1}^{(1)}=\{h_{2,1},\ldotp\ldotp,h_{2,6}\}$ and $\mathcal{C}_{1,2}^{(2)}=\varnothing$.

Lastly, we apply again a horizontal clustering step, now to level $r=2$. Since we have $\widetilde{n}_{1}=2$, we have two subsets of classes to cluster, i.e.~$\mathcal{C}^{(2)}_{1,1}=\{h_{2,1},\ldotp\ldotp,h_{2,6}\}$ and $\mathcal{C}^{(2)}_{1,2}=\varnothing$. For the latter, we can immediately set $\widetilde{\mathcal{H}}_{1,2}=\varnothing$, which denotes the set of descendants of $\widetilde{h}_{1,2}$ in the structure of $\boldsymbol{\widetilde{h}}$. As a result of this horizontal clustering step, we find $\widetilde{n}_2=2$ clusters, i.e.~$C_{2,1}=\{h_{2,1},h_{2,2},h_{2,4},h_{2,5}\}$ and $C_{2,2}=\{h_{2,3},h_{2,6}\}$. Because both clusters have the same parent, we define $\widetilde{H}_{2}=\{\widetilde{h}_{2,1},\widetilde{h}_{2,2} \}$ and $\widetilde{\mathcal{H}}_{1,1}=\{\widetilde{h}_{2,1},\widetilde{h}_{2,2}\}$. We do not apply the vertical clustering step to level $r=2$ as there are no descendant classes at the lowest level in the hierarchy (here: $R=2$). Figure~\ref{fig:collapse:d} shows the resulting reduced representation $\boldsymbol{\widetilde{h}}=(\widetilde{h}_{1},\widetilde{h}_{2})$ with $\widetilde{R}=2$, $\widetilde{h}_{1} \in \widetilde{H}_{1}=\{\widetilde{h}_{1,1},\widetilde{h}_{1,2} \}$, $\widetilde{h}_{2} \in \widetilde{H}_{2}=\{\widetilde{h}_{2,1},\widetilde{h}_{2,2} \}$, $\widetilde{\mathcal{H}}_{1,1}=\{\widetilde{h}_{2,1},\widetilde{h}_{2,2}\}$ and $\widetilde{\mathcal{H}}_{1,2}=\varnothing$.
\paragraph{Clustering technique and distance metric} In each within-level as well as between-level step, a clustering technique is applied multiple times to different subsets of classes. Classes that are close to each other in the embedding space, or classes that are similar to their parent class, are clustered together. We opt for a hard clustering technique that results in non-overlapping clusters. \added{Among these, the k-medoids algorithm \citep{kaufman2009finding} is computationally less expensive than hierarchical clustering analysis (HCA) methods and less sensitive to outliers compared to k-means, making it well-suited for our approach. Specifically, we use the \texttt{clara} implementation \citep{kaufman2009finding} of the k-medoids algorithm.} To measure the (dis)similarity between two embedding vectors $\boldsymbol{e}_1$ and $\boldsymbol{e}_2$, we need a suitable metric $D(\boldsymbol{e}_1,\boldsymbol{e}_2)$. \added{When dealing with textual representations, the cosine similarity is often considered because the embeddings are sparse, i.e. they are highly dimensional with a lot of zero entries. However, in our case, the embedding dimension $q_e$ in general is relatively low since our original dimension depends on the number of classes in $h_R$.} The embedding space constructed in Section~\ref{sec:embedding} is Euclidean and the embedding vectors can be considered as quantitative attributes. Therefore, we use the Euclidean distance, 
\begin{align*}
    D(\boldsymbol{e}_1,\boldsymbol{e}_2) = || \boldsymbol{e}_1 - \boldsymbol{e}_2 ||_{2},
\end{align*}
to measure the distance between two embedding vectors, i.e.~$\boldsymbol{e}_1$ and $\boldsymbol{e}_2$. 

\paragraph{Clustering validation criterion} To apply a clustering technique, we need to select the number of clusters \citep{hastie2009elements}. \cite{wierzchon2018modern} state that the optimal number of clusters can be obtained via visual methods, heuristics or the optimisation of a criterion. Because of the multitude of clustering tasks, we opt to use a criterion that is straightforward and can be evaluated efficiently. \cite{rousseeuw1987silhouettes} proposed the silhouette index, which considers the similarity within a cluster as well as the dissimilarity between clusters and maps this to a number in $[-1,1]$. A higher silhouette index indicates a better clustering solution. The comparative study of \cite{vendramin2010relative} found the silhouette index to be the most robust and one of the best performing clustering validation metrics. Suppose we have a clustering solution $E=\{E^{(1)},\ldotp\ldotp,E^{(K)}\}$ for the set $\{\boldsymbol{e}_{1},\ldotp\ldotp,\boldsymbol{e}_{J}\}$, where we have that $\bigcup^{K}_{k=1}E^{(k)}=\{\boldsymbol{e}_{1},\ldotp\ldotp,\boldsymbol{e}_{J}\}$. Let the embedding vector $e_{j}$ be a member of cluster $E^{(k)}$, which consists of $c_{k}$ embeddings. The silhouette index for that clustering solution can then be defined as
\begin{align*}
    &SI = \frac{1}{J} \sum_{j=1}^{J} SI(\boldsymbol{e}_{j}) \,\,\text{where}\,\, SI(\boldsymbol{e}_{j}) = \frac{b(\boldsymbol{e}_{j})-a(\boldsymbol{e}_{j})}{\max (a(\boldsymbol{e}_{j}),b(\boldsymbol{e}_{j}))}, \\ 
    &a(\boldsymbol{e}_{j}) = \frac{1}{c_{k}-1} \sum_{e_{j^{*}} \in E^{(k)},j^{*} \ne j}D(\boldsymbol{e}_{j},\boldsymbol{e}_{j^{*}}) \,\,\text{and}\,\, b(\boldsymbol{e}_{j}) = \min_{E^{(k^{'})} \ne E^{(k)}} \frac{1}{c_{j^{'}}} \sum_{\boldsymbol{e}_{j^{'}} \in E^{(k^{'})}} D(\boldsymbol{e}_{j},\boldsymbol{e}_{j^{'}}).
\end{align*}
\added{The overall silhouette index $SI$ is the average of the silhouette indices for each embedding $\boldsymbol{e}_j$ in the set $\{\boldsymbol{e}_{1},\ldotp\ldotp,\boldsymbol{e}_{J}\}$. This is useful as it allows to inspect the overall quality of a clustering solution as well as for a specific class. For embedding $\boldsymbol{e}_j$, the within-cluster cohesion and the between-cluster separation are measured via the $a(\boldsymbol{e}_{j})$ and $b(\boldsymbol{e}_{j})$ term, respectively.} We select the number of clusters $K^*$ as
\begin{align*}
    K^{*} = \argmax_{K} \,\, SI_{K},
\end{align*}
where $SI_{K}$ is the silhouette index for the $K$-cluster solution.  

\paragraph{Additional rules} 

As detailed in Appendix~\ref{append:algo}, we impose two additional decision rules in our algorithm connected to the use of the silhouette index. \added{A first rule pertains to the overall quality of the clustering solution.} The silhouette index is not defined for $K=1$ clusters. In Figure~\ref{fig:collapse:b}, for example, all descendants of $\widetilde{h}_{1,2}$ have the same effect on their parent class and should be collapsed (i.e.~$K^*=1$). To account for this, we impose an additional rule: if the silhouette index $SI_{K^{*}}$ is below some threshold $SI^{*}$, i.e.~$SI_{K^{*}} < SI^{*}$, we opt for $K^{*}=1$. The value $-1 \le SI^{*} \le 1$ is a tuning parameter for the algorithm. \added{By varying $SI^{*}$, we can control the degree to which the hierarchical structure is reduced.} A higher value for $SI^{*}$ will result in a more collapsed hierarchical structure. \added{Intuitively, $SI^{*}$ is a lower bound on the quality of the clustering solution, i.e.~how well a given set is partitioned in terms of within-cluster cohesion and between-cluster separation, when more than one cluster is considered.} \added{A second rule is specific to the between-level clustering step and imposes an additional criterion that needs to be met in order for classes to be merged with their parent class.} Suppose we have a cluster $E^{(p)}$ that contains the embedding vector $e^{p}$ of a parent class and the embedding vectors of its descendant classes that are close to their parent in the embedding space. \added{Via $SI(\boldsymbol{e}_{p})$, we can assess the degree to which $e^{p}$ belongs in the cluster $E^{(p)}$.} Intuitively, if the parent is at the edge of the cluster, it might indicate that the descendants were in the same cluster while having a different effect on the response variable. Therefore if we have that $SI(\boldsymbol{e}_{p}) = \min_{\boldsymbol{e}_{j} \in E^{(p)}} SI(\boldsymbol{e}_{j})$, i.e.~the embedding of the parent class is the least similar object in cluster $E^{(p)}$, 
it might indicate that those descendant classes (of which the embeddings are in $E^{(p)}$) are sufficiently different in their effect on the response compared to their parent class. Consequently, we do not collapse those classes into their parent class. If not, we do collapse the descendant classes of which the embeddings are in $E^{(p)}$.


\section{Simulation experiments} \label{sec:simulating}

With simulation experiments we aim to assess the algorithm's performance in reducing the hierarchical structure of a categorical variable $\boldsymbol{h}$ (almost) correctly. First, we define a hierarchical structure for $\boldsymbol{h}$ and fix which classes have the same effect on the response variable, resulting in a reduced representation $\boldsymbol{\widetilde{h}}$ of that hierarchy. Subsequently, we embed the hierarchy as laid out in Section~\ref{sec:embedding} and apply our proposed algorithm to examine how accurately we can retrieve the reduced representation $\boldsymbol{\widetilde{h}}$. Lastly, we compare the performance of a predictive model including $\boldsymbol{\widetilde{h}}$ with a model incorporating $\boldsymbol{h}$.  

\subsection{Data}

\subsubsection{Covariates}
\begin{figure}[H]
	\tikzset{every picture/.style={line width=0.75pt}} 

\begin{tikzpicture}[x=0.60pt,y=0.60pt,yscale=-0.8,xscale=0.8]

\draw    (190.23,124.38) -- (125.93,250.42) ;
\draw    (220.23,124.38) -- (285.93,250.76) ;
\draw    (110.73,286.95) -- (46.43,412.99) ;
\draw    (126.09,292.58) -- (126.09,413.66) ;
\draw    (140.73,286.95) -- (206.43,413.33) ;
\draw    (529.73,123.88) -- (465.43,249.92) ;
\draw    (559.73,123.88) -- (625.43,250.26) ;
\draw    (449.73,286.38) -- (385.43,412.42) ;
\draw    (479.73,286.38) -- (545.43,412.76) ;
\draw    (375.11,10.18) -- (205.59,87.76) ;
\draw    (375.11,10.18) -- (545.09,87.26) ;
\draw  [fill={rgb, 255:red, 25; green, 169; blue, 69 }  ,fill opacity=1 ] (174.77,117.62) .. controls (174.77,101.13) and (188.57,87.76) .. (205.59,87.76) .. controls (222.61,87.76) and (236.41,101.13) .. (236.41,117.62) .. controls (236.41,134.11) and (222.61,147.48) .. (205.59,147.48) .. controls (188.57,147.48) and (174.77,134.11) .. (174.77,117.62) -- cycle ;
\draw  [fill={rgb, 255:red, 175; green, 143; blue, 81 }  ,fill opacity=1 ] (514.27,117.12) .. controls (514.27,100.63) and (528.07,87.26) .. (545.09,87.26) .. controls (562.11,87.26) and (575.91,100.63) .. (575.91,117.12) .. controls (575.91,133.61) and (562.11,146.98) .. (545.09,146.98) .. controls (528.07,146.98) and (514.27,133.61) .. (514.27,117.12) -- cycle ;
\draw  [fill={rgb, 255:red, 247; green, 21; blue, 184 }  ,fill opacity=1 ] (434.61,279.79) .. controls (434.61,263.29) and (448.41,249.92) .. (465.43,249.92) .. controls (482.45,249.92) and (496.25,263.29) .. (496.25,279.79) .. controls (496.25,296.28) and (482.45,309.65) .. (465.43,309.65) .. controls (448.41,309.65) and (434.61,296.28) .. (434.61,279.79) -- cycle ;
\draw  [fill={rgb, 255:red, 252; green, 166; blue, 62 }  ,fill opacity=1 ] (594.61,280.12) .. controls (594.61,263.63) and (608.41,250.26) .. (625.43,250.26) .. controls (642.45,250.26) and (656.25,263.63) .. (656.25,280.12) .. controls (656.25,296.61) and (642.45,309.98) .. (625.43,309.98) .. controls (608.41,309.98) and (594.61,296.61) .. (594.61,280.12) -- cycle ;
\draw  [fill={rgb, 255:red, 97; green, 246; blue, 252 }  ,fill opacity=1 ] (354.61,442.29) .. controls (354.61,425.79) and (368.41,412.42) .. (385.43,412.42) .. controls (402.45,412.42) and (416.25,425.79) .. (416.25,442.29) .. controls (416.25,458.78) and (402.45,472.15) .. (385.43,472.15) .. controls (368.41,472.15) and (354.61,458.78) .. (354.61,442.29) -- cycle ;
\draw  [fill={rgb, 255:red, 198; green, 150; blue, 233 }  ,fill opacity=1 ] (514.61,442.62) .. controls (514.61,426.13) and (528.41,412.76) .. (545.43,412.76) .. controls (562.45,412.76) and (576.25,426.13) .. (576.25,442.62) .. controls (576.25,459.11) and (562.45,472.48) .. (545.43,472.48) .. controls (528.41,472.48) and (514.61,459.11) .. (514.61,442.62) -- cycle ;
\draw  [fill={rgb, 255:red, 254; green, 234; blue, 111 }  ,fill opacity=1 ] (175.61,443.19) .. controls (175.61,426.7) and (189.41,413.33) .. (206.43,413.33) .. controls (223.45,413.33) and (237.25,426.7) .. (237.25,443.19) .. controls (237.25,459.68) and (223.45,473.05) .. (206.43,473.05) .. controls (189.41,473.05) and (175.61,459.68) .. (175.61,443.19) -- cycle ;
\draw  [fill={rgb, 255:red, 153; green, 233; blue, 150 }  ,fill opacity=1 ] (95.27,443.52) .. controls (95.27,427.03) and (109.07,413.66) .. (126.09,413.66) .. controls (143.11,413.66) and (156.91,427.03) .. (156.91,443.52) .. controls (156.91,460.01) and (143.11,473.38) .. (126.09,473.38) .. controls (109.07,473.38) and (95.27,460.01) .. (95.27,443.52) -- cycle ;
\draw  [fill={rgb, 255:red, 97; green, 164; blue, 252 }  ,fill opacity=1 ] (15.61,442.85) .. controls (15.61,426.36) and (29.41,412.99) .. (46.43,412.99) .. controls (63.45,412.99) and (77.25,426.36) .. (77.25,442.85) .. controls (77.25,459.35) and (63.45,472.72) .. (46.43,472.72) .. controls (29.41,472.72) and (15.61,459.35) .. (15.61,442.85) -- cycle ;
\draw  [fill={rgb, 255:red, 105; green, 98; blue, 237 }  ,fill opacity=1 ] (95.11,280.29) .. controls (95.11,263.79) and (108.91,250.42) .. (125.93,250.42) .. controls (142.95,250.42) and (156.75,263.79) .. (156.75,280.29) .. controls (156.75,296.78) and (142.95,310.15) .. (125.93,310.15) .. controls (108.91,310.15) and (95.11,296.78) .. (95.11,280.29) -- cycle ;
\draw  [fill={rgb, 255:red, 237; green, 98; blue, 98 }  ,fill opacity=1 ] (255.11,280.62) .. controls (255.11,264.13) and (268.91,250.76) .. (285.93,250.76) .. controls (302.95,250.76) and (316.75,264.13) .. (316.75,280.62) .. controls (316.75,297.11) and (302.95,310.48) .. (285.93,310.48) .. controls (268.91,310.48) and (255.11,297.11) .. (255.11,280.62) -- cycle ;

\draw (185.64-1,100) node [anchor=north west][inner sep=0.75pt]  [font=\large] [align=left] {$\tilde{h}_{1,1}$};
\draw (525.64-2,100) node [anchor=north west][inner sep=0.75pt]  [font=\large] [align=left] {$\tilde{h}_{1,2}$};
\draw (104,264) node [anchor=north west][inner sep=0.75pt]  [font=\large] [align=left] {$\tilde{h}_{2,1}$};
\draw (264,264) node [anchor=north west][inner sep=0.75pt]  [font=\large] [align=left] {$\tilde{h}_{2,2}$};
\draw (444,264) node [anchor=north west][inner sep=0.75pt]  [font=\large] [align=left] {$\tilde{h}_{2,3}$};
\draw (604,264) node [anchor=north west][inner sep=0.75pt]  [font=\large] [align=left] {$\tilde{h}_{2,4}$};
\draw (24.64,426) node [anchor=north west][inner sep=0.75pt]  [font=\large] [align=left] {$\tilde{h}_{3,1}$};
\draw (105,426) node [anchor=north west][inner sep=0.75pt]  [font=\large] [align=left] {$\tilde{h}_{3,2}$};
\draw (186,426) node [anchor=north west][inner sep=0.75pt]  [font=\large] [align=left] {$\tilde{h}_{3,3}$};
\draw (364,426) node [anchor=north west][inner sep=0.75pt]  [font=\large] [align=left] {$\tilde{h}_{3,4}$};
\draw (524,426) node [anchor=north west][inner sep=0.75pt]  [font=\large] [align=left] {$\tilde{h}_{3,5}$};

\end{tikzpicture}
    \centering
	\centering
        \caption{True reduced representation $\boldsymbol{\widetilde{h}}$ of the hierarchical variable $\boldsymbol{h}$ specified in Figure~\ref{fig:3factfull} in Appendix~\ref{append:simul} when there is an effect on the response. The colour encoding in these figures indicates whether classes have the same effect on the response variable.}
        \label{fig:3fact_true}
\end{figure}
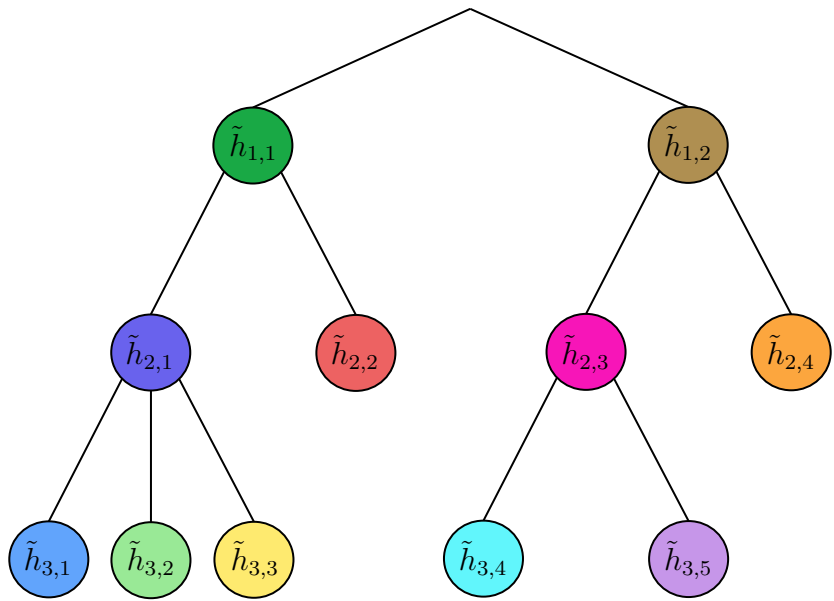

\paragraph{Hierarchical covariate} We simulate a hierarchical categorical variable $\boldsymbol{h}$ with $R=3$ levels. Figure~\ref{fig:3factfull} in Appendix~\ref{append:simul} pictures the original granular specification of the hierarchy, the colours indicate which classes are assumed to have the same effect on the response variable. The first level consists of five classes, the second level consists of 20 classes and the third level consists of 86 classes. \added{We consider two reduced representations, which are known upfront. If we include an effect of $\boldsymbol{h}$ on the response, the true reduced representation $\boldsymbol{\widetilde{h}}$ is pictured in Figure~\ref{fig:3fact_true}.} The number of classes in the first level is reduced to two \added{($\widetilde{n}_1=2$)}, the second level consists of four classes \added{($\widetilde{n}_2=4$)} and the third level consists of five classes \added{($\widetilde{n}_3=5$)}. \added{If there is no effect on the response, the resulting structure is completely collapsed, i.e. $\widetilde{R}=1$ and $\widetilde{n}_1=1$ } \added{For each class in $h_R$, a unique vector of indicator variables (one for each class at every level) encodes the path from that class to the upper level in the hierarchy. For each observation $i$, we store the information about $\boldsymbol{h}$ by assigning the vector of indicator variables corresponding to its class in $h_R$.}

\paragraph{Other covariates} In addition to a hierarchical categorical variable $\boldsymbol{h}$, we simulate a vector $\boldsymbol{x}=(x_{1},x_{2},x_{3})$ consisting of three non-hierarchical covariates. We construct $x_{1}=\sin(a_{1})$ with $a_{1} \sim U(0,5)$, $x_{2} \sim N(0,1)$ and $x_{3}=a_{3}^2$ with $a_{3} \sim U(1,2)$.

\subsubsection{Response} \label{sec:response} We use a generalised linear model structure when simulating the response variable. More specifically, the conditional mean of the distribution of the target variable $y$ is defined as
\begin{equation} \label{eq:simul}
    \text{E}(y|\boldsymbol{h},\boldsymbol{x}) = g^{-1}\Bigg( \mu + \underbrace{\sum^{n_{1}-1}_{s=1} \beta_{1,s}  X_{1,s} + \sum^{R-1}_{r=1} \sum^{n_{r}}_{s=1} \sum^{\text{dim}(\mathcal{H}_{r,s})-1}_{k=1} \beta_{r+1,d+k}  X_{r+1,d+k}}_{\text{Effect} \: \boldsymbol{h}} +  \boldsymbol{\beta}^{(x)} \boldsymbol{x}^{\prime}\Bigg),
\end{equation}
where $d= \sum^{s-1}_{z=1} \text{dim}(\mathcal{H}_{r,z})$, $\mu$ is the common component and $g(.)$ is a link function that connects the linear predictor to the conditional mean and depends on the choice of distribution. The coefficient related to $h_{r,s}$ is denoted by $\beta_{r,s}$, while $\boldsymbol{\beta}^{(x)}$ refers to the coefficients corresponding to the covariate vector $\boldsymbol{x}$. Following \cite{neter1996applied}, $X_{r+1,d+k}$ is an indicator variable that takes a value of one if $h_{r+1} = h_{r+1,d+k}$, a value of negative one if $h_{r+1} = h_{r+1,d+ \text{dim}(\mathcal{H}_{r,s})}$ and zero in all other cases. In order for the model to be identifiable, we must have that 
\begin{align*}
    \sum^{n_{1}}_{k=1} \beta_{1,k} = 0 \,\, \text{and} \,\, \sum^{\text{dim}(\mathcal{H}_{r,s})}_{k=1} \beta_{r+1,d+k} = 0 \:\:\:\: \text{for} \, r=1,\ldotp\ldotp,R-1 \:,\: 
 \, s=1,\ldotp\ldotp,n_{r}.
\end{align*}
By construction, the indicator variables ensure that Equation~\eqref{eq:simul} satisfies the identification constraints.

\subsubsection{Experiments}
\vspace{-0.1cm}
\paragraph{Balanced data} We consider three distinct settings: including no effects on the response, including an effect of only $\boldsymbol{h}$ or including an effect of both $\boldsymbol{h}$ and $\boldsymbol{x}$. For each setting, we consider both a normally distributed response as well as a Poisson response\added{, allowing us to assess the impact of the chosen distributional assumption on the effectiveness of our methodology}. As a result, we have six simulation experiments. For each experiment, we simulate 100 datasets consisting of a balanced design with 1\ 000 observations of every class at the lowest level, i.e.~for $h_{3,s} \in H_{3}$ in Figure~\ref{fig:3factfull}. The response variable follows a distribution with mean specified as in Equation~\eqref{eq:experiment} in Appendix~\ref{append:simul} where the link function and parameter values are set according to Table~\ref{tab:equation} in Appendix~\ref{append:simul}, depending on the specific experiment. The effect of $\boldsymbol{h}$'s classes is set so that classes with the same colour encoding in Figure~\ref{fig:3factfull} have the same effect on the response. When we consider a normal distribution, we specify the standard deviation $\sigma=1.5$.

\vspace{-0.1cm}
\paragraph{Unbalanced data} We have four unbalanced experiments consisting of 100 datasets for the case of a normally distributed response including an effect of both $\boldsymbol{h}$ and $\boldsymbol{x}$. For each of the unbalanced experiments, we simulate a different number of observations for each class of $h_R$, i.e.~a random number between 50 and 100, 50 and 150, 50 and 200 or 50 and 250 observations, respectively. As such, we can assess how well our approach is able to approximate the original structure when the number of observations differs across classes at the lowest level\added{, i.e.~when some classes are more sparse than others}.

\vspace{-0.2cm}
\subsection{Network architecture}

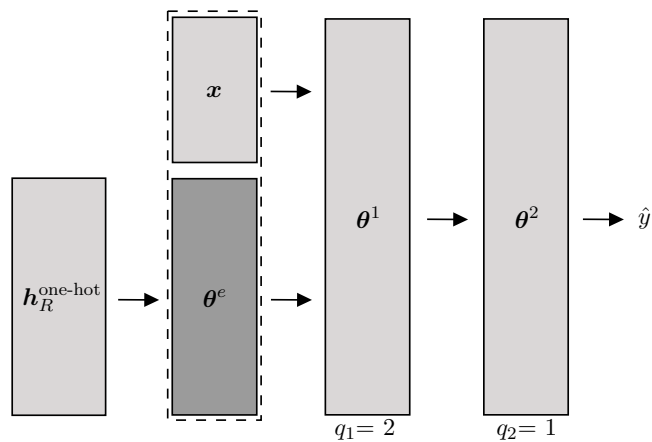
\begin{figure}[H]

\tikzset{every picture/.style={line width=0.75pt}} 

\begin{adjustbox}{width=0.53\textwidth}
    \begin{tikzpicture}[x=0.75pt,y=0.75pt,yscale=-1,xscale=1]

\draw  [fill={rgb, 255:red, 218; green, 215; blue, 215 }  ,fill opacity=1 ][line width=0.75]  (198,23) -- (248.57,23) -- (248.57,262.14) -- (198,262.14) -- cycle ;
\draw    (165.4,192.6) -- (186.97,192.73) ;
\draw [shift={(189.97,192.74)}, rotate = 180.33] [fill={rgb, 255:red, 0; green, 0; blue, 0 }  ][line width=0.08]  [draw opacity=0] (8.93,-4.29) -- (0,0) -- (8.93,4.29) -- cycle    ;
\draw    (259,144) -- (280.57,144.13) ;
\draw [shift={(283.57,144.14)}, rotate = 180.33] [fill={rgb, 255:red, 0; green, 0; blue, 0 }  ][line width=0.08]  [draw opacity=0] (8.93,-4.29) -- (0,0) -- (8.93,4.29) -- cycle    ;
\draw  [fill={rgb, 255:red, 218; green, 215; blue, 215 }  ,fill opacity=1 ][line width=0.75]  (293,23) -- (343.57,23) -- (343.57,262.14) -- (293,262.14) -- cycle ;
\draw  [fill={rgb, 255:red, 218; green, 215; blue, 215 }  ,fill opacity=1 ][line width=0.75]  (106.4,22) -- (156.97,22) -- (156.97,109.71) -- (106.4,109.71) -- cycle ;
\draw  [color={rgb, 255:red, 0; green, 0; blue, 0 }  ,draw opacity=1 ][fill={rgb, 255:red, 155; green, 155; blue, 155 }  ,fill opacity=1 ] (106.22,119.51) -- (156.88,119.51) -- (156.88,262.2) -- (106.22,262.2) -- cycle ;
\draw  [color={rgb, 255:red, 0; green, 0; blue, 0 }  ,draw opacity=1 ][fill={rgb, 255:red, 218; green, 215; blue, 215 }  ,fill opacity=1 ] (10.28,119.2) -- (66.28,119.2) -- (66.28,262.4) -- (10.28,262.4) -- cycle ;
\draw    (165.2,66.8) -- (186.77,66.93) ;
\draw [shift={(189.77,66.94)}, rotate = 180.33] [fill={rgb, 255:red, 0; green, 0; blue, 0 }  ][line width=0.08]  [draw opacity=0] (8.93,-4.29) -- (0,0) -- (8.93,4.29) -- cycle    ;
\draw    (73.8,192.6) -- (95.37,192.73) ;
\draw [shift={(98.37,192.74)}, rotate = 180.33] [fill={rgb, 255:red, 0; green, 0; blue, 0 }  ][line width=0.08]  [draw opacity=0] (8.93,-4.29) -- (0,0) -- (8.93,4.29) -- cycle    ;
\draw  [dash pattern={on 4.5pt off 4.5pt}] (103.49,18.33) -- (159.49,18.33) -- (159.49,265.33) -- (103.49,265.33) -- cycle ;
\draw    (352.4,144) -- (373.97,144.13) ;
\draw [shift={(376.97,144.14)}, rotate = 180.33] [fill={rgb, 255:red, 0; green, 0; blue, 0 }  ][line width=0.08]  [draw opacity=0] (8.93,-4.29) -- (0,0) -- (8.93,4.29) -- cycle    ;

\draw (125,62) node [anchor=north west][inner sep=0.75pt]   [align=left] {$\displaystyle \boldsymbol{x}$};
\draw (384,135) node [anchor=north west][inner sep=0.75pt]   [align=left] {$\displaystyle \hat{y}$};
\draw (123,185) node [anchor=north west][inner sep=0.75pt]   [align=left] {$\displaystyle \boldsymbol{\theta }^{e}$};
\draw (215,135) node [anchor=north west][inner sep=0.75pt]   [align=left] {$\displaystyle \boldsymbol{\theta }^{1}$};
\draw (310,135) node [anchor=north west][inner sep=0.75pt]   [align=left] {$\displaystyle \boldsymbol{\theta }^{2}$};
\draw (15,181) node [anchor=north west][inner sep=0.75pt]   [align=left] {$\displaystyle \boldsymbol{h}_{R}^{\text{one-hot}}$};
\draw (202,263.5) node [anchor=north west][inner sep=0.75pt]   [align=left] {$\displaystyle q_{1}$= 2};
\draw (299,263.5) node [anchor=north west][inner sep=0.75pt]   [align=left] {$\displaystyle q_{2}$= 1};

\end{tikzpicture}
\end{adjustbox}
    \centering
    \caption{Network architecture used in the simulation experiments. The two-dimensional embedding layer $\boldsymbol{\theta}^e$ is concatenated with the covariate vector $\boldsymbol{x}$ to form the input layer. The hidden layer $\boldsymbol{\theta}^1$ and the output layer $\boldsymbol{\theta}^2$ consist of $q_1=2$ and $q_2=1$ neurons, respectively.  }
    \label{fig:simulnetwork}
\end{figure}

To learn the embeddings as laid out in Section~\ref{sec:embedding}, we opt for a network architecture consisting of a single hidden layer with $q_{1}=2$ neurons as pictured in Figure~\ref{fig:simulnetwork}. For a normally distributed response, we use the identity function as the activation function in the output layer $\boldsymbol{\theta}^2$. Conversely, when we have a Poisson response, we use the exponential function. The network parameters are optimised using the \texttt{adam} algorithm \citep{kingma2014adam} with the mean squared error as the loss function $\mathcal{L}_{NN}(.,.)$ for a normal response or the Poisson deviance in the case of a Poisson response. In our examples, we set the embedding dimension $q_{e}=2$, as two-dimensional embeddings allow a direct visualisation of the embedding space while maintaining good performance. We compare the model performance when using different embedding dimensions in Table~\ref{tab:embedtune} of Appendix~\ref{append:simul}.

\begin{figure}[H] 
\centering
\subfloat[$\boldsymbol{h}$ has effect on $y$]{\includegraphics[width = 0.41\textwidth]{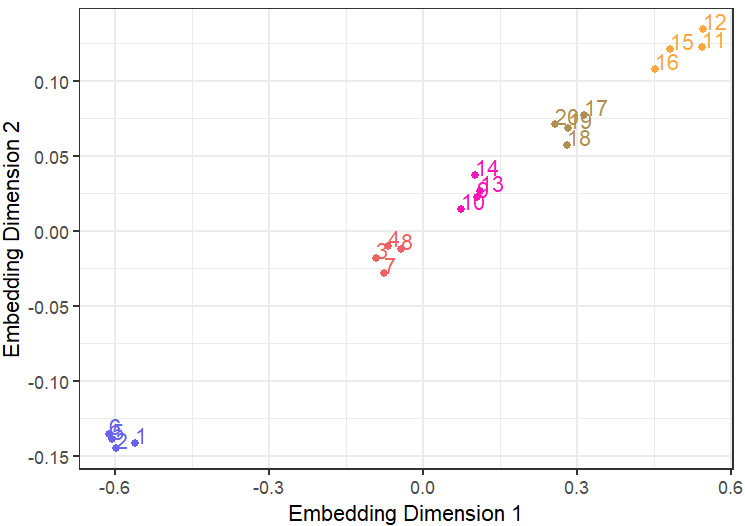}} 
\subfloat[$\boldsymbol{h}$ has no effect on $y$]{\includegraphics[width = 0.41\textwidth]{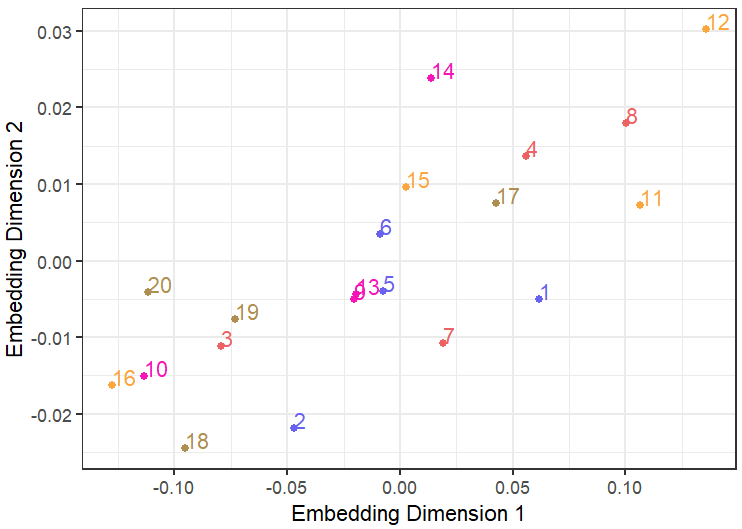}}
\caption{Examples of the two-dimensional embedding vectors constructed for the second level's classes, i.e.~$h_{2,1},\ldotp\ldotp,h_{2,20}$ in the balanced experiment with a normally distributed response. The numbering refers to the class index, e.g.~5 refers to $h_{2,5}$. The left-hand side pictures an example where $y$ was simulated including an effect of $\boldsymbol{h}$. Conversely, the right-hand side shows an example where we did not include an effect of $\boldsymbol{h}$. The colour encoding corresponds to Figure~\ref{fig:3factfull}.} 
\label{fig:example_normal}
\end{figure}

\begin{figure}[H] 
\centering
\subfloat{\includegraphics[width = 0.43\textwidth]{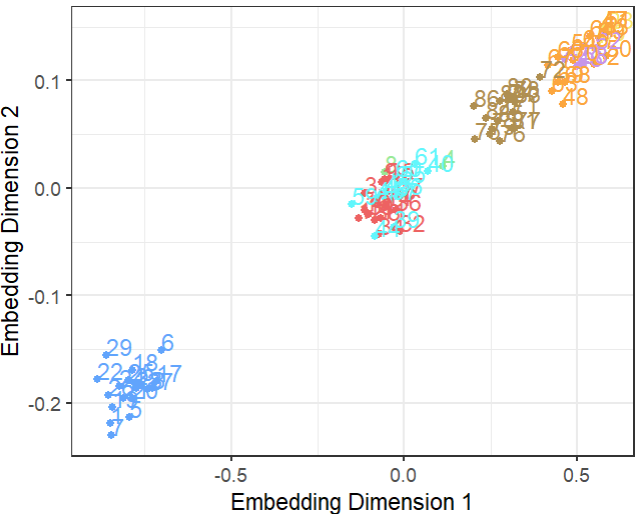}} 
\subfloat{\includegraphics[width = 0.43\textwidth]{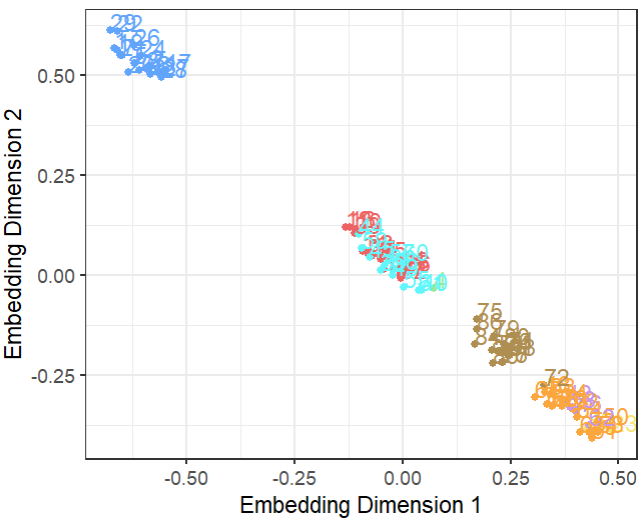}}
\caption{Examples of the embedding vectors for the third level's classes, i.e.~$h_{3,1},\ldotp\ldotp,h_{3,86}$ for a normally distributed response where we included an effect of $\boldsymbol{h}$ on the response. The left- and the right-hand side picture the embedding vectors obtained with two different initialisations of the network parameters.} 
\label{fig:space}
\end{figure}

\vspace{-0.1cm}
Figure~\ref{fig:example_normal} shows an example of the embedding space as obtained for the second level of the hierarchy. The left-hand side pictures the embeddings learned for a dataset where only $\boldsymbol{h}$ has an effect on the response variable $y$ while the right-hand side illustrates the case where the response variable does not depend on $\boldsymbol{h}$ nor $\boldsymbol{x}$. In the latter case, the hierarchy can be completely collapsed as $\boldsymbol{h}$ does not have any predictive power for $y$. We can clearly observe a clustering of the embeddings on the left-hand side, while this is not the case when the hierarchical categorical variable has no effect on the response.
%
\vspace{-0.2cm}
For each simulated dataset, we consider five different sets of initialisation values for the network parameters to assess whether the performance of our methodology is impacted by these starting values. As pictured in Figure~\ref{fig:space}, the embedding vectors can take different values depending on the initialisation but the relative positioning of the embedding vectors within the embedding space mostly persists. Table~\ref{tab:seeds} in Appendix~\ref{append:simul} shows that the performance of the algorithm is stable over the different sets of initialisation values for the network parameters.

%

\subsection{Results} \label{sec:simulexample}

\subsubsection{Balanced experiments} Table~\ref{tab:3fact} summarises the results of applying our algorithm, see Section~\ref{sec:collapsing}, on the balanced simulation experiments. We fix $SI^*$ at 0.7. For each experiment, Table~\ref{tab:3fact} includes how often (as a percentage) the true structure is retrieved as well as the number of different reduced structures. We find that our algorithm is able to retrieve the true structure reliably. Figure~\ref{fig:normalstructures} in Appendix~\ref{append:simul} shows the retrieved structures when we include only an effect of $\boldsymbol{h}$ on a normally distributed response. We observe that all retrieved structures closely resemble the true structure pictured in Figure~\ref{fig:3fact_true}. Our methodology seems to perform slightly better on the Poisson data compared to the normally distributed data. Moreover, in the event were the true structure is not fully retrieved, there are fewer different structures found for a Poisson response. When $\boldsymbol{h}$ has no effect on the response, the true structure is completely collapsed and consists of a single level with one class, i.e. $\widetilde{R}=1$ and $\widetilde{H}_1=\{\widetilde{h}_{1,1}\}$. For these experiments, we find a higher number of different structures. This is because the embeddings are essentially randomly distributed across the embedding space, as illustrated by Figure~\ref{fig:example_normal}. Depending on the specific dataset considered, the classes that are not collapsed correctly differ randomly, leading to a variety of structures. However, most structures are still close to the true underlying structure where all classes at each level are collapsed. For example, in the case of a Poisson response, we find that 11 out of the 19 structures have $\widetilde{R}=1$ and $\widetilde{n}_1 \leq 2$, accounting for 98.4\% of cases.


\begin{table}[H]
\centering
\begin{tabular}{m{10.5em} m{3.5em}m{3.5em}m{3.5em} m{0.3em} m{3.5em}m{3.5em}m{3.5em}}
  \hline

 & \multicolumn{3}{c}{Normal distribution} && \multicolumn{3}{c}{Poisson distribution}\\
 \cline{2-4} 
 \cline{6-8}
 &none&only $\boldsymbol{h}$&both $\boldsymbol{h}$ and $\boldsymbol{x}$&&none&only $\boldsymbol{h}$&both $\boldsymbol{h}$ and $\boldsymbol{x}$\\
 \hline
 True structure retrieved&96.6\%&90.4\%&92.8\%&&92.6\%&97.4\%&99\%\\
 Different structures &9&8&8&&19&2&2\\
 $\text{AIC}(\boldsymbol{\widetilde{h}}) < \text{AIC}(\boldsymbol{h})$&100\%&99.8\%&100\%&&100\%&100\%&100\%\\
 $\text{BIC}(\boldsymbol{\widetilde{h}}) < \text{BIC}(\boldsymbol{h})$&100\%&100\%&100\%&&100\%&100\%&100\%\\
 $\text{RMSE}(\boldsymbol{\widetilde{h}}) < \text{RMSE}(\boldsymbol{h})$&99\%&96.4\%&99\%&&99\%&96\%&80\%\\

\hline 
 
\end{tabular}
\caption{Summary of the results obtained for the balanced simulation experiments with $SI^{*}=0.7$. The first row shows how often (as a percentage) the true structure is recovered over the 100 datasets generated for each experiment. The second row indicates the number of different structures. The third and fourth rows show the comparison of model fit in terms of the $\text{AIC}$ and $\text{BIC}$ for the models incorporating $\boldsymbol{h}$ and $\boldsymbol{\widetilde{h}}$.}
\label{tab:3fact}
\end{table}


\vspace{-0.18cm}
To assess whether the new categorical variable $\boldsymbol{\widetilde{h}}$ improves the model fit, we compare a (generalised) linear model incorporating the most granular level in $\boldsymbol{h}$ with a similar model including the information of $\boldsymbol{\widetilde{h}}$ that is obtained by grouping the classes of $h_R$ that were merged or collapsed. To compare the models, we prefer to use a criterion that takes into account the model complexity as well as the predictive accuracy\added{, allowing a minor decrease in model accuracy for the benefit of obtaining a sparser model}. Therefore, we opt to use the Akaike Information Criterion ($\text{AIC}$) \citep{akaike1974new} as well as the Bayesian Information Criterion ($\text{BIC}$) \citep{schwarz1978estimating}. In both cases, a lower value indicates a preferred model. We denote the $\text{AIC}$ and $\text{BIC}$ for the model incorporating the information of $\boldsymbol{h}$ as $\text{AIC}(\boldsymbol{h})$ and $\text{BIC}(\boldsymbol{h})$, respectively. Analogously, we use $\text{AIC}(\boldsymbol{\widetilde{h}})$ and $\text{BIC}(\boldsymbol{\widetilde{h}})$ when including information of the reduced representation $\boldsymbol{\widetilde{h}}$. As shown in Table~\ref{tab:3fact}, we observe that in almost all cases, incorporating the reduced hierarchy leads to an improved model in terms of the $\text{AIC}$ and $\text{BIC}$ even when $\boldsymbol{\widetilde{h}}$ does not exactly match the true structure pictured in Figure~\ref{fig:3fact_true}. \added{To assess the out-of-sample performance, we simulate additional data for each of the experiments. Similarly to before, Table~\ref{tab:3fact} indicates that the out-of-sample root mean squared error (RMSE) in general improves when including the reduced representation $\boldsymbol{\widetilde{h}}$. Consequently, we find that the reduced structure generalises better to unseen data.}

\subsubsection{Unbalanced experiments} Table~\ref{tab:3factunbalanced} summarises the results of the unbalanced experiment. Overall, as the number of observations per class decreases, the true structure is retrieved in fewer instances and the resulting number of different structures for $\boldsymbol{\widetilde{h}}$ is higher. However, these structures all closely resemble the true structure of Figure~\ref{fig:3fact_true}. This is evidenced by the fact that, in almost all cases, the model incorporating $\boldsymbol{\widetilde{h}}$ is preferred over the model using $\boldsymbol{h}$ in terms of the $\text{AIC}$ and $\text{BIC}$. \added{Table~\ref{tab:3factunbalanced} also shows the out-of-sample RMSE on additional simulated data. Overall, the RMSE is lower when including the reduced structure, indicating that it generalises better to unseen data.}

\begin{table}[H]
\centering
\begin{tabular}{m{12em} m{4.3em}m{4.3em}m{4.3em}m{4.3em}}
  \hline

 & \multicolumn{4}{c}{Number of observations in each class of $h_R$} \\
 \cline{2-5}
 &50-100&50-150&50-200&50-250\\
 \hline
 True structure retrieved&43.2\%&52.8\%&60.8\%&68.4\%\\
 Different structures &51&32&39&27\\
 $\text{AIC}(\boldsymbol{\widetilde{h}}) < \text{AIC}(\boldsymbol{h})$&99.4\%&100\%&100\%&100\%\\
 $\text{BIC}(\boldsymbol{\widetilde{h}}) < \text{BIC}(\boldsymbol{h})$&100\%&100\%&100\%&100\%\\
 $\text{RMSE}(\boldsymbol{\widetilde{h}}) < \text{RMSE}(\boldsymbol{h})$&85.8\%&89.4\%&92.8\%&89.4\%\\

\hline 
 
\end{tabular}
\caption{Summary of the results obtained for the unbalanced simulation experiments with $SI^{*}=0.7$. The first row shows how often (as a percentage) the true structure is recovered over the 100 datasets generated for each experiment, the second row indicates the number of different structures and the third and fourth rows include the comparison of the model fit in terms of the $\text{AIC}$ and $\text{BIC}$ for the models incorporating $\boldsymbol{h}$ and $\boldsymbol{\widetilde{h}}$. }
\label{tab:3factunbalanced}
\end{table}
With Figures~\ref{fig:truefalse} and~\ref{fig:boxplot} in Appendix~\ref{append:simul}, we further substantiate our claim that the retrieved structures closely resemble the true structure. Figure~\ref{fig:truefalse} pictures the two most common structures for the unbalanced simulation experiment with 50-100 observations for each class of $h_R$. The predominant structure is shown on the left-hand side and corresponds to the true structure pictured in Figure~\ref{fig:3fact_true}. We observe that the structure on the right-hand side, which is found the second most, is the true structure except for the fact that $\widetilde{h}_{3,4}$ was collapsed incorrectly. Figure~\ref{fig:boxplot} shows boxplots of the number of classes retrieved for the unbalanced simulation experiments at all three levels. We find that in almost all cases and for all levels, the number of classes is close to that of the true underlying structure.

\begin{figure}[H] 
\centering
\subfloat[]{

\begin{adjustbox}{width=0.45\textwidth}
    \begin{tikzpicture}[x=0.60pt,y=0.60pt,yscale=-0.8,xscale=0.8]

\draw    (190.23,124.38) -- (125.93,250.42) ;
\draw    (220.23,124.38) -- (285.93,250.76) ;
\draw    (110.73,286.95) -- (46.43,412.99) ;
\draw    (126.09,292.58) -- (126.09,413.66) ;
\draw    (140.73,286.95) -- (206.43,413.33) ;
\draw    (529.73,123.88) -- (465.43,249.92) ;
\draw    (559.73,123.88) -- (625.43,250.26) ;
\draw    (449.73,286.38) -- (385.43,412.42) ;
\draw    (479.73,286.38) -- (545.43,412.76) ;
\draw    (375.11,10.18) -- (205.59,87.76) ;
\draw    (375.11,10.18) -- (545.09,87.26) ;
\draw  [fill={rgb, 255:red, 25; green, 169; blue, 69 }  ,fill opacity=1 ] (174.77,117.62) .. controls (174.77,101.13) and (188.57,87.76) .. (205.59,87.76) .. controls (222.61,87.76) and (236.41,101.13) .. (236.41,117.62) .. controls (236.41,134.11) and (222.61,147.48) .. (205.59,147.48) .. controls (188.57,147.48) and (174.77,134.11) .. (174.77,117.62) -- cycle ;
\draw  [fill={rgb, 255:red, 175; green, 143; blue, 81 }  ,fill opacity=1 ] (514.27,117.12) .. controls (514.27,100.63) and (528.07,87.26) .. (545.09,87.26) .. controls (562.11,87.26) and (575.91,100.63) .. (575.91,117.12) .. controls (575.91,133.61) and (562.11,146.98) .. (545.09,146.98) .. controls (528.07,146.98) and (514.27,133.61) .. (514.27,117.12) -- cycle ;
\draw  [fill={rgb, 255:red, 247; green, 21; blue, 184 }  ,fill opacity=1 ] (434.61,279.79) .. controls (434.61,263.29) and (448.41,249.92) .. (465.43,249.92) .. controls (482.45,249.92) and (496.25,263.29) .. (496.25,279.79) .. controls (496.25,296.28) and (482.45,309.65) .. (465.43,309.65) .. controls (448.41,309.65) and (434.61,296.28) .. (434.61,279.79) -- cycle ;
\draw  [fill={rgb, 255:red, 252; green, 166; blue, 62 }  ,fill opacity=1 ] (594.61,280.12) .. controls (594.61,263.63) and (608.41,250.26) .. (625.43,250.26) .. controls (642.45,250.26) and (656.25,263.63) .. (656.25,280.12) .. controls (656.25,296.61) and (642.45,309.98) .. (625.43,309.98) .. controls (608.41,309.98) and (594.61,296.61) .. (594.61,280.12) -- cycle ;
\draw  [fill={rgb, 255:red, 97; green, 246; blue, 252 }  ,fill opacity=1 ] (354.61,442.29) .. controls (354.61,425.79) and (368.41,412.42) .. (385.43,412.42) .. controls (402.45,412.42) and (416.25,425.79) .. (416.25,442.29) .. controls (416.25,458.78) and (402.45,472.15) .. (385.43,472.15) .. controls (368.41,472.15) and (354.61,458.78) .. (354.61,442.29) -- cycle ;
\draw  [fill={rgb, 255:red, 198; green, 150; blue, 233 }  ,fill opacity=1 ] (514.61,442.62) .. controls (514.61,426.13) and (528.41,412.76) .. (545.43,412.76) .. controls (562.45,412.76) and (576.25,426.13) .. (576.25,442.62) .. controls (576.25,459.11) and (562.45,472.48) .. (545.43,472.48) .. controls (528.41,472.48) and (514.61,459.11) .. (514.61,442.62) -- cycle ;
\draw  [fill={rgb, 255:red, 254; green, 234; blue, 111 }  ,fill opacity=1 ] (175.61,443.19) .. controls (175.61,426.7) and (189.41,413.33) .. (206.43,413.33) .. controls (223.45,413.33) and (237.25,426.7) .. (237.25,443.19) .. controls (237.25,459.68) and (223.45,473.05) .. (206.43,473.05) .. controls (189.41,473.05) and (175.61,459.68) .. (175.61,443.19) -- cycle ;
\draw  [fill={rgb, 255:red, 153; green, 233; blue, 150 }  ,fill opacity=1 ] (95.27,443.52) .. controls (95.27,427.03) and (109.07,413.66) .. (126.09,413.66) .. controls (143.11,413.66) and (156.91,427.03) .. (156.91,443.52) .. controls (156.91,460.01) and (143.11,473.38) .. (126.09,473.38) .. controls (109.07,473.38) and (95.27,460.01) .. (95.27,443.52) -- cycle ;
\draw  [fill={rgb, 255:red, 97; green, 164; blue, 252 }  ,fill opacity=1 ] (15.61,442.85) .. controls (15.61,426.36) and (29.41,412.99) .. (46.43,412.99) .. controls (63.45,412.99) and (77.25,426.36) .. (77.25,442.85) .. controls (77.25,459.35) and (63.45,472.72) .. (46.43,472.72) .. controls (29.41,472.72) and (15.61,459.35) .. (15.61,442.85) -- cycle ;
\draw  [fill={rgb, 255:red, 105; green, 98; blue, 237 }  ,fill opacity=1 ] (95.11,280.29) .. controls (95.11,263.79) and (108.91,250.42) .. (125.93,250.42) .. controls (142.95,250.42) and (156.75,263.79) .. (156.75,280.29) .. controls (156.75,296.78) and (142.95,310.15) .. (125.93,310.15) .. controls (108.91,310.15) and (95.11,296.78) .. (95.11,280.29) -- cycle ;
\draw  [fill={rgb, 255:red, 237; green, 98; blue, 98 }  ,fill opacity=1 ] (255.11,280.62) .. controls (255.11,264.13) and (268.91,250.76) .. (285.93,250.76) .. controls (302.95,250.76) and (316.75,264.13) .. (316.75,280.62) .. controls (316.75,297.11) and (302.95,310.48) .. (285.93,310.48) .. controls (268.91,310.48) and (255.11,297.11) .. (255.11,280.62) -- cycle ;

\draw (185.64-1,100) node [anchor=north west][inner sep=0.75pt]  [font=\large] [align=left] {$\tilde{h}_{1,1}$};
\draw (525.64-2,100) node [anchor=north west][inner sep=0.75pt]  [font=\large] [align=left] {$\tilde{h}_{1,2}$};
\draw (104,264) node [anchor=north west][inner sep=0.75pt]  [font=\large] [align=left] {$\tilde{h}_{2,1}$};
\draw (264,264) node [anchor=north west][inner sep=0.75pt]  [font=\large] [align=left] {$\tilde{h}_{2,2}$};
\draw (444,264) node [anchor=north west][inner sep=0.75pt]  [font=\large] [align=left] {$\tilde{h}_{2,3}$};
\draw (604,264) node [anchor=north west][inner sep=0.75pt]  [font=\large] [align=left] {$\tilde{h}_{2,4}$};
\draw (24.64,426) node [anchor=north west][inner sep=0.75pt]  [font=\large] [align=left] {$\tilde{h}_{3,1}$};
\draw (105,426) node [anchor=north west][inner sep=0.75pt]  [font=\large] [align=left] {$\tilde{h}_{3,2}$};
\draw (186,426) node [anchor=north west][inner sep=0.75pt]  [font=\large] [align=left] {$\tilde{h}_{3,3}$};
\draw (364,426) node [anchor=north west][inner sep=0.75pt]  [font=\large] [align=left] {$\tilde{h}_{3,4}$};
\draw (524,426) node [anchor=north west][inner sep=0.75pt]  [font=\large] [align=left] {$\tilde{h}_{3,5}$};

\end{tikzpicture}
\end{adjustbox}

}
\hspace{1.5em}
\subfloat[]{

\begin{adjustbox}{width=0.45\textwidth}
\begin{tikzpicture}[x=0.60pt,y=0.60pt,yscale=-0.8,xscale=0.8]

\draw    (190.23,124.38) -- (125.93,250.42) ;
\draw    (220.23,124.38) -- (285.93,250.76) ;
\draw    (110.73,286.95) -- (46.43,412.99) ;
\draw    (126.09,292.58) -- (126.09,413.66) ;
\draw    (140.73,286.95) -- (206.43,413.33) ;
\draw    (529.73,123.88) -- (465.43,249.92) ;
\draw    (559.73,123.88) -- (625.43,250.26) ;
\draw    (479.73,286.38) -- (545.43,412.76) ;
\draw    (375.11,10.18) -- (205.59,87.76) ;
\draw    (375.11,10.18) -- (545.09,87.26) ;
\draw  [fill={rgb, 255:red, 25; green, 169; blue, 69 }  ,fill opacity=1 ] (174.77,117.62) .. controls (174.77,101.13) and (188.57,87.76) .. (205.59,87.76) .. controls (222.61,87.76) and (236.41,101.13) .. (236.41,117.62) .. controls (236.41,134.11) and (222.61,147.48) .. (205.59,147.48) .. controls (188.57,147.48) and (174.77,134.11) .. (174.77,117.62) -- cycle ;
\draw  [fill={rgb, 255:red, 175; green, 143; blue, 81 }  ,fill opacity=1 ] (514.27,117.12) .. controls (514.27,100.63) and (528.07,87.26) .. (545.09,87.26) .. controls (562.11,87.26) and (575.91,100.63) .. (575.91,117.12) .. controls (575.91,133.61) and (562.11,146.98) .. (545.09,146.98) .. controls (528.07,146.98) and (514.27,133.61) .. (514.27,117.12) -- cycle ;
\draw  [fill={rgb, 255:red, 247; green, 21; blue, 184 }  ,fill opacity=1 ] (434.61,279.79) .. controls (434.61,263.29) and (448.41,249.92) .. (465.43,249.92) .. controls (482.45,249.92) and (496.25,263.29) .. (496.25,279.79) .. controls (496.25,296.28) and (482.45,309.65) .. (465.43,309.65) .. controls (448.41,309.65) and (434.61,296.28) .. (434.61,279.79) -- cycle ;
\draw  [fill={rgb, 255:red, 252; green, 166; blue, 62 }  ,fill opacity=1 ] (594.61,280.12) .. controls (594.61,263.63) and (608.41,250.26) .. (625.43,250.26) .. controls (642.45,250.26) and (656.25,263.63) .. (656.25,280.12) .. controls (656.25,296.61) and (642.45,309.98) .. (625.43,309.98) .. controls (608.41,309.98) and (594.61,296.61) .. (594.61,280.12) -- cycle ;
\draw  [fill={rgb, 255:red, 198; green, 150; blue, 233 }  ,fill opacity=1 ] (514.61,442.62) .. controls (514.61,426.13) and (528.41,412.76) .. (545.43,412.76) .. controls (562.45,412.76) and (576.25,426.13) .. (576.25,442.62) .. controls (576.25,459.11) and (562.45,472.48) .. (545.43,472.48) .. controls (528.41,472.48) and (514.61,459.11) .. (514.61,442.62) -- cycle ;
\draw  [fill={rgb, 255:red, 254; green, 234; blue, 111 }  ,fill opacity=1 ] (175.61,443.19) .. controls (175.61,426.7) and (189.41,413.33) .. (206.43,413.33) .. controls (223.45,413.33) and (237.25,426.7) .. (237.25,443.19) .. controls (237.25,459.68) and (223.45,473.05) .. (206.43,473.05) .. controls (189.41,473.05) and (175.61,459.68) .. (175.61,443.19) -- cycle ;
\draw  [fill={rgb, 255:red, 153; green, 233; blue, 150 }  ,fill opacity=1 ] (95.27,443.52) .. controls (95.27,427.03) and (109.07,413.66) .. (126.09,413.66) .. controls (143.11,413.66) and (156.91,427.03) .. (156.91,443.52) .. controls (156.91,460.01) and (143.11,473.38) .. (126.09,473.38) .. controls (109.07,473.38) and (95.27,460.01) .. (95.27,443.52) -- cycle ;
\draw  [fill={rgb, 255:red, 97; green, 164; blue, 252 }  ,fill opacity=1 ] (15.61,442.85) .. controls (15.61,426.36) and (29.41,412.99) .. (46.43,412.99) .. controls (63.45,412.99) and (77.25,426.36) .. (77.25,442.85) .. controls (77.25,459.35) and (63.45,472.72) .. (46.43,472.72) .. controls (29.41,472.72) and (15.61,459.35) .. (15.61,442.85) -- cycle ;
\draw  [fill={rgb, 255:red, 105; green, 98; blue, 237 }  ,fill opacity=1 ] (95.11,280.29) .. controls (95.11,263.79) and (108.91,250.42) .. (125.93,250.42) .. controls (142.95,250.42) and (156.75,263.79) .. (156.75,280.29) .. controls (156.75,296.78) and (142.95,310.15) .. (125.93,310.15) .. controls (108.91,310.15) and (95.11,296.78) .. (95.11,280.29) -- cycle ;
\draw  [fill={rgb, 255:red, 237; green, 98; blue, 98 }  ,fill opacity=1 ] (255.11,280.62) .. controls (255.11,264.13) and (268.91,250.76) .. (285.93,250.76) .. controls (302.95,250.76) and (316.75,264.13) .. (316.75,280.62) .. controls (316.75,297.11) and (302.95,310.48) .. (285.93,310.48) .. controls (268.91,310.48) and (255.11,297.11) .. (255.11,280.62) -- cycle ;

\draw (185.64-1,100) node [anchor=north west][inner sep=0.75pt]  [font=\large] [align=left] {$\tilde{h}_{1,1}$};
\draw (525.64-2,100) node [anchor=north west][inner sep=0.75pt]  [font=\large] [align=left] {$\tilde{h}_{1,2}$};
\draw (104,264) node [anchor=north west][inner sep=0.75pt]  [font=\large] [align=left] {$\tilde{h}_{2,1}$};
\draw (264,264) node [anchor=north west][inner sep=0.75pt]  [font=\large] [align=left] {$\tilde{h}_{2,2}$};
\draw (444,264) node [anchor=north west][inner sep=0.75pt]  [font=\large] [align=left] {$\tilde{h}_{2,3}$};
\draw (604,264) node [anchor=north west][inner sep=0.75pt]  [font=\large] [align=left] {$\tilde{h}_{2,4}$};
\draw (24.64,426) node [anchor=north west][inner sep=0.75pt]  [font=\large] [align=left] {$\tilde{h}_{3,1}$};
\draw (105,426) node [anchor=north west][inner sep=0.75pt]  [font=\large] [align=left] {$\tilde{h}_{3,2}$};
\draw (186,426) node [anchor=north west][inner sep=0.75pt]  [font=\large] [align=left] {$\tilde{h}_{3,3}$};
\draw (524,426) node [anchor=north west][inner sep=0.75pt]  [font=\large] [align=left] {$\tilde{h}_{3,4}$};

\end{tikzpicture}
    
\end{adjustbox}

}
\caption{Two most found structures for the unbalanced simulation experiment with 50-100 observations for each class of $h_R$. The structure pictured on the left-hand side accounts for 43.2\% of instances while the right-hand side shows the second most retrieved structure, which accounts for 19\% of instances.} 
\label{fig:truefalse}
\end{figure}
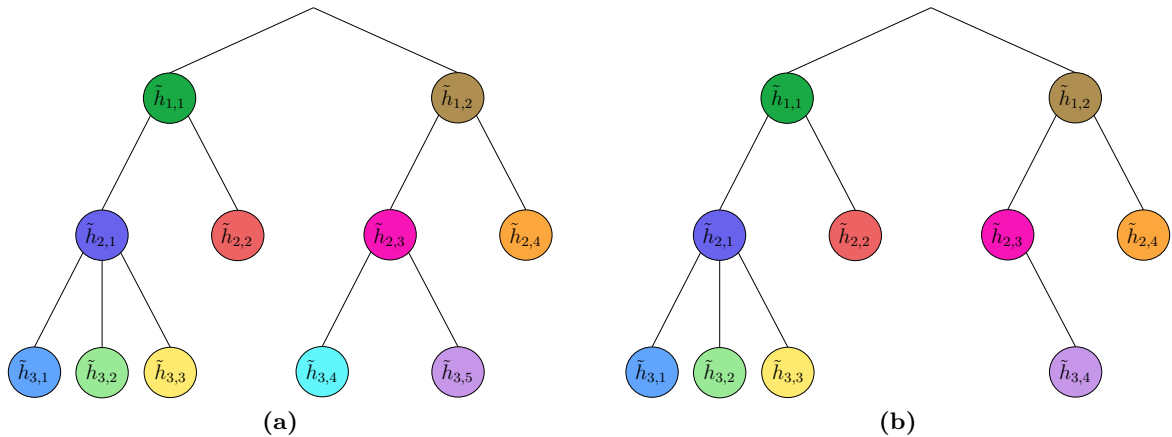
%

\section{Application to a real dataset} \label{reallife}
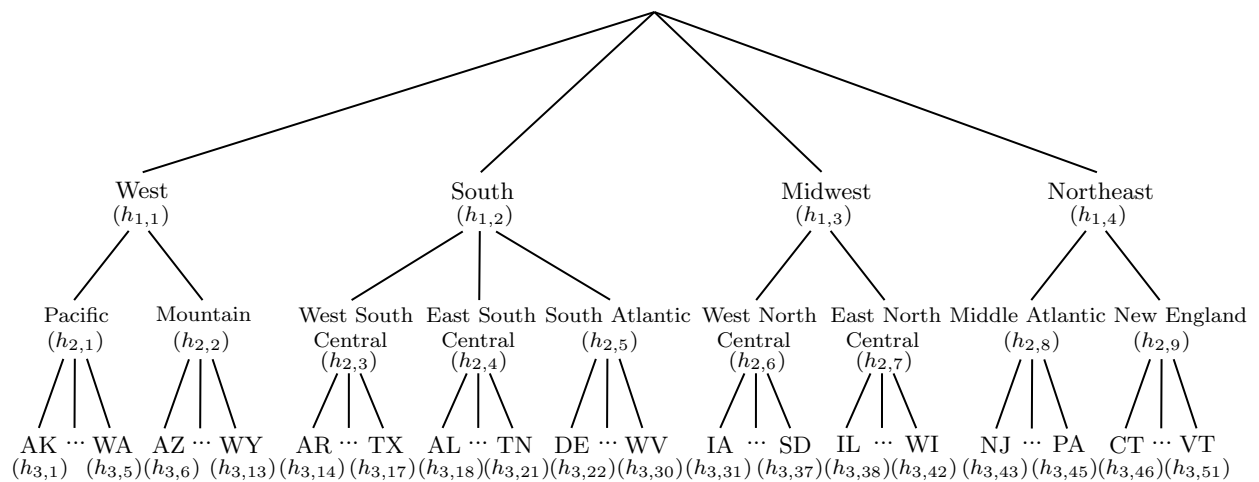
\begin{figure}[h]
\tikzset{every picture/.style={line width=0.75pt}} 

\begin{tikzpicture}[x=0.53pt,y=0.53pt,yscale=-0.7,xscale=0.7]

\draw    (50.8,363.96) -- (26.09,434.14) ;
\draw    (62.4,363.76) -- (62.09,434.14) ;
\draw    (74.6,363.86) -- (97.69,434.14) ;
\draw    (177.6,364.36) -- (152.89,434.54) ;
\draw    (189.2,364.16) -- (188.89,434.54) ;
\draw    (201.4,364.26) -- (224.49,434.54) ;
\draw    (588.2,363.96) -- (563.49,434.14) ;
\draw    (599.8,363.76) -- (599.49,434.14) ;
\draw    (612,363.86) -- (635.09,434.14) ;
\draw    (1016.83,364.56) -- (992.11,434.74) ;
\draw    (1028.43,364.36) -- (1028.11,434.74) ;
\draw    (1040.63,364.46) -- (1063.71,434.74) ;
\draw    (1147.43,364.36) -- (1122.71,434.54) ;
\draw    (1159.03,364.16) -- (1158.71,434.54) ;
\draw    (1171.23,364.26) -- (1194.31,434.54) ;
\draw    (327.3,378.75) -- (302.89,434.14) ;
\draw    (338.94,378.59) -- (338.89,434.14) ;
\draw    (350.55,378.75) -- (373.49,434.39) ;
\draw    (457.9,378.75) -- (433.49,434.14) ;
\draw    (469.54,378.59) -- (469.49,434.14) ;
\draw    (481.15,378.75) -- (504.09,434.39) ;
\draw    (738.1,378.75) -- (713.69,434.14) ;
\draw    (749.74,378.59) -- (749.69,434.14) ;
\draw    (761.35,378.75) -- (784.29,434.39) ;
\draw    (865.3,378.75) -- (840.89,434.14) ;
\draw    (876.94,378.59) -- (876.89,434.14) ;
\draw    (888.55,378.75) -- (911.49,434.39) ;
\draw    (116.8,233.76) -- (61.69,304.14) ;
\draw    (136.8,233.76) -- (191.29,304.14) ;
\draw    (805.06,233.76) -- (749.94,304.14) ;
\draw    (825.06,233.76) -- (879.54,304.14) ;
\draw    (1082.66,233.76) -- (1027.54,304.14) ;
\draw    (1102.66,233.76) -- (1157.14,304.14) ;
\draw    (471.06,233.96) -- (470.49,304.14) ;
\draw    (453.86,234.36) -- (341.69,304.14) ;
\draw    (487.46,234.36) -- (602.49,303.74) ;
\draw    (647.26,11.25) -- (130.89,174.14) ;
\draw    (647.26,11.25) -- (472.09,174.14) ;
\draw    (647.26,11.25) -- (815.97,173.77) ;
\draw    (647.26,11.25) -- (1093.8,174.14) ;

\draw (4,440) node [anchor=north west][inner sep=0.75pt]  [font=\footnotesize] [align=left] {AK};
\draw (50,445) node [anchor=north west][inner sep=0.75pt]  [font=\footnotesize] [align=left] {...};
\draw (-5,462) node [anchor=north west][inner sep=0.75pt]  [font=\scriptsize] [align=left] {$\displaystyle ( h_{3,1})$};
\draw (77,440) node [anchor=north west][inner sep=0.75pt]  [font=\footnotesize] [align=left] {WA};
\draw (70,462) node [anchor=north west][inner sep=0.75pt]  [font=\scriptsize] [align=left] {$\displaystyle ( h_{3,5})$};
\draw (137,440) node [anchor=north west][inner sep=0.75pt]  [font=\footnotesize] [align=left] {AZ};
\draw (177,445) node [anchor=north west][inner sep=0.75pt]  [font=\footnotesize] [align=left] {...};
\draw (128,462) node [anchor=north west][inner sep=0.75pt] [font=\scriptsize]  [align=left] {$\displaystyle ( h_{3,6})$};
\draw (205,440) node [anchor=north west][inner sep=0.75pt]  [font=\footnotesize] [align=left] {WY};
\draw (195,462) node [anchor=north west][inner sep=0.75pt]  [font=\scriptsize] [align=left] {$\displaystyle ( h_{3,13})$};
\draw (282,440) node [anchor=north west][inner sep=0.75pt]  [font=\footnotesize] [align=left] {AR};
\draw (325.46,445) node [anchor=north west][inner sep=0.75pt]  [font=\footnotesize] [align=left] {...};
\draw (265,462) node [anchor=north west][inner sep=0.75pt]  [font=\scriptsize] [align=left] {$\displaystyle ( h_{3,14})$};
\draw (355,440) node [anchor=north west][inner sep=0.75pt]  [font=\footnotesize] [align=left] {TX};
\draw (336,462) node [anchor=north west][inner sep=0.75pt]  [font=\scriptsize] [align=left] {$\displaystyle ( h_{3,17})$};
\draw (415,440) node [anchor=north west][inner sep=0.75pt]  [font=\footnotesize] [align=left] {AL};
\draw (456,445) node [anchor=north west][inner sep=0.75pt]  [font=\footnotesize] [align=left] {...};
\draw (405,462) node [anchor=north west][inner sep=0.75pt]  [font=\scriptsize] [align=left] {$\displaystyle ( h_{3,18})$};
\draw (485,440) node [anchor=north west][inner sep=0.75pt]  [font=\footnotesize] [align=left] {TN};
\draw (471,462) node [anchor=north west][inner sep=0.75pt]  [font=\scriptsize] [align=left] {$\displaystyle ( h_{3,21})$};
\draw (543,440) node [anchor=north west][inner sep=0.75pt] [font=\footnotesize]  [align=left] {DE};
\draw (587,445) node [anchor=north west][inner sep=0.75pt]  [font=\footnotesize] [align=left] {...};
\draw (538,462) node [anchor=north west][inner sep=0.75pt]  [font=\scriptsize] [align=left] {$\displaystyle ( h_{3,22})$};
\draw (614,440) node [anchor=north west][inner sep=0.75pt]  [font=\footnotesize] [align=left] {WV};
\draw (606,462) node [anchor=north west][inner sep=0.75pt]  [font=\scriptsize] [align=left] {$\displaystyle ( h_{3,30})$};
\draw (697,440) node [anchor=north west][inner sep=0.75pt] [font=\footnotesize]  [align=left] {IA};
\draw (737,445) node [anchor=north west][inner sep=0.75pt]  [font=\footnotesize] [align=left] {...};
\draw (675,462) node [anchor=north west][inner sep=0.75pt]  [font=\scriptsize] [align=left] {$\displaystyle ( h_{3,31})$};
\draw (770,440) node [anchor=north west][inner sep=0.75pt] [font=\footnotesize]  [align=left] {SD};
\draw (747,462) node [anchor=north west][inner sep=0.75pt] [font=\scriptsize]  [align=left] {$\displaystyle ( h_{3,37})$};
\draw (828,439) node [anchor=north west][inner sep=0.75pt]  [font=\footnotesize] [align=left] {IL};
\draw (865,445) node [anchor=north west][inner sep=0.75pt]  [font=\footnotesize] [align=left] {...};
\draw (815,462) node [anchor=north west][inner sep=0.75pt]  [font=\scriptsize] [align=left] {$\displaystyle ( h_{3,38})$};
\draw (897,439) node [anchor=north west][inner sep=0.75pt]  [font=\footnotesize] [align=left] {WI};
\draw (882,462) node [anchor=north west][inner sep=0.75pt] [font=\scriptsize]  [align=left] {$\displaystyle ( h_{3,42})$};
\draw (973,440) node [anchor=north west][inner sep=0.75pt] [font=\footnotesize]  [align=left] {NJ};
\draw (1015,445) node [anchor=north west][inner sep=0.75pt]  [font=\footnotesize] [align=left] {...};
\draw (954,462) node [anchor=north west][inner sep=0.75pt] [font=\scriptsize]  [align=left] {$\displaystyle ( h_{3,43})$};
\draw (1046,439) node [anchor=north west][inner sep=0.75pt] [font=\footnotesize]  [align=left] {PA};
\draw (1024,462) node [anchor=north west][inner sep=0.75pt]  [font=\scriptsize] [align=left] {$\displaystyle ( h_{3,45})$};
\draw (1103,440) node [anchor=north west][inner sep=0.75pt] [font=\footnotesize]  [align=left] {CT};
\draw (1146,445) node [anchor=north west][inner sep=0.75pt]  [font=\footnotesize] [align=left] {...};
\draw (1091,462) node [anchor=north west][inner sep=0.75pt]  [font=\scriptsize] [align=left] {$\displaystyle ( h_{3,46})$};
\draw (1174,439) node [anchor=north west][inner sep=0.75pt]  [font=\footnotesize] [align=left] {VT};
\draw (1158,462) node [anchor=north west][inner sep=0.75pt]  [font=\scriptsize] [align=left] {$\displaystyle ( h_{3,51})$};
\draw (28,308) node [anchor=north west][inner sep=0.75pt] [font=\scriptsize]  [align=left] {Pacific};
\draw (285,308) node [anchor=north west][inner sep=0.75pt] [font=\scriptsize]  [align=left] {West South\\ \ \ Central};
\draw (141,308) node [anchor=north west][inner sep=0.75pt]  [font=\scriptsize] [align=left] {Mountain};
\draw (414,308) node [anchor=north west][inner sep=0.75pt]  [font=\scriptsize] [align=left] {East South\\ \ \ Central};
\draw (534,308) node [anchor=north west][inner sep=0.75pt]  [font=\scriptsize] [align=left] {South Atlantic};
\draw (692,308) node [anchor=north west][inner sep=0.75pt]  [font=\scriptsize] [align=left] {West North\\ \ \ Central};
\draw (822,308) node [anchor=north west][inner sep=0.75pt] [font=\scriptsize]  [align=left] {East North\\ \ \ Central};
\draw (942,308) node [anchor=north west][inner sep=0.75pt]  [font=\scriptsize] [align=left] {Middle Atlantic};
\draw (1108,308) node [anchor=north west][inner sep=0.75pt]  [font=\scriptsize] [align=left] {New England};
\draw (100,181) node [anchor=north west][inner sep=0.75pt] [font=\footnotesize]  [align=left] {West};
\draw (439,181) node [anchor=north west][inner sep=0.75pt] [font=\footnotesize]  [align=left] {South};
\draw (772,181) node [anchor=north west][inner sep=0.75pt]  [font=\footnotesize] [align=left] {Midwest};
\draw (1041,181) node [anchor=north west][inner sep=0.75pt]  [font=\footnotesize] [align=left] {Northeast};
\draw (31,332.5) node [anchor=north west][inner sep=0.75pt]  [font=\scriptsize] [align=left] {$\displaystyle ( h_{2,1})$};
\draw (158,332.5) node [anchor=north west][inner sep=0.75pt]  [font=\scriptsize] [align=left] {$\displaystyle ( h_{2,2})$};
\draw (307,352.5) node [anchor=north west][inner sep=0.75pt] [font=\scriptsize]  [align=left] {$\displaystyle ( h_{2,3})$};
\draw (438,352.5) node [anchor=north west][inner sep=0.75pt] [font=\scriptsize]  [align=left] {$\displaystyle ( h_{2,4})$};
\draw (570,332.5) node [anchor=north west][inner sep=0.75pt]  [font=\scriptsize] [align=left] {$\displaystyle ( h_{2,5})$};
\draw (719,352.5) node [anchor=north west][inner sep=0.75pt] [font=\scriptsize]  [align=left] {$\displaystyle ( h_{2,6})$};
\draw (847,352.5) node [anchor=north west][inner sep=0.75pt]  [font=\scriptsize] [align=left] {$\displaystyle ( h_{2,7})$};
\draw (996,332.5) node [anchor=north west][inner sep=0.75pt]  [font=\scriptsize] [align=left] {$\displaystyle ( h_{2,8})$};
\draw (1127,332.5) node [anchor=north west][inner sep=0.75pt] [font=\scriptsize]  [align=left] {$\displaystyle ( h_{2,9})$};
\draw (97,205) node [anchor=north west][inner sep=0.75pt]  [font=\scriptsize] [align=left] {$\displaystyle ( h_{1,1})$};
\draw (444,205) node [anchor=north west][inner sep=0.75pt]  [font=\scriptsize] [align=left] {$\displaystyle ( h_{1,2})$};
\draw (786,205) node [anchor=north west][inner sep=0.75pt] [font=\scriptsize]  [align=left] {$\displaystyle ( h_{1,3})$};
\draw (1063,205) node [anchor=north west][inner sep=0.75pt] [font=\scriptsize]  [align=left] {$\displaystyle ( h_{1,4})$};

\end{tikzpicture}

	\centering
        \caption{Visualisation of the variable \texttt{geography} in the \texttt{cancer\_reg} dataset. The first level has the lowest granularity and consists of the four major U.S. regions. The second level denotes the nine U.S. subregions. The lowest and most granular level consists of the 50 U.S. states and the District of Columbia.}
        \label{fig:geography}
\end{figure}

\begin{figure}[H] 
\centering
\subfloat[$r=1$]{\includegraphics[width = 0.33\textwidth]{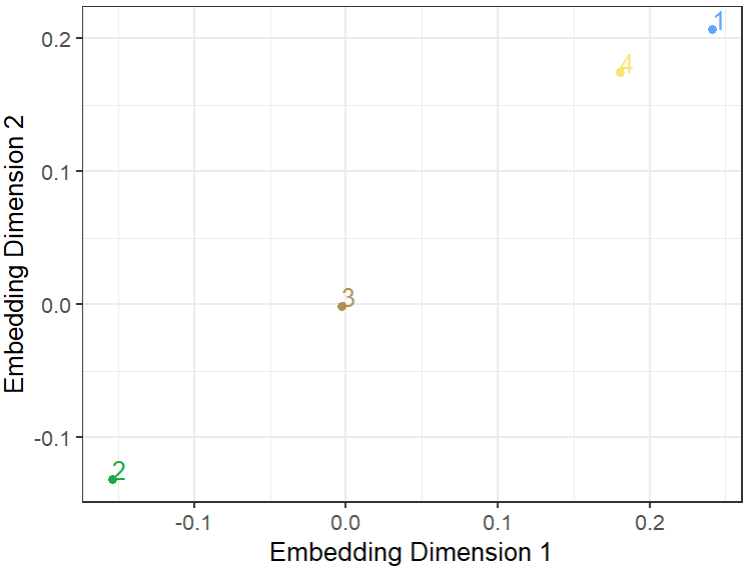}} 
\subfloat[$r=2$]{\includegraphics[width = 0.33\textwidth]{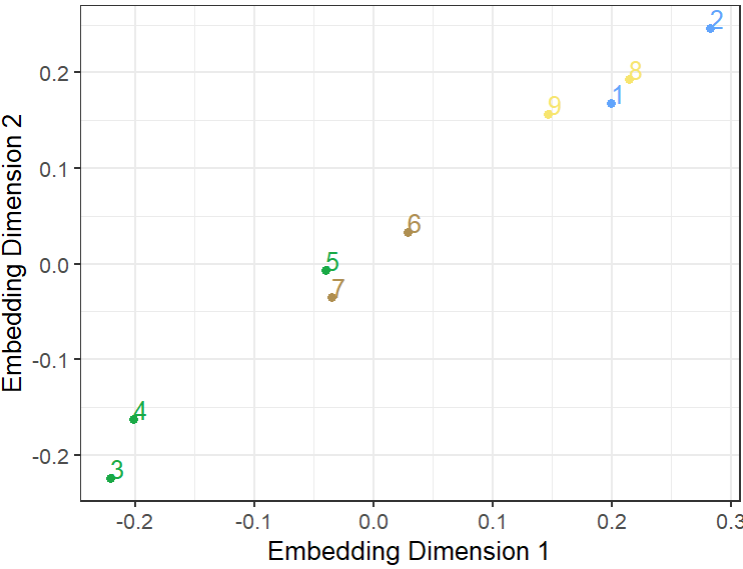}}
\subfloat[$r=3$]{\includegraphics[width = 0.33\textwidth]{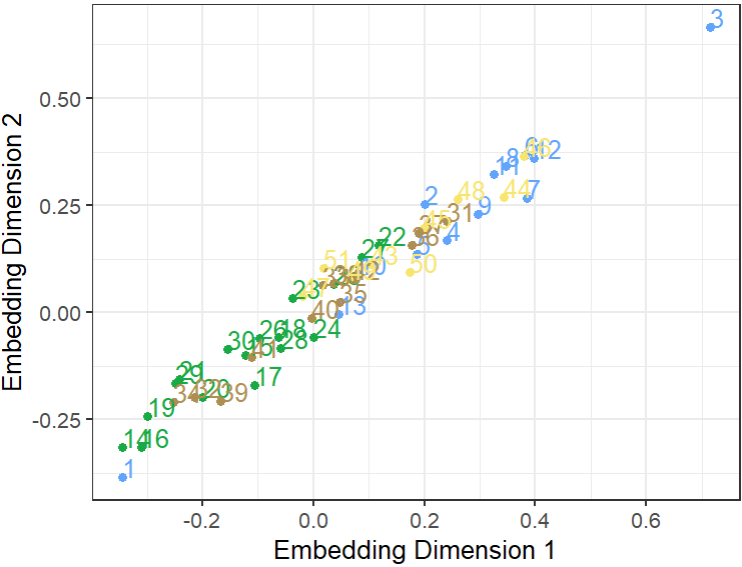}}
\caption{Visualisation of the embedding vectors obtained for the classes at each level of the hierarchical categorical variable \texttt{geography}. The colour encoding refers to which US region a state or subregion belongs. Blue indicates the West, green the South, brown the Midwest and yellow the Northeast. } 
\label{fig:embedus}
\end{figure}

In this section, we apply our method to the \texttt{cancer\_reg} dataset which is used by \cite{carrizosa2022tree} \added{and was compiled using data from \texttt{clinicaltrials.gov}, \texttt{cancer.gov} and \texttt{census.gov} \citep{rippner_2017}}. The dataset contains 31 variables describing socio-economic information of 3\ 047 U.S. counties as well as information about the number of cancer mortalities per one hundred thousand inhabitants, which is the response variable. The dataset also includes a hierarchical categorical variable \texttt{geography} pictured in Figure~\ref{fig:geography}, which encodes the location of the county at the level of the U.S. region, subregion and state in order of increasing granularity. \added{We split the data into a training and test set, stratified at the state  level, accounting for 80\% and 20\% of the data respectively.  }
To learn the embedding vectors, we use the network architecture pictured in Figure~\ref{fig:simulnetwork} of Section~\ref{sec:simulating}. Figure~\ref{fig:embedus} shows the embedding values for each of the levels of the hierarchical categorical variable. On the one hand, we observe that the states in the South (green) and Midwest (brown), and, on the other hand, the states in the West (blue) and Northeast (yellow) tend to lie close in the embedding space. 
We consider a grid of values for the tuning parameter $SI^{*}$ introduced in Section~\ref{sec:collapsing}. To find a preferred value for $SI^{*}$, we fit a linear model where we include a categorical variable that groups the states that were merged or collapsed together. We do this for each value in the grid, using the hierarchical structure $\boldsymbol{\widetilde{h}}$ obtained for the specific value of $SI^{*}$. We select the value of $SI^{*}$ resulting in the model with the lowest AIC and/or BIC. We standardise the non-hierarchical continuous variables and exclude the covariates with missing values, i.e. \texttt{pctsomecol18\_24}, \texttt{pctemployed16\_over} and \texttt{pctprivatecoveragealone}. Table~\ref{tab:BICgeograph} summarises the $\text{AIC}$ and $\text{BIC}$ values for four different values of $SI^{*}$. As indicated in Section~\ref{sec:collapsing}, a higher value for $SI^{*}$ results in a more collapsed representation $\boldsymbol{\widetilde{h}}$ of the hierarchical categorical variable \texttt{geography}. We also compare our reduced representations to the solutions proposed by \cite{carrizosa2022tree}. Figure~\ref{fig:carri} in Appendix~\ref{append:applic} pictures their reduced representations as preferred by the $\text{AIC}$ and $\text{BIC}$. We refer to the resulting models as \texttt{Carrizosa\_AIC} and \texttt{Carrizosa\_BIC}, respectively. In terms of $\text{AIC}$ and $\text{BIC}$, we observe that our method improves the model fit and outperforms the models incorporating the structures preferred by the $\text{AIC}$ and $\text{BIC}$ of \cite{carrizosa2022tree}. The resulting structure is the same for $SI^{*}=0.1$ or $SI^{*}=0.3$. The $\text{AIC}$ and $\text{BIC}$ values both indicate that the resulting structure obtained for $SI^{*}=0.5$ is preferred. When we opt for $SI^{*}=0.7$, we obtain a completely collapsed structure which clearly performs worse in terms of AIC and BIC compared to the other considered model specifications. \added{Table~\ref{tab:BICgeograph} also includes the out-of-sample performance on the test set in terms of the RMSE. The structure for $SI^{*}=0.5$ results in the best out-of-sample predictive accuracy.} 
\begin{table}[H]
\centering
\begin{tabular}{c c c c c}
  \hline
 &&$\text{AIC}$&$\text{BIC}$& $\text{RMSE}$ \\
\hline 
$\boldsymbol{h}$&&4880.48&5391.47&0.3357\\[5pt]
$SI^{*}$&0.1&4854.17&5121.28&0.3352\\
&0.3&4854.17&5121.28&0.3352\\
&0.5&\textbf{4841.44}&\textbf{5114.35}&\textbf{0.3341}\\
&0.7&5178.03&5398.684&0.3515\\
\texttt{Carrizosa\_AIC}&&4875.40&5345.74&0.3357\\
\texttt{Carrizosa\_BIC}&&4946.48&5236.801&0.3399\\

\hline 
 
\end{tabular}
\caption{$\text{AIC}$, $\text{BIC}$ \added{and the out-of-sample $\text{RMSE}$} for the linear models incorporating $\boldsymbol{\widetilde{h}}$ for different values of $SI^{*}$ as well as the model including \texttt{geography}. The last two rows show the $\text{AIC}$, $\text{BIC}$ and the out-of-sample $\text{RMSE}$ for the models including the reduced representations preferred by the $\text{AIC}$ and $\text{BIC}$ proposed by \cite{carrizosa2022tree}, referred to as \texttt{Carrizosa\_AIC} and \texttt{Carrizosa\_BIC}, respectively.} 
\label{tab:BICgeograph}
\end{table}
Figure~\ref{fig:truegeo} pictures the resulting reduced representation $\boldsymbol{\widetilde{h}}$ with $SI^{*}=0.5$, i.e.~the structure for which we achieve a minimal $\text{AIC}$ and $\text{BIC}$ value. The complexity of the hierarchical structure is significantly reduced, as we have $\widetilde{n}_1=2$, $\widetilde{n}_2=2$ and $\widetilde{n}_3=8$ compared to $n_1=4$, $n_2=9$ and $n_3=51$. States that are not explicitly shown in Figure~\ref{fig:truegeo} are collapsed upstream, e.g. Tennessee (TN) and Mississipi (MS) are collapsed into West South Central \& East South Central.
\begin{figure}[H]
        \centering
        \scalebox{0.55}{

\tikzset{every picture/.style={line width=0.75pt}} 

\tikzset{every picture/.style={line width=0.75pt}} 

\begin{tikzpicture}[x=0.75pt,y=0.75pt,yscale=-1,xscale=1]

\draw    (610.83,375.42) -- (563.1,524.26) ;
\draw    (641.33,375.67) -- (695.82,524.34) ;
\draw    (179.93,190.33) -- (23.53,523.62) ;
\draw    (740.5,186.15) -- (629.83,315.42) ;
\draw    (784.5,186.15) -- (896.08,315.42) ;
\draw    (492.43,5.26) -- (762.37,154.34) ;
\draw    (492.43,5.26) -- (222.62,154.59) ;
\draw    (876.95,397.67) -- (829.6,524.26) ;
\draw    (916.15,396.87) -- (962.32,524.34) ;
\draw    (269.93,190.33) -- (415.13,523.22) ;
\draw    (210.33,189.93) -- (159.6,523.62) ;
\draw    (240.33,190.33) -- (284.8,523.62) ;

\draw (140.75,159) node [anchor=north west][inner sep=0.75pt]  [font=\Large] [align=left] {West \& Northeast};
\draw (674.25,156.5) node [anchor=north west][inner sep=0.75pt]  [font=\Large] [align=left] {South \& Midwest};
\draw (526,315.75) node [anchor=north west][inner sep=0.75pt]  [font=\Large] [align=left] {West South Central \&\\ East South Central};
\draw (806,312.5) node [anchor=north west][inner sep=0.75pt]  [font=\Large] [align=left] {South Atlantic, \\West North Central \\\& East North Central};
\draw (8.25,527.39) node [anchor=north west][inner sep=0.75pt]  [font=\Large] [align=left] {AK};
\draw (272.5,527.39) node [anchor=north west][inner sep=0.75pt]  [font=\Large] [align=left] {HI};
\draw (358,526.89) node [anchor=north west][inner sep=0.75pt]  [font=\Large] [align=left] {AR, CO, ID,\\NM, UT, NY\\\& CT};
\draw (97.65,527.39) node [anchor=north west][inner sep=0.75pt]  [font=\Large] [align=left] {NV, WY, NJ,\\ME, NH \& VT};
\draw (526.5,526.89) node [anchor=north west][inner sep=0.75pt]  [font=\Large] [align=left] {AR, OK \\\& KY};
\draw (662,527.14) node [anchor=north west][inner sep=0.75pt]  [font=\Large] [align=left] {LA, TX \\\& AL};
\draw (764.5,526.64) node [anchor=north west][inner sep=0.75pt]  [font=\Large] [align=left] {DE, GA, NC,\\IA, MN, ND,\\NE, SD, IL, \\MI \& WI};
\draw (916,526.61) node [anchor=north west][inner sep=0.75pt]  [font=\Large] [align=left] {DC, FL, MD, \\SC, VA, WV, \\KS, MO, IN \\\& OH};

\end{tikzpicture}}
        \caption{Reduced representation of the hierarchical categorical variable \texttt{geography} for $SI^{*}=0.5$. }
        \label{fig:truegeo}
\end{figure}
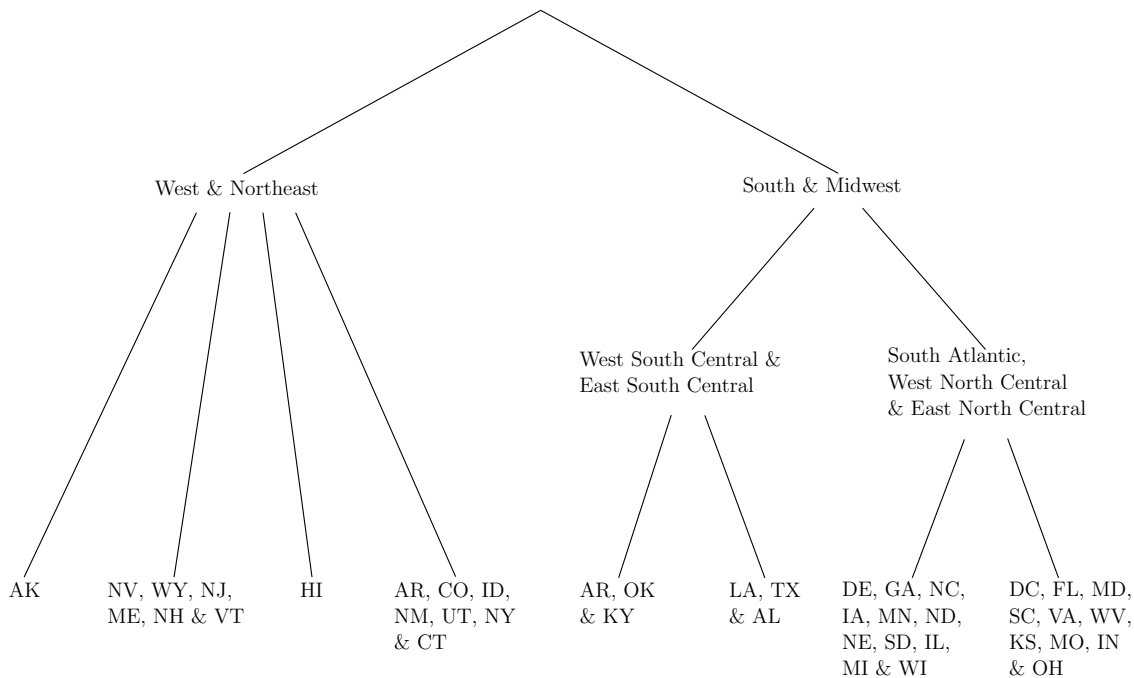

\begin{figure}[H] 
\centering
\includegraphics[width = 0.8\textwidth]{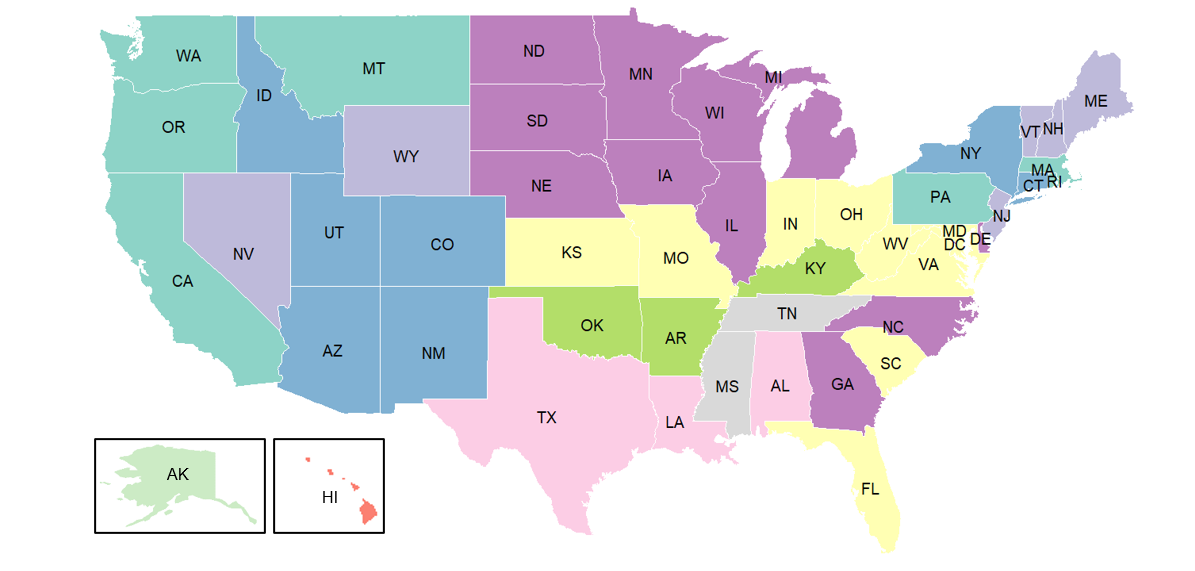}
\caption{Map of the United States of America including all 50 states as well as the District of Columbia, states that are merged together according to the reduced structure pictured in Figure~\ref{fig:truegeo} are visualised using the same colour.} 
\label{fig:usa}
\end{figure}

Figure~\ref{fig:usa} pictures the classes that are merged (or collapsed) together at the state level according to the solution preferred by the $\text{AIC}$ as shown in Figure~\ref{fig:truegeo}. We observe that progressive states, e.g. Pennsylvania and California, tend to be merged together and this is also the case for more conservative states like Texas and Louisiana. We also note that, in general, neighbouring states exhibit a tendency to be merged or collapsed together and that the two states (Hawaii and Alaska) isolated from the mainland are not merged with other states.

Compared to the structures pictured in Figure~\ref{fig:carri} in Appendix~\ref{append:applic}, we observe that our solution clusters states that were not originally within the same region like Kansas (Midwest) and Florida (South). In terms of the number of clusters and cluster size, our solution falls in between the structures proposed by \cite{carrizosa2022tree}.


\section{Discussion} \label{sec:conclus}

In this paper, we propose a novel methodology relying on entity embeddings and clustering techniques to reduce the dimensionality and granularity of a hierarchical categorical variable. We propose a top-down clustering algorithm that clusters entity embedding vectors, which encode the classes within a hierarchy, to obtain a reduced representation of a hierarchical categorical variable. The algorithm extends the current literature by allowing for both the merging of classes within a level of a hierarchy as well as the (partial) collapse of a level. The result of our methodology is a reduced hierarchical categorical variable, which can subsequently be used in a predictive model of choice. The resulting model is more sparse while maintaining (or improving) the \added{balance between model fit and complexity} compared to including the original hierarchy. We show the effectiveness of the proposed algorithm in capturing the true structure of a simulated hierarchical categorical variable with respect to a response variable. We also verify our methodology on a real dataset and find that the resulting structure is an improvement over the original hierarchical structure and outperforms existing solutions \added{for this dataset}. Future research can investigate alternatives to obtain the class embeddings. \added{Specifically, random effects entity embedding \citep{avanzi2024machine,richman2024high} can lead to improved performance when dealing with sparse classes.} \added{A second direction for future research is the extension of the methodology to different hierarchical structures, relaxing the assumption that a class can only have a single parent class. This would generalise our approach to the set of directed acyclic graphs. In this multi-parent setting, the between-level clustering step needs to be adapted to take into account the multitude of parent-descendant relations within the hierarchical structure.}

\biblist

\pagebreak
\appendix
\section{Top-down clustering algorithm for embedding vectors representing a hierarchical categorical variable } \label{append:algo}

Algorithm~\ref{algo:pseudo} gives a summary of the top-down algorithm. Our algorithm consists out of two major steps. In the first (horizontal) step, we merge classes within a given level and in the second (vertical) step, we merge classes at a given level with their parent class at the previous level. The steps are performed sequentially at each level in the hierarchy in a top-down approach, starting with the upper level. The second step is not performed when considering the lowest level of the hierarchy, i.e.~$r=R$.

\begin{algorithm}[H] 
    \caption{Pseudocode summary top-down algorithm\label{algo:pseudo}}
    
    \For{$r=1,\ldotp\ldotp,R$}{
    \vspace{0.2em}
        Apply within-level clustering step (see Algorithm~\ref{algo:horizontal}) to level $r$. \\
        \vspace{0.2em}
        \If{$r<R$}{
            Apply between-level clustering step (see Algorithm~\ref{algo:vertical}) to level $r$ and $r+1$. \\
        
        }
        \vspace{0.2em}
    }
\end{algorithm}

\subsection{Within-level clustering step}
The first step, i.e.~the merging of classes within a given level, is detailed in Algorithm~\ref{algo:horizontal}. When applying this step to level $r$ with $r>1$, we already have $\widetilde{H}_{r-1}=\{\widetilde{h}_{r-1,1},\ldotp\ldotp,\widetilde{h}_{r-1,\widetilde{n}_{r-1}}\}$. Additionally, for $\widetilde{h}_{r-1,s} \in \widetilde{H}_{r-1}$, we know the set of classes that was clustered to construct $\widetilde{h}_{r-1,s}$, which we refer to as cluster $C_{r-1,s}$. We use $\mathcal{C}^{(1)}_{r-1,s}$ to denote the descendant classes of the merged classes. Since we already applied the second step to level $r-1$, we also know which classes from $H_{r-1}$ are collapsed. Subsequently, for $\widetilde{h}_{r-1,s}$, we denote the descendant classes excluding the collapsed classes as $\mathcal{C}^{(2)}_{r-1,s}$.

For each class $\widetilde{h}_{r-1,s} \in \widetilde{H}_{r-1}$, we apply k-medoids to the embeddings of the classes in $\mathcal{C}^{(2)}_{r-1,s}$, i.e.~to $\{ \boldsymbol{e}_{r,l} \: | \:  h_{r,l} \in \mathcal{C}_{r-1,s}^{(2)} \}$. The resulting clustering solution $E = \{ E^{(1)}, \ldotp \ldotp, E^{(K^{*})}\}$ is determined by choosing the solution with the lowest silhouette index while taking into account $SI^{*}$, as discussed in Section~\ref{sec:collapsing}. Assume we have already constructed $c_r$ clusters at level $r$. For the $k$th cluster in $E$, we can then define $C_{r,c_r+k}$, $\mathcal{C}^{(1)}_{r,c_r+k}$ and $\boldsymbol{e}^{c}_{r,c_r+k}$ as shown in Algorithm~\ref{algo:horizontal}, where the embedding $\boldsymbol{e}^{c}_{r,c_r+k}$ is defined as the average of the embeddings of the classes in $C_{r,c_r+k}$. Consequently, we use $\mathcal{\widetilde{H}}_{r-1,s} = \{ \widetilde{h}_{r,k} | k=c_r+1,\ldotp \ldotp,c_r+K^* \}$ to denote the descendant classes of $\widetilde{h}_{r-1,s}$ within the new hierarchical structure. 

The steps laid out above are not only applied to each class $\widetilde{h}_{r-1,s} \in \widetilde{H}_{r-1}$, but also to $F_{r-1,s} \in \mathfrak{F}_{r-1}$. We use $F_{r-1,s}$ to denote a set of classes in $H_{r-1}$ having the same parent class that is collapsed with level $r-2$ and refer to this as a pseudocluster. Analogously to the notation $\mathcal{C}^{(1)}_{r,s}$, $\mathcal{C}^{(2)}_{r,s}$ and $\boldsymbol{e}^{c}_{r,s}$ for cluster $C_{r,s}$, we use $\mathcal{F}^{(1)}_{r,s}$, $\mathcal{F}^{(2)}_{r,s}$ and $\boldsymbol{e}^{f}_{r,s}$ for pseudocluster $F_{r,s}$. By also applying the first step of the algorithm to $\{ \boldsymbol{e}_{r,l} \: | \:  h_{r,l} \in \mathcal{F}_{r-1,s}^{(2)} \}$, for $F_{r-1,s} \in \mathfrak{F}_{r-1}$, we allow for more flexible structures as illustrated by Figure~\ref{fig:truegeo}. After performing all the intermediary clustering tasks, we define $\widetilde{H}_{r}$ as $\{ \widetilde{h}_{r,s}|s=1,\ldotp\ldotp,\widetilde{n}_{r} \}$ where $\widetilde{n}_{r}$ is the sum of the number of clusters found in each intermediary step.

\begin{algorithm}[H] \label{algo:horizontal}
    \caption{\textbf{Pseudocode within-level (horizontal) clustering step}}

    \eIf{$r=1$}{
            Apply k-medoids to the set $\{ \boldsymbol{e}_{r,s} \: | \:  h_{r,s} \in H_{1} \}$. Denote $E = \{ E^{(1)}, \ldotp \ldotp, E^{(K^{*})}\}$ as the clustering solution for which the corresponding silhouette index $SI_{K^{*}}$ is the lowest.\\
            
            \If{$SI_{K^{*}} < SI^{*}$}{
            $E = \{ E^{(1)}\}$ with $E^{(1)} = \bigcup_{k=1}^{K^{*}} E^{(k)}$\\
            $K^{*}=1$
            }

            \For{$k=1,\ldotp,K^{*}$}{
            $C_{r,k} = \{  h_{r,s} \: | \:   \boldsymbol{e}_{r,s} \in E^{(k)} \}$\\
            $\mathcal{C}_{r,k}^{(1)} =  \bigcup_{s|h_{r,s} \in C_{r,c_{r}+k}} \mathcal{H}_{r,s}$\\
            $\boldsymbol{e}^{c}_{r,k} = \frac{1}{\text{dim}(E^{(k)}) 
            }\sum_{s|\boldsymbol{e}_{r,s} \in E^{(k)}} \boldsymbol{e}_{r,s}$
        
        }
        $c_{1} = K^{*}$

        }{
        $c_{r}=0$\\
        \For{$\widetilde{h}_{r-1,s} \in \widetilde{H}_{r-1}$, $F_{r-1,s} \in \mathfrak{F}_{r-1}$}{
        Apply k-medoids to the set $\{ \boldsymbol{e}_{r,l} \: | \:  h_{r,l} \in \mathcal{C}_{r-1,s}^{(2)} \}$ or the set $\{ \boldsymbol{e}_{r,l} \: | \:  h_{r,l} \in \mathcal{F}_{r-1,s}^{(2)} \}$  when considering $\widetilde{h}_{r-1,s}$ or $F_{r-1,s}$, respectively. Denote $E = \{ E^{(1)}, \ldotp \ldotp, E^{(K^{*})}\}$ as the clustering solution for which the corresponding silhouette index $SI_{K^{*}}$ is the lowest.\\

        \If{$SI_{K^{*}} < SI^{*}$}{
        $E = \{ E^{(1)}\}$ with $E^{(1)} = \bigcup_{k=1}^{K^{*}} E^{(k)}$\\
        $K^{*}=1$
        }

        \For{$k=1,\ldotp,K^{*}$}{
        $C_{r,c_{r}+k} = \{  h_{r,s} \: | \:   \boldsymbol{e}_{r,s} \in E^{(k)} \}$\\
        $\mathcal{C}_{r,c_{r}+k}^{(1)} =  \bigcup_{s|h_{r,s} \in C_{r,c_{r}+k}} \mathcal{H}_{r,s}$\\
        $\boldsymbol{e}^{c}_{r,c_{r}+k} = \frac{1}{\text{dim}(E^{(k)}) 
     }\sum_{s|\boldsymbol{e}_{r,s} \in E^{(k)}} \boldsymbol{e}_{r,s}$
        
        }
        $\mathcal{\widetilde{H}}_{r-1,s} = \{ \widetilde{h}_{r,k} | k=c_r+1,\ldotp \ldotp,c_r+K^* \}$ or $\mathcal{\widetilde{F}}_{r-1,s} = \{ \widetilde{h}_{r,k} | k=c_r+1,\ldotp \ldotp,c_r+K^* \}$, when considering $\widetilde{h}_{r-1,s}$ or $F_{r-1,s}$,respectively.\\
        $c_{r}=c_{r}+K^{*}$
        }
        }
        $\widetilde{n}_{r}=c_{r}$\\
        $\mathfrak{C}_{r} =\{ C_{r,s}|s=1,\ldotp\ldotp,\widetilde{n}_{r} \} $\\
        $\widetilde{H}_{r} =\{ \widetilde{h}_{r,s}|s=1,\ldotp\ldotp,\widetilde{n}_{r} \} $

\end{algorithm}
\subsection{Between-level clustering step}
The second step of the algorithm is laid out in Algorithm~\ref{algo:vertical}. From applying the first step to level $r$ and the second step to level $r-1$, we know each class $\widetilde{h}_{r,s} \in \widetilde{H}_{r}$ and pseudocluster $F_{r,s} \in \mathfrak{F}_{r}$, respectively. We also know the corresponding embedding vectors $\boldsymbol{e}_{r,s}^{c}$ and $\boldsymbol{e}_{r,s}^{f}$ as well as the set of descendants within $H_{r+1}$, i.e.~$\mathcal{C}_{r,s}^{(1)}$ and $\mathcal{F}_{r,s}^{(1)}$ for $\widetilde{h}_{r,s}$ and $F_{r,s}$, respectively.

For each class $\widetilde{h}_{r,s} \in \widetilde{H}_{r}$, we merge its descendant classes that are sufficiently close in the embedding space. We do this by applying k-medoids to the set $\boldsymbol{e}_{r,s}^{c} \cup \{ \boldsymbol{e}_{r+1,l} \: | \:  h_{r+1,l} \in \mathcal{C}_{r,s}^{(1)} \}$. As in the first step of the algorithm, we denote the clustering solution as $E = \{ E^{(1)}, \ldotp \ldotp, E^{(K^{*})}\}$, where the number of clusters $K^*$ is determined by the silhouette index. In this case, we are only interested in the classes that are clustered together with their parent class $\widetilde{h}_{r,s}$. We denote the cluster containing the embedding of the parent class excluding the parent embedding itself, as $E^{(p)}$. If the silhouette index corresponding to the clustering solution $E$ is below $SI^*$, we consider all descendant classes in $\mathcal{C}_{r,s}^{(1)}$ as being clustered with $\widetilde{h}_{r,s}$. As discussed in Section~\ref{sec:collapsing}, we impose an additional constraint where we do not collapse any of the considered classes in $E^{}(p)$ with their parent if we have that $SI(\boldsymbol{e}_{r,s}^{c}) = \min_{\boldsymbol{e} \in E^{(p)}} SI(\boldsymbol{e})$. If the classes are collapsed, we define $F_{r+1,f_{r+1}}$, $\mathcal{F}^{(1)}_{r+1,f_{r+1}}$ and $\boldsymbol{e}_{r+1,f_{r+1}}^{f}$ as indicated in Algorithm~\ref{algo:vertical}, where $f_{r+1}$ denotes the number of pseudoclusters we already constructed on level $r+1$. We denote the set of descendant classes of $\widetilde{h}_{r,s}$ excluding the collapsed descendants, as $\mathcal{C}_{r,s}^{(2)}$. This set is again required when applying the first step of the algorithm to level $r+1$.

The steps laid out above are also applied to each pseudocluster $F_{r,s}$ on the previous level. In that case, the set of embeddings considered for clustering is $\boldsymbol{e}_{r,s}^{f} \cup  \{\boldsymbol{e}_{r+1,l} \: | \:  h_{r+1,l} \in \mathcal{F}_{r,s}^{(1)} \}$. Instead of $\mathcal{C}_{r,s}^{(2)}$, we then construct $\mathcal{F}_{r,s}^{(2)}$ in the last step.

\begin{algorithm}[H] \label{algo:vertical}
    \caption{\textbf{Pseudocode between-level (vertical) clustering step}}

    $f_{r+1}=0$
            
            \For{$\widetilde{h}_{r,s} \in \widetilde{H}_{r}$, $F_{r,s} \in \mathfrak{F}_{r}$}{

            Apply k-medoids to the set $\boldsymbol{e}_{r,s}^{c} \cup \{ \boldsymbol{e}_{r+1,l} \: | \:  h_{r+1,l} \in \mathcal{C}_{r,s}^{(1)} \}$ or the set $\boldsymbol{e}_{r,s}^{f} \cup  \{\boldsymbol{e}_{r+1,l} \: | \:  h_{r+1,l} \in \mathcal{F}_{r,s}^{(1)} \}$  when considering $\widetilde{h}_{r,s}$ or $F_{r,s}$, respectively. Denote $E = \{ E^{(1)}, \ldotp \ldotp, E^{(K^{*})}\}$ as the clustering solution for which the corresponding silhouette index $SI_{K^{*}}$ is the lowest.\\

            \If{$SI_{K^{*}} < SI^{*}$}{
            $E = \{ E^{(1)}\}$ with $E^{(1)} = \bigcup_{k=1}^{K^{*}} E^{(k)}$\\
            $K^{*}=1$
            }

            $E^{(p)} = \{ E^{(k)} | \boldsymbol{e}_{r,s}^{c} \in E^{k} \}$

            \If{$E^{(p)} \ne \varnothing$, $SI(\boldsymbol{e}_{r,s}^{c} \text{ or } \boldsymbol{e}_{r,s}^{f}) \ne \min_{\boldsymbol{e} \in E^{(p)}} SI(\boldsymbol{e})$}{
            $f_{r+1} = f_{r+1} + 1$\\
            $F_{r+1,f_{r+1}} = \{ h_{r+1,k} | \boldsymbol{e}_{r+1,k} \in E^{(p)} \}$\\
            $\mathcal{F}^{(1)}_{r+1,f_{r+1}} = \bigcup_{k|h_{r+1,k} \in F_{r+1,f_{r+1}}} \mathcal{H}_{r+1,k} $\\

            $\boldsymbol{e}_{r+1,f_{r+1}}^{f} = \boldsymbol{e}_{r,s}^{c}$ or $\boldsymbol{e}_{r+1,f_{r+1}}^{f} = \boldsymbol{e}_{r,s}^{f}$, for $\widetilde{h}_{r,s}$ and $F_{r,s}$, respectively.\\

            $\mathcal{C}_{r,s}^{(2)} = \mathcal{C}_{r,s}^{(1)} \backslash F_{r+1,f_{r+1}}$ or $\mathcal{F}_{r,s}^{(2)} = \mathcal{F}_{r,s}^{(1)} \backslash F_{r+1,f_{r+1}}$, for $\widetilde{h}_{r,s}$ and $F_{r,s}$, respectively.
            }
            
            }

            $\mathfrak{F}_{r+1} = \{ F_{r+1,s} | s=1,\ldotp\ldotp,f_{r+1} \}$
    
\end{algorithm}
%


\section{Simulation experiments} \label{append:simul}
\begin{landscape}
    \begin{figure}[H]
       \centering
       \includegraphics[width=1.26\textwidth]{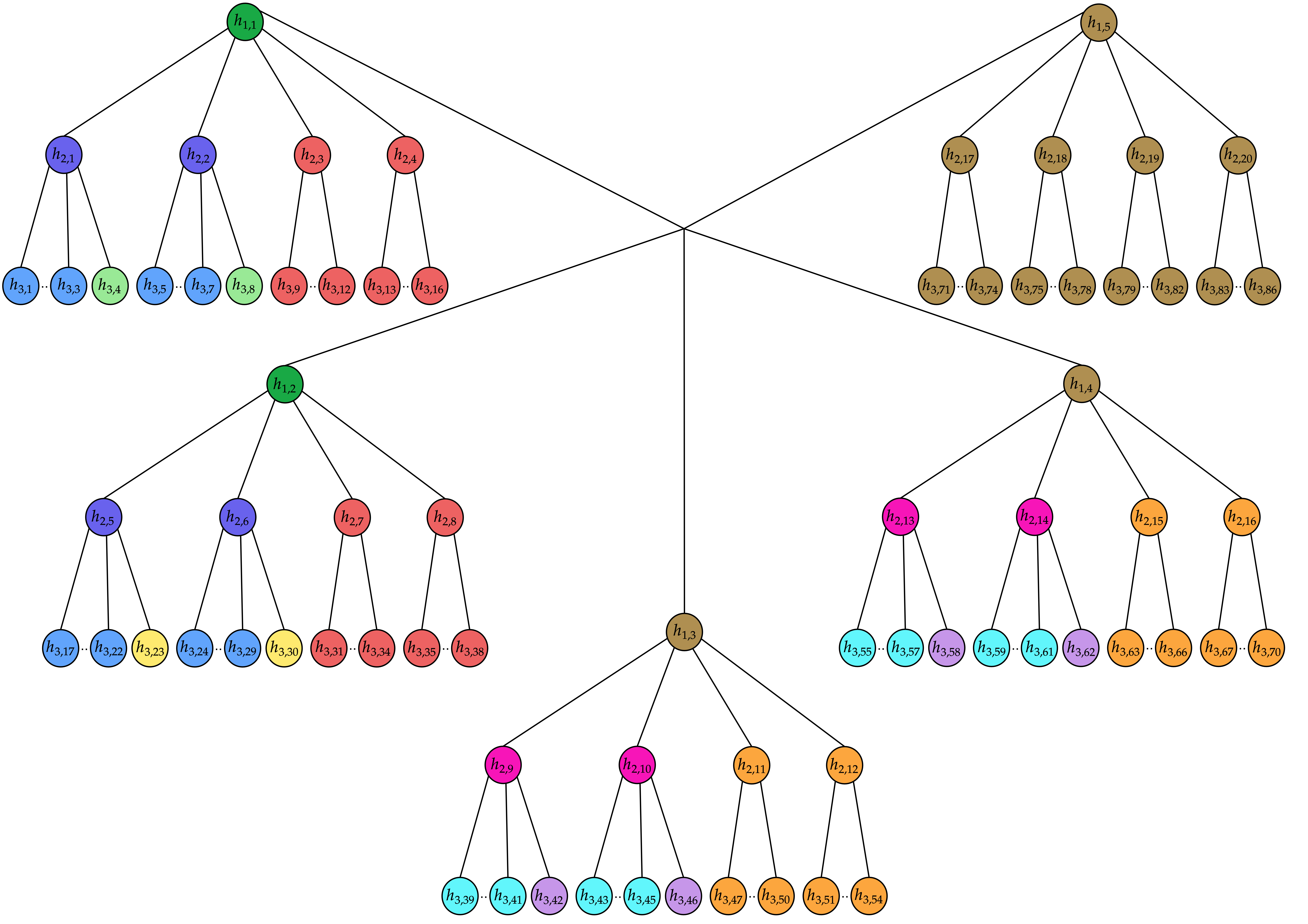}
        \caption{Hierarchical categorical variable $\boldsymbol{h}$ used in the simulation experiments in Section~\ref{sec:simulating}. There are three levels with $n_1=5$, $n_2=20$ and $n_3=86$ classes, respectively. Classes that are assumed to have the same effect on the response are pictured using the same colour.}
        \label{fig:3factfull}
\end{figure}
\end{landscape}

\paragraph{Response} For the simulation experiments in Section~\ref{sec:simulating}, the conditional mean of the distribution of the response variable is defined as:
\begin{equation} \label{eq:experiment}
    \begin{aligned}
        \text{E}(y|\boldsymbol{h}) = g^{-1}(\mu &+
    \gamma_{1} X_{1,1} + \gamma_{1} X_{1,2} - \frac{2}{3} \gamma_{1} X_{1,3} - \frac{2}{3} \gamma_{1} X_{1,4} +
    \gamma_{2} X_{2,1} + \gamma_{2} X_{2,2} - \gamma_{2} X_{2,3}
    + \gamma_{2} X_{2,5}\\ &+ \gamma_{2} X_{2,6} - \gamma_{2} X_{2,7}
    + \gamma_{3} X_{2,9} + \gamma_{3} X_{2,10}- \gamma_{3} X_{2,11}
    + \gamma_{3} X_{2,13}+ \gamma_{3} X_{2,14}- \gamma_{3} X_{2,15}
    \\ &+ \gamma_{4} X_{3,1}+ \gamma_{4} X_{3,2}+ \gamma_{4} X_{3,3}
    + \gamma_{4} X_{3,5}+ \gamma_{4} X_{3,6}+ \gamma_{4} X_{3,7}
    + \gamma_{4} X_{3,17}+ \gamma_{4} X_{3,18}\\ &+ \gamma_{4} X_{3,19}+ \gamma_{4} X_{3,20}+ \gamma_{4} X_{3,21}+ \gamma_{4} X_{3,22}
    + \gamma_{4} X_{3,24}+ \gamma_{4} X_{3,25}+ \gamma_{4} X_{3,26}+ \gamma_{4} X_{3,27}\\ &+ \gamma_{4} X_{3,28}+ \gamma_{4} X_{3,29}
    + \gamma_{5} X_{3,39}+ \gamma_{5} X_{3,40}+ \gamma_{5} X_{3,41}
    + \gamma_{5} X_{3,43}+ \gamma_{5} X_{3,44}+ \gamma_{5} X_{3,45}
    \\ &+ \gamma_{5} X_{3,55}+ \gamma_{5} X_{3,56}+ \gamma_{5} X_{3,57}
    + \gamma_{5} X_{3,59}+ \gamma_{5} X_{3,60}+ \gamma_{5} X_{3,61} + \boldsymbol{\beta}^{(x)} \boldsymbol{x}^{\prime}),
    \end{aligned}
\end{equation}
where the parameter values are set according to Table~\ref{tab:equation} depending on the specific experiment. The indicator variables $X_{r,s}$ are defined as laid out in Section~\ref{sec:response}. Equation~\eqref{eq:experiment} is constructed so that it matches the colour encoding of Figure~\ref{fig:3factfull}. For example, classes $h_{3,1}$ and $h_{3,2}$ have the same effect $\gamma_1+\gamma_2+\gamma_4$ on the response, while $h_{3,4}$ has a different effect, i.e.~$\gamma_1+\gamma_2-3\gamma_4$.
\begin{table}[H]
\centering
\begin{tabular}{m{9.1em} c c c c c c c c}
  \hline
  \rule{0pt}{17pt}
 &$\mu$&$\gamma_1$&$\gamma_2$&$\gamma_3$&$\gamma_4$&$\gamma_5$&$\boldsymbol{\beta}^{(x)}$&$g^-1(.)$\\[-2pt] 
\hline 
Normal distribution&&&&&&&&\\
\quad none&20&0&0&0&0&0&(0,0,0)&identity\\
\quad only $\boldsymbol{h}$ &20&0.6&0.4&0.3&0.3&0.2&(0,0,0)&identity\\
\quad both $\boldsymbol{h}$ and $\boldsymbol{x}$ &20&0.6&0.4&0.3&0.3&0.2&(6,0.8,0.5)&identity\\[5pt]
Poisson distribution&&&&&&&&\\
\quad none&0&0&0&0&0&0&(0,0,0)&exponential\\
\quad only $\boldsymbol{h}$ &0&0.6&0.4&0.3&0.3&0.2&(0,0,0)&exponential\\
\quad both $\boldsymbol{h}$ and $\boldsymbol{x}$ &0&0.6&0.4&0.3&0.3&0.2&(6,0.8,0.5)&exponential\\

\hline 
 
\end{tabular}
\caption{Parameter values for the conditional mean specified in Equation~\eqref{eq:experiment} for the simulation experiments in Section~\ref{sec:simulating}.}
\label{tab:equation}
\end{table}
%


\paragraph{Results simulation experiments}
Table~\ref{tab:seeds} summarises the results of the simulation experiment in Section~\ref{sec:simulexample} disaggregated for each initialisation seed of the neural network. 
\begin{table}[H] 
\centering
\begin{tabular}{m{11em} c c c c c}
  \hline
 &Seed 1&Seed 2&Seed 3&Seed 4&Seed 5 \\ 
\hline 
Normal distribution&&\\
\quad none&96&94&97&97&99\\
\quad only $\boldsymbol{h}$ &90&89&93&85&95\\
\quad both $\boldsymbol{h}$ and $\boldsymbol{x}$ &92&92&95&93&92\\[5pt]
Poisson distribution&&\\
\quad none&95&96&78&97&97\\
\quad only $\boldsymbol{h}$ &100&100&100&100&87\\
\quad both $\boldsymbol{h}$ and $\boldsymbol{x}$ &100&100&100&100&95\\

\hline 
 
\end{tabular}
\caption{Number of times the true structure according to Figure~\ref{fig:3fact_true} is retrieved for each set of initialisation values for the network parameters in the balanced simulation experiments.}
\label{tab:seeds}
\end{table}

\begin{table}[H]
\centering
\begin{tabular}{m{14em} m{3em} m{3em} m{3em}}
  \hline
  \rule{0pt}{17pt}
 &$q_e=1$& $q_e=2$&$q_e=3$\\[-2pt] 
\hline 
True structure retrieval rate&22\%&43.2\%&37.4\%\\
Different reduced structures&228&51&28\\

\hline 
 
\end{tabular}
\caption{Results of the unbalanced simulation experiment where 50 to 100 observations are simulated for each class of $h_R$ for $SI^*=0.7$. The first row shows how often (as a percentage) the true structure is recovered. The second row indicates the number of different structures. The results are shown for embedding dimension $q_e=1,2,3$.}
\label{tab:embedtune}
\end{table}

\begin{figure}[H] 
\centering
\subfloat[$r=1$]{\includegraphics[width = 0.5\textwidth]{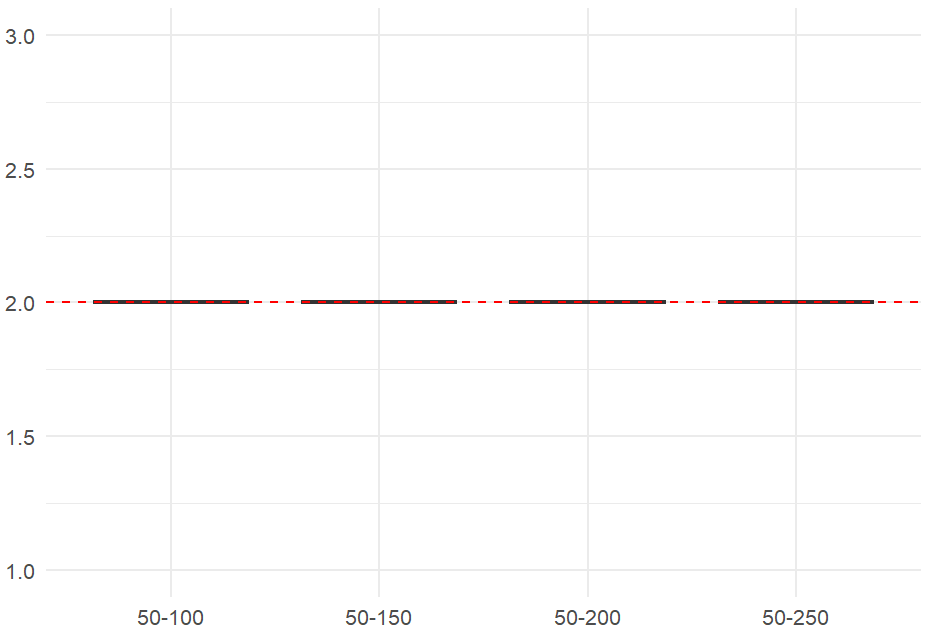}} 
\subfloat[$r=2$]{\includegraphics[width = 0.5\textwidth]{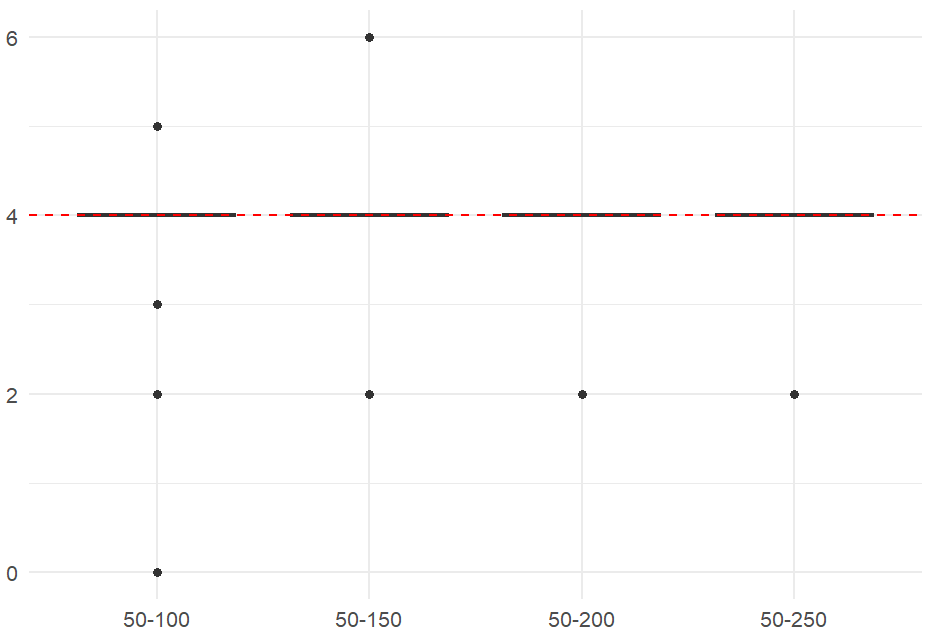}}\\
\subfloat[$r=3$]{\includegraphics[width = 0.5\textwidth]{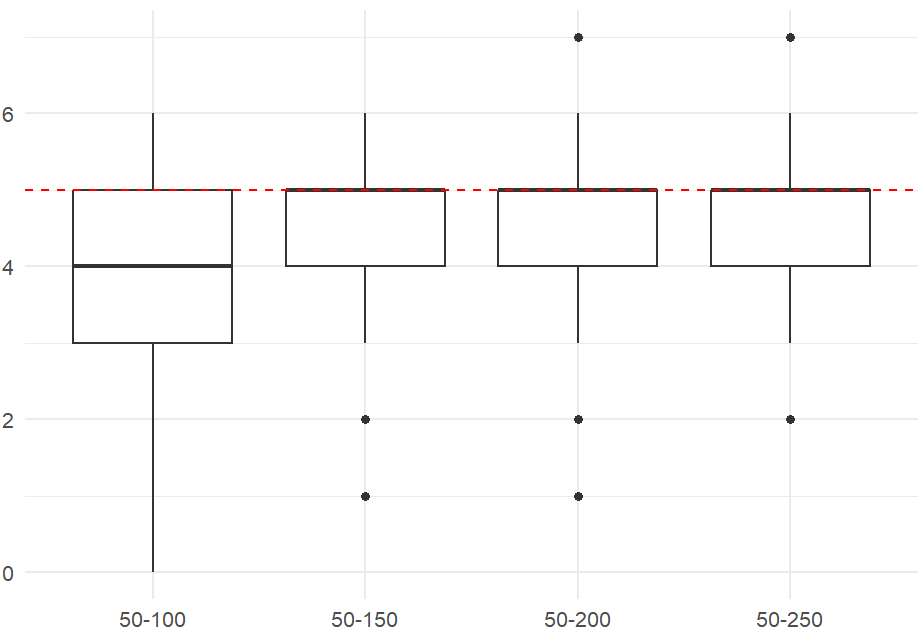}}
\caption{Boxplots of the number of classes in the reduced representation for the unbalanced simulation experiments in Section~\ref{sec:simulating}. For each of the three levels $r=1,2,3$, the red line indicates the number of classes in the true underlying structure, which are 2, 4 and 5, respectively. } 
\label{fig:boxplot}
\end{figure}
\begin{landscape}
    
\begin{figure}
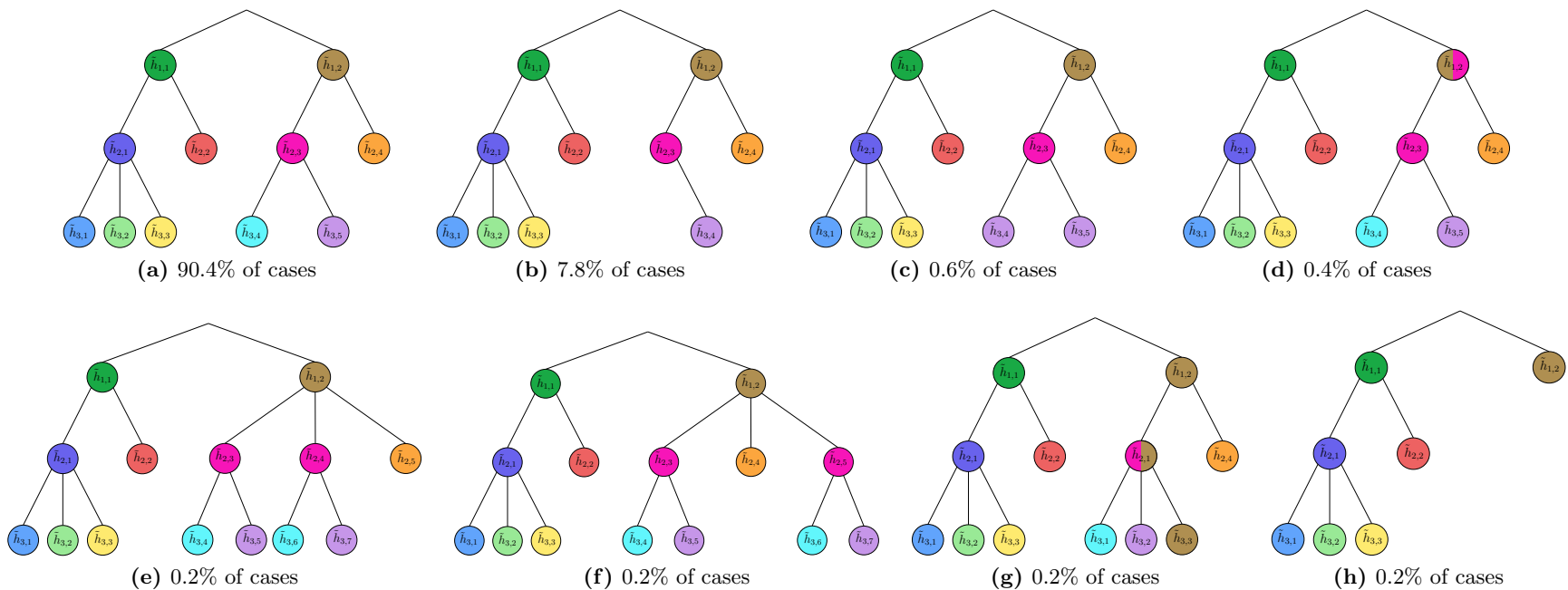

    \centering

    \subfloat[90.4\% of cases]{
\begin{adjustbox}{width=0.3\textwidth}



\end{adjustbox}
}
\caption{The different reduced structures obtained for the balanced experiment including only an effect of $\boldsymbol{h}$ on the normally distributed response. If classes of $\boldsymbol{h}$ with a different effect on the response are merged or collapsed together, the resulting class of $\boldsymbol{\widetilde{h}}$ is represented using multiple colours according to the colour encoding in Figure~\ref{fig:3factfull}.}
\label{fig:normalstructures}
\end{figure}
\end{landscape}
\section{Application to a real dataset: additional plots} \label{append:applic}

\begin{figure}[H] 
\centering 
\subfloat[Reduced representation preferred by $\text{AIC}$.]{\includegraphics[width = 0.8\textwidth]{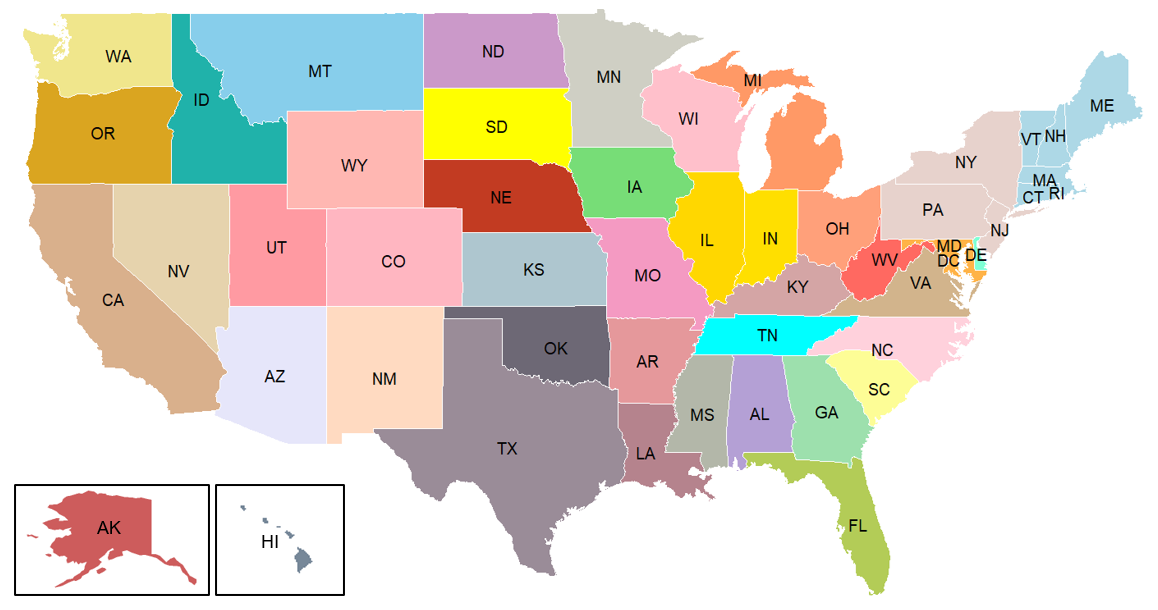}}\\
\subfloat[Reduced representation preferred by $\text{BIC}$.]{\includegraphics[width = 0.8\textwidth]{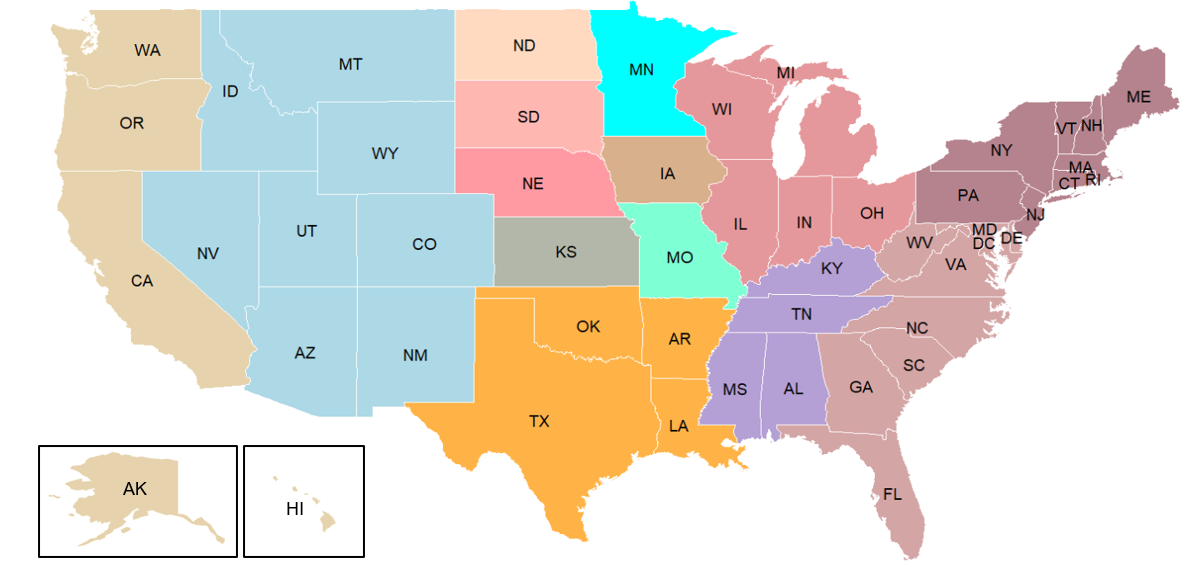}}
\caption{Map of the United States of America including all 50 states as well as the District of Columbia, states that are merged together according to the reduced structures preferred by the $\text{AIC}$ and $\text{BIC}$ proposed by \cite{carrizosa2022tree}, are visualised using the same colour. The upper figure pictures the structure preferred by the $\text{AIC}$ while the lower figure visualises the structure preferred by the $\text{BIC}$.} 
\label{fig:carri}
\end{figure}

\end{document}